\documentclass[12pt]{article}
\usepackage[margin=0.9in]{geometry}
\usepackage{amsthm,amsmath,amssymb,amscd,enumerate, graphicx,caption,subcaption,bm,booktabs,longtable,colortbl,array,makecell}
\usepackage{setspace,array}
\usepackage{thmtools}
\usepackage{thm-restate}
\usepackage{color}
\usepackage{xcolor}
\usepackage{natbib}
\usepackage[mathscr]{euscript}
\usepackage[shortlabels]{enumitem}
\usepackage[colorlinks=true, urlcolor=blue, linkcolor=blue, citecolor=blue]{hyperref}
\usepackage{algpseudocode}
\usepackage{algorithm}
\algrenewcommand\algorithmicindent{0.4em}
\usepackage{natbib}
\usepackage[T1]{fontenc}
\usepackage[utf8]{inputenc}
\usepackage{authblk}

\declaretheorem[name=Theorem]{theorem}

\DeclareMathOperator*{\argmax}{argmax}
\DeclareMathOperator*{\argmin}{argmin}
\DeclareMathOperator{\sign}{sgn}

\newcommand{\sms}[1]{{\textcolor{purple}}}

\title{Inference on summaries of a model-agnostic longitudinal variable importance trajectory with application to suicide prevention}
\author[1,2,3,*]{Brian~D. Williamson}
\author[4]{Erica~E.M. Moodie}
\author[5,6,7]{Gregory~E. Simon}
\author[8]{Rebecca~C. Rossom}
\author[1,3]{Susan~M. Shortreed}
\affil[1]{Biostatistics Division, Kaiser Permanente Washington Health Research Institute}
\affil[2]{Vaccine and Infectious Disease Division, Fred Hutchinson Cancer Center}
\affil[3]{Department of Biostatistics, University of Washington}
\affil[4]{Department of Epidemiology, Biostatistics, and Occupational Health, McGill University}
\affil[5]{Investigative Science Division, Kaiser Permanente Washington Health Research Institute}
\affil[6]{Department of Health Systems Science, Kaiser Permanente Bernard~J. Tyson School of Medicine}
\affil[7]{Department of Psychiatry and Behavioral Sciences, University of Washington}
\affil[8]{Division of Research, HealthPartners Institute}
\affil[*]{Corresponding Author: Biostatistics Division, Kaiser Permanente Washington Health Research Institute, 1730 Minor Ave Ste 1600, Seattle, WA 98101. Email: brian.d.williamson@kp.org}

\begin{document}

\maketitle

\begin{abstract}
Risk of suicide attempt varies over time. Understanding the importance of risk factors measured at a mental health visit can help clinicians evaluate future risk and provide appropriate care during the visit. In prediction settings where data are collected over time, such as in mental health care, it is often of interest to understand both the importance of variables for predicting the response at each time point and the importance summarized over the time series. Building on recent advances in estimation and inference for variable importance measures, we define summaries of variable importance trajectories and corresponding estimators. Under common regularity conditions, the same approaches for inference can be applied to these measures regardless of the choice of the algorithm(s) used to estimate the prediction function under standard convergence conditions. We propose a nonparametric efficient estimation and inference procedure as well as a null hypothesis testing procedure that are valid even when complex machine learning tools are used for prediction. Through simulations, we demonstrate that our proposed procedures have good operating characteristics. We use these approaches to analyze electronic health records data from two large health systems to investigate the longitudinal importance of risk factors (i.e., the importance of risk factors over time) for suicide attempt to inform future suicide prevention research and clinical workflow.

  \begin{center}{\small \textbf{Keywords:} intrinsic variable importance; longitudinal data; machine learning; prediction; risk factors of self-harm; suicide prevention}.\end{center}
\end{abstract}

\doublespacing

\section{Introduction}\label{sec:intro}

It is medically important to understand how different clinical data contribute to accurately predicting risk of self-harm or suicidal behavior resulting from statistical or machine learning models \citep{simon2018,shortreed2023}, as it can help clinicians understand why an individual is estimated to be at highest risk and guide the collection of impactful data for understanding risk and identifying individuals for additional interventions. Several research teams have developed or validated, and some health systems have implemented, suicide risk prediction models based on health records data \citep[see, e.g.,][]{jacobs2010,kessler2015predicting,kessler2017predicting,barak2017predicting,walsh2018,simon2018,gradus2020prediction,chen2020predicting,sanderson2020predicting,zheng2020development,tsui2021natural,penfold2021predicting,walker2021evaluation,coley2021racial,haroz2021designing,bayramli2022temporally,rossom2022connecting,nock2022prediction,shaw2022validating,matarazzo2023veterans,shortreed2023,papini2023performance,papini2024validation,haroz2024performance,adams2024developing}.

Models developed in a range of health care systems and patient populations have all shown overall area under the receiver operating characteristic curve approaching or exceeding 85\%. Data types used in those models include encounter or billing diagnoses, medications prescribed, use of specific mental health services, and self-report questionnaires completed during routine care, such as the patient health questionnaire 9-item depression questionnaire (PHQ-9) \citep{kroenke2001}. The total of the first eight items of the PHQ-9 can be summed for a measure of depressive symptoms (PHQ-8) and the ninth item is often used as a separate measure of suicidal ideation (PHQi9), which has been identified as a strong predictor of suicide attempt (fatal and non-fatal) \citep{kroenke2001,kroenke2010patient,simon2013}, and that those who repeatedly endorse suicidal ideation are at the highest risk of suicide attempt \citep{simon2016risk}. These risk prediction models can be used at a health care visit to alert the clinician to conduct additional risk assessments and provide appropriate care \citep{rossom2022connecting}.

While all of those categories of clinical information contribute to overall accuracy of prediction, the importance of any specific data type or data element may vary across time, being more or less important or influential if recorded more recently or more remotely. That variation in prediction importance is of interest for both theoretical reasons, to identify more variable and more stable risk factors for suicidal behavior; and practically, because specific types of clinical data may not be available for more recent or more remote time periods for some patients or in some settings. 

As part of prediction model development in many scientific areas, the impact of the predictors on prediction performance is often assessed \citep[see, e.g.,][]{gromping2015variable,wei2015}. There are typically two goals to these analyses: to make the model more interpretable \citep[see, e.g.,][]{murdoch2019} and to understand important predictors of the outcome \citep[see, e.g.,][]{williamson2021}. The latter notion of variable importance, while typically not explicitly causal, can provide some understanding of mechanisms that can aid in developing new interventions. Additionally, some variables may be either expensive (in time or monetary cost; e.g., a biomarker that is not measured as part of routine care) or possibly harmful to measure (e.g., causing pain; for example, endoscopic ultrasound). If such expensive or harmful variables are found to have low variable importance, or if the importance of these variables wanes over time, then a clinician can use this information when determining whether to measure these variables at a given point in the patient care trajectory.

In previous suicide risk prediction work, simple measures of variable importance have been reported: for example, the estimated coefficient from a lasso regression \citep{tibshirani1996} model \citep{simon2018,simon2019health,simon2024stability}; the decrease in accuracy \citep{vogt2024well} or impurity \citep{walsh2017predicting} measures from random forests \citep{breiman2001}; or the variables selected in a variable selection algorithm \citep{kessler2015predicting,kessler2017predicting}. These variable importance measures (both estimate and estimand) are fundamentally different from each other, sometimes depend on the scale of the variable, and are typically not accompanied by confidence intervals, rendering their interpretation across modeling strategy difficult. While these measures can be difficult to compare across algorithms, they are still useful towards understanding how a risk prediction algorithm makes use of the features in terms of either prediction performance (e.g., decrease in accuracy) or structure (e.g., regression coefficient). It is of further scientific interest to understand the trajectory of the importance of the PHQi9 in predicting risk of suicide or self-harm to guide its use in clinical care \citep[see, e.g.,][]{simon2016risk,simon2017between,simon2024stability}. In this paper, we propose a model-free method for quantifying and summarizing variable importance measures (VIMs) over time and an approach for developing confidence intervals for the estimated longitudinal VIMs. This method provides complementary information to the algorithm-specific VIMs described above.

Most existing variable importance approaches for longitudinal data are motivated by the causal inference literature. 
Several approaches exist that do not admit statistical inference; further, these approaches are only interpretable in the presence of a treatment or exposure variable, defining the importance of a variable by whether it can be used to select an optimal treatment \citep{gunter2011,fan2016,wu2022,ma2023}. One approach for assessing the importance of these tailoring variables does yield statistical inference on importance \citep{bian2021}, but relies on parametric assumptions. More general approaches that do not rely on parametric assumptions and yield inference are defined only for a single measure of variable importance and are valid under standard causal inference assumptions \citep{vanderlaan2006}. \citet{hubbard2013} and \citet{diaz2015} proposed measures for continuous and binary outcomes that can be interpreted as the mean difference in the outcome given an increase of $\delta > 0$ in the covariate of interest at a specified time point. Other approaches are based on the conditional average treatment effect, defined as $E(Y^1 - Y^0 \mid X = x)$, where $Y^a$ denotes the counterfactual outcome that would be observed if treatment $A$ were set to the value $a \in \{0,1\}$ and $X$ is a vector of pre-treatment covariates \citep{robins2008,chen2017,hines2022,boileau2022}. In general prediction settings, where there is no explicit treatment or exposure variable, these approaches require the analyst to define meaningful levels of a covariate of interest to remain interpretable. Finally, the relatively simple approach of fitting a longitudinal prediction model using generalized estimating equations \citep{liang1986longitudinal} or generalized linear mixed models \citep{breslow1993approximate} is unsatisfactory due to the reliance on parametric models for either the outcome-covariate relationship or covariance structure (or both).

Here, we focus on longitudinal VIM inference which can be implemented in the setting of either prediction or treatment effect estimation. Our proposed method provides inference at individual time points as well as on well-defined single-number summaries over longitudinal VIMs. The VIM at each timepoint can be defined in several ways \citep{williamson2021}, including the difference in the mean outcome value, difference in $R^2$, or difference in classification accuracy. 
Given this cross-sectional VIM, we then define population-level summaries of the variable importance trajectory over time, facilitating the development of valid confidence intervals and hypothesis tests even if flexible machine learning tools are used in estimating the prediction functions and the data are correlated over time. This framework allows us to investigate the trajectory of the importance of the PHQi9 for predicting suicide attempt; we present the VIM results in this article.

The remainder of the manuscript is organized as follows. In Section~\ref{sec:methods}, we describe our proposed class of longitudinal VIM summary measures and procedures for valid inference. We evaluate the performance of our proposed methods in numerical experiments in Section~\ref{sec:sims}. We conclude by identifying and summarizing longitudinal VIMs for the PHQi9 and other predictors of suicide attempt in Section~\ref{sec:data}. Technical details, including proofs of theorems, and additional results are provided in the Supplementary Material.

\section{Summarizing a variable importance trajectory}\label{sec:methods}

\subsection{Data structure and notation}

Suppose we observe longitudinal data  $L_1, \ldots, L_T$, where $T$ denotes the total number of observation time points. For a given participant, let $\mathbf{L} := (L_1, \ldots, L_T) \sim P_0$, allowing data for a given participant to  be correlated across time. We observe $n$ iid copies $\mathbf{L}_1, \ldots, \mathbf{L}_n$ and the $i$th observation at time $t$ is $\mathbf{L}_{i,t}$. At each time point $t$, we observe variables $\mathbf{X}_t$ and outcome $\mathbf{Y}_t$; i.e., $\mathbf{L}_t := (\mathbf{X}_t, \mathbf{Y}_t) \sim P_{0,t}$. 
We define a \textit{base set} of variables, denoted $\mathbf{W}_t \subset X_t$, to be a set of variables (possibly the null set) that we will always include in the outcome model. 
Then let $\mathbf{S}_t=\{X_t\} \setminus \{\mathbf{W}_t\}$ be the variables for which we have interest in estimating VIMs.
In our motivating example, age and sex could be our base set, information readily available in most settings. We could then focus on estimating the VIM of the PHQ-8, PHQi9, and prior diagnoses of self-harm, evaluating how these VIMs change over time. 

The definition of the variables and the set of variables considered for prediction at each timepoint determines the interpretation of any resulting VIMs. In addition to considering the type of predictors included in the variable set, the amount of historical information to include is an important consideration. In some prediction settings, only information collected at the given time point is used. In other settings, historical information may be part of the data at each time point: for example, separate variables could capture the number of times a person reported a PHQi9 of 3 in the past 6 months, 1 year, or 5 years. Each of these settings can be analyzed using the approach we describe here. 

Our work applies both to general prediction settings and settings with a treatment or exposure variable of special interest. When estimating the VIM for variables which might inform treatment selection, we further subdivide $\mathbf{L_t}$ into $(\mathbf{X}_t, \mathbf{A}_t, \mathbf{Y}_t)$, where $\mathbf{A}_t$ denotes observed treatment assignment or exposure at time $t$. We could then, for example, define the base set, $\mathbf{W}_t$, to be confounders of the exposure-outcome relationship, and $\mathbf{S}_t$ could be a set of potential tailoring variables for treatment. Under traditional causal inference assumptions (consistency, non-interference, and no unmeasured confounders \citep{rosenbaum1983}) all results of this work apply to estimation and inference of the impact of tailoring variables. For ease of notation and to improve the focus of the manuscript, we focus on the prediction setting from here on. We revisit the task of estimating longitudinal importance of tailoring variables in the discussion.  

We further define $s$, the set of indices corresponding to the variable(s) for which it is of scientific interest to make inference on variable importance. For a vector $\mathbf{x} \in \mathbb{R}^p$, we use the notation $\mathbf{x}_{p,t} \in \mathbb{R}^p$ to refer to the entire vector; $\mathbf{x}_{w,t} \in \mathbb{R}^{\lVert W_t \rVert}$ refers to the variables in the base set alone; $\mathbf{x}_{s \cup w, t} \in \mathbb{R}^{\lVert s \rVert + \lVert W_t \rVert}$ refers to the base set and the variables with index in $s$; and $\mathbf{x}_{p \setminus s, t} \in \mathbb{R}^{p - \lVert s \rVert}$ refers to the variables with index not in $s$. We use $\lvert \cdot \rvert$ as the absolute value and $\lVert \cdot \rVert$ to measure the length of a vector. We define rich classes of functions $\mathcal{F}_{p,t}$ that makes use of all variables at timepoint $t$; a subset $\mathcal{F}_{p \setminus s, t}$ that does not make use of the variables with index in $s$ at timepoint $t$; a subset $\mathcal{F}_{s \cup w, t}$ that makes use of the variables with index in $s$ and the base set variables; and a subset $\mathcal{F}_{w,t}$ that makes use of the base set alone. 

\subsection{Variable importance trajectories}
In a cross-sectional setting, a VIM provides a score that can be used to rank or measure how valuable a given feature is to accurately predicting the health outcome. Variable importance may be \textit{extrinsic}, capturing the contribution of variables to a particular fitted algorithm \citep[see, e.g.,][]{lundberg2017,fisher2019}; or \textit{intrinsic}, a quantification of the population prediction potential of a variable or group of variables \citep{williamson2021} that is not tied to a particular algorithm. Examples of intrinsic VIMs include the impact of adding a variable into an algorithm on $R^2$ values or area under the receiver operating curve (AUC). Both $R^2$ and AUC do not rely on distributional assumptions and are defined when flexible algorithms (e.g., machine learning tools) are used in estimation and inference \citep{vanderlaan2006,lei2017,williamson2020a,williamson2021}. We focus on intrinsic VIMs that describe population quantities, allowing the development of confidence intervals with correct coverage and hypothesis tests with correct type I error rates.

\begin{figure}
  \centering
  \includegraphics[width=4in]{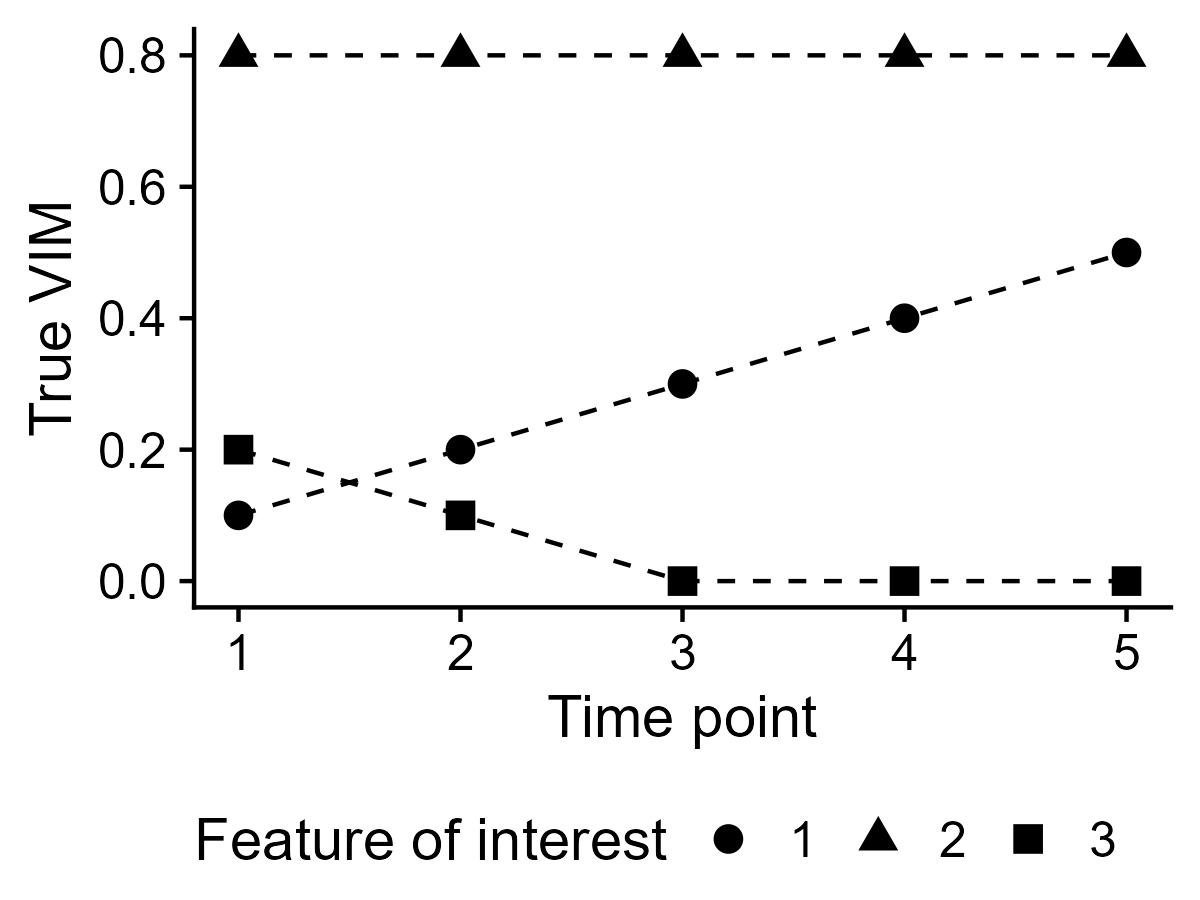}
  \caption{Example variable importance measure (VIM) trajectories for three hypothetical features of interest over five points in time.}
  \label{fig:lvim_example}
\end{figure}

In a longitudinal setting, VIMs may be calculated at each time point separately and summarized across timepoints using longitudinal measures. Consider, for example, the hypothetical example given in Figure \ref{fig:lvim_example}, with three features: PHQi9, age, and prior self-harm. Qualitatively, variable 1 has low initial importance but increasing importance over time, variable 2 has a large and constant importance over time, and variable 3 is useful at the start of the trajectory but quickly loses predictive importance. It is useful to quantitatively summarize the trajectories of VIMs for these variables and conduct inference on these quantities to understand the general importance of collecting and using a variable to model the outcome over time, distinguishing true differences in predictive power from expected sampling variability. In this example, not collecting variable 3 after timepoint two could conserve resources, as it no longer impacts prediction performance. A simple example is the mean VIM value over the time series; this measure captures some information but may not serve to fully characterize the importance of each variable, particularly if the importance exhibits a strong temporal pattern (e.g., variable 3). In the next subsection, we provide several examples of possible (longitudinal) summary measures, $m : \mathbb{R}^{T} \mapsto \mathbb{R}$, of VIM trajectories. We first make concrete the notions of variable importance on which we rely for the remainder of this paper.

For a given prediction function $f_t$ defined at time point $t \in \{1, \ldots, T\}$ and a data-generating mechanism $P_t$ corresponding to the same time point, where as above $P_t$ can include historical information, we define $V(f_t, P_t)$ as a measure of the predictiveness of $f_t$ under distribution $P_t$ (e.g., $R^2$ or AUC)  \citep{williamson2021} or a causal inference-inspired parameter, which describes the impact a variable has on the expected treatment outcome, such as the conditional average treatment effect \citep[see, e.g.,][]{diaz2015,hines2022}. The predictiveness-maximizing functions that make use of each set of variables, 
$\mathcal{F}_{p,t}$,  $\mathcal{F}_{p \setminus s, t}$,  $\mathcal{F}_{s \cup w, t}$, and $\mathcal{F}_{w,t}$ are:
\begin{align*}
  f_{0,p,t} \in& \ \argmax_{f \in \mathcal{F}_{p,t}} V(f, P_{0,t}), \ f_{0,p\setminus s, t} \in \ \argmax_{f \in \mathcal{F}_{p \setminus s, t}} V(f, P_{0,t}), \\
  f_{0, s \cup w, t} \in& \ \argmax_{f \in \mathcal{F}_{s \cup w, t}} V(f, P_{0,t}), \text{ and } f_{0, w, t} \in \ \argmax_{f \in \mathcal{F}_{w, t}} V(f, P_{0,t}),
\end{align*}
where the middle subscript refers to the set of variables the prediction function uses. These prediction functions are equivalent to the conditional mean outcome given covariates for many choices of $V$ \citep{williamson2021}. Based on these predictiveness-maximizing functions, we can define several values of predictiveness that will be useful in defining variable importance. The \textit{total predictiveness} $v_{0,p,t}:= V(f_{0,p,t}, P_{0,t})$ provides a measure of the total prediction potential under $P_{0,t}$. The \textit{residual predictiveness} $v_{0,p\setminus s, t}:= V(f_{0,p\setminus s, t}, P_{0,t})$ and the \textit{marginal predictiveness} $v_{0,s\cup w, t} := V(f_{0, s\cup w, t}, P_{0,t})$ are used to help quantify the prediction potential of the covariate features with index in $s$, through quantifying predictiveness of a model including all variables except those in $s$, or the predictiveness of a model using just those in $s$ and the base set of predictors. Finally, the \textit{irreducible predictiveness} $v_{0,w,t} := V(f_{0, w, t}, P_{0,t})$ quantifies the minimal prediction potential using the base set of variables $W$ alone; if $W$ is empty, then the irreducible predictiveness is equivalent to the prediction potential ignoring all features. In the Supplementary Material (Section~\ref{sec:simple_example}), we show both these prediction functions and their associated predictiveness values in a simple setting. 

We consider two definitions of the population-level variable importance at time $t$, which vary in terms of the maximal set of variables under consideration. In both cases, the predictiveness of models using the set of variables of interest is compared to the predictiveness of models not using the variables. The \textit{add-in} VIM is defined as the difference between the marginal and irreducible predictiveness, $\theta_{0, s, t} := v_{0,s\cup w,t} - v_{0,w,t}$. The \textit{leave-out} VIM is defined as the difference between the total and residual predictiveness, $\psi_{0,s,t} := v_{0,p,t} - v_{0, p\setminus s, t}$. The choice of VIM depends on the scientific goal: add-in importance is defined relative only to the base set, while leave-out importance is defined relative to all other variables. Add-in importance may be more scientifically relevant in settings where resources are limited and there are a small number of pre-established important covariates; a goal of estimating VIMs in this setting might be to determine if additional variables can improve prediction performance and if so by how much. In contrast, leave-out importance may be more pertinent when there is no clear base set and a prediction with many predictors can be implemented. Both of these VIMs provide complementary and useful information; they may be used together or separately, depending on the scientific goal. With these cross-sectional definitions of VIMs, trajectories may now be defined as the collection of predictiveness or VIM values over time. 

Both our theoretical results and the examples of longitudinal summaries that we discuss next apply to both the predictiveness values (total, residual, marginal, and irreducible) and VIM values (add-in and leave-out). As mentioned above, our interest is in inference on a summary measure $m: \mathbb{R}^T \mapsto \mathbb{R}$; this summary can be of either a predictiveness or VIM trajectory, since we define VIM as a contrast in predictiveness values. However, to simplify the presentation, we will focus our examples in the next subsection on the add-in VIM $\theta_{0,s,t}$ and a summary of the add-in VIM trajectory, $m_{0,p,\tau} := m(\theta_{0,s,\tau})$, where $\tau = [t_0, t_1]$ represents a contiguous subset of the time series. Each longitudinal summary that we consider is defined as a summary function applied to the cross-sectional VIM values.

\subsection{Examples of longitudinal summaries}\label{sec:examples}

Longitudinal summaries can be computed over the entire longitudinal trajectory or based on a subset of the timepoints. We provide here four examples of longitudinal summaries, and discuss the relative merits and utility of each of these different summaries. 

\noindent\textit{Example 1: mean over a contiguous subset of the time series.} For a given time series of add-in VIM values, $\{\theta_{0,s,t}\}_{t = 0}^T$, the mean value over the contiguous subset $\tau$ is defined as $\mu_{0,s,\tau} := \lVert \tau \rVert^{-1}\sum_{t \in \tau} \theta_{0,s,t}$. In the context of Figure~\ref{fig:lvim_example}, the mean VIM values over the entire time series are 0.3, 0.8, and 0.06, for variables 1, 2, and 3, respectively.

\noindent\textit{Example 2:  linear trend over a contiguous subset of the time series.} We can define the linear trend as the linear regression of $\theta_{0,s,t}$ on $\tau$, by defining the matrix
\begin{align*}
  U = \begin{bmatrix} 1 & 1 & \cdots & 1 \\
  \tau_{0,1} & \tau_{0,2} & \cdots & \tau_{0, \lVert \tau \rVert}\end{bmatrix}^\top.
\end{align*}
Set $\beta_0 = \argmin_{\beta \in \mathbb{R}^2} \lVert \theta_{0,s,\tau} - \beta_1 - t\beta_2 \rVert_2^2 = (U^\top U)^{-1}U^\top \theta_{0,s,\tau}$. Then $\beta_0$ can be interpreted as identifying the linear trend in the $\theta_{0,s}$ trajectory over the time period $\tau$. In the example of Figure~\ref{fig:lvim_example}, the slope of the linear add-in VIM trend over the entire time series for variables 1, 2, and 3 is 0.1, 0, and -0.05, respectively. Values of the slope summary measure that are close to zero do not necessarily indicate low variable importance; instead, they indicate a flat linear trend in variable importance.

\noindent\textit{Example 3: area under a subset of the trajectory curve.} To allow for a more flexible modelling of the trajectory function, we define $h: (t, \theta) \in (\tau \times \mathbb{R}) \mapsto \mathbb{R}$ that interpolates the VIM values $\{\theta_{0,s,t}\}_{t \in \tau}$. Two example trajectory functions are a piecewise linear interpolator and a spline interpolator. Then the area under the trajectory curve (AUTC) for the time period $\tau$ can be written as $\alpha_0 = \int_{t_0}^{t_1} h(t, \theta_{0,s}) dt$. Importantly, the interpretation of AUTC depends on the trajectory function class. The AUTC based on a piecewise linear interpolator over the entire add-in VIM trajectory for variables 1, 2, and 3 in Figure~\ref{fig:lvim_example} is 1.2, 3.2, and 0.2, respectively; based on a cubic spline interpolator, the AUTC is 1.2, 3.2, and 0.198, respectively. Unlike the area under the receiver operating characteristic curve, which lies in $[0,1]$, the AUTC takes value in $\mathbb{R}^+$. While larger values are indicative of greater VIM, it is not straightforward to compare AUTC values for variables with very differently shaped trajectory curves. We comment below in more depth on interpreting comparisons across different covariates.

\textit{Example 4: geometric mean rate of change over a subset of the trajectory curve.} As with the AUTC, suppose that the trajectory function $h$ interpolates the VIM values $\{\theta_{0,s,t}\}_{t \in \tau}$. Suppose further that $h$ is differentiable, defining its derivative with respect to $t$ by $h'$. Then the geometric mean rate of change (GMRC) over the time period $\tau$ can be written as $\gamma_0 = \exp\left[\lVert \tau \rVert^{-1}\sum_{t \in \tau}\log \{\lvert h'(t, \xi_{0,c}) \rvert\}\right]$. As with the AUTC, the interpretation of $\gamma_0$ depends on the trajectory function class. The GMRC based on a piecewise linear interpolator over the entire add-in VIM trajectory for variables 1, 2, and 3 in Figure~\ref{fig:lvim_example} is 0.1, 0, and 0, respectively; based on a cubic spline interpolator, the GMRC is 0.1, 0, and 0.035, respectively. The GMRC, which takes values in $\mathbb{R}^+$, can be interpreted as a nonlinear summary of the trajectory \citep{link1997}, as opposed to the slope of the linear trend.

Each of the VIM summaries listed above can be computed for each variable in the set $S$, although we note that the AUTC and GMRC (Examples 3 and 4) additionally depend on the given interpolator function. In many applications, it may be desirable to use these summaries to compare (or rank) VIMs over the time series. These different example summaries can provide complementary information, and as with most longitudinal summary measures they should be used in conjunction with each other. While the mean VIM and linear trend are easily compared across variables, it is likely that each alone is not fully informative: a zero slope of the linear trend can indicate that the VIMs do not change over time, but the magnitude of that importance is critical. In contrast, while the AUTC and GMRC are in general difficult to compare across all variables, unless the trajectory curves being compared are all monotone, they can be useful summary measures of the general importance of a variable. For example, in Figure~\ref{fig:lvim_example}
variable 1 will have the largest AUTC, which reflects that it consistently has the largest variable importance. The varying importance of variable 3 means that using multiple summaries is essential for appropriate interpretation. In particular, when comparing VIMs across variables, we recommend testing for monotonicity of the trajectory for each variable first. Then the mean, AUTC, and GMRC can be used to further summarize the variable's impact over the entire time period. Note that the test of monotonicity involves checking to see if the first derivative of the interpolator is always the same sign. If so, then the sign of the slope of the linear trend in the VIM trajectory can be used to describe the trajectory as increasing if the slope is positive and decreasing otherwise.

As a final point, the choice of summary measures should depend on the scientific purpose of the analysis under question. For example, in a clinical setting where resources are often highly constrained, researchers may want to choose only to measure a small number of variables with high average predictiveness, regardless of trend, or may choose to measure variables whose VIM decreases with time only early on, dropping those variables at later times when the burden of data collection is not balanced by the improvement in predictiveness.
 
\subsection{Estimation and inference}

Our goal is to use the data $\mathbf{L}_{1,\tau}, \ldots, \mathbf{L}_{n,\tau}$ to make inference on a summary of the VIM trajectory. Suppose that at each time point, we obtain prediction functions $f_{n,p,t}$, $f_{n,p\setminus s, t}$, $f_{n,s\cup w, t}$, and $f_{n,w,t}$; these are estimators of the true predictiveness-maximizing functions $f_{0,p,t}$, $f_{0,p\setminus s, t}$, $f_{0,s\cup w, t}$, and $f_{0,w,t}$, respectively. Here, the subscript $n$ serves as a reminder that these estimators of the population truth (denoted with a subscript 0) are obtained on our sample of size $n$. For many choices of $V$, these will be estimators of the population conditional mean outcome given covariates. We further obtain plug-in estimators $v_{n,p,t}:=V(f_{n,p,t},P_{n,t})$, $v_{n,p\setminus s, t}:=V(f_{n,p\setminus s, t},P_{n,t})$, $v_{n,s\cup w, t} := V(f_{n,s\cup w, t}, P_{n,t})$, $v_{n,w,t}:= V(f_{n,w,t}, P_{n,t})$, $\psi_{n,s,t}:= v_{n,p,t} - v_{n,p\setminus s, t}$ and $\theta_{n,s,t}:=v_{n,s\cup w, t} - v_{n,w,t}$ of the total predictiveness $v_{0,p,t}$, the residual predictiveness $v_{0,p\setminus s, t}$, the marginal predictiveness $v_{0,s\cup w, t}$, the irreducible predictiveness $v_{0,w,t}$, the add-in VIM $\theta_{0,s,t}$, and the leave-out VIM $\psi_{0,s,t}$, respectively. 

We focus our presentation of theoretical results on an estimator $m_{n,p,\tau} = m(v_{n,p,\tau})$ of the true population measure $m_{0,p,\tau} = m(v_{0,p,\tau})$, the summary of the total predictiveness trajectory over time. As in the definition of the true longitudinal VIM summary, the estimator is defined by estimating the cross-sectional VIM at each timepoint and applying the summary function of interest. All results can be readily extended to summaries of the residual, marginal, and irreducible oracle predictiveness, and thus to the add-in and leave-out VIM values, by modifying the technical conditions provided in the Supplementary Material (Section~\ref{sec:proofs}) to apply to the necessary collection of variables for each parameter. 

We first demonstrate that $m_{n,p,\tau}$ is a consistent and efficient estimator of $m_{0,p,\tau}$, under conditions on the data-generating process, function classes $\mathcal{F}_{p,t}$, and estimators of the prediction functions $f_{0,p,t}$. We list the detailed conditions in Section~\ref{sec:proofs}. Conditions (A1)--(A4) and (B1)--(B3) were presented in \citet{williamson2021} for a single time point. For the results in this paper, we require them to hold at each time point. Further, the new condition (A5) restricts the types of longitudinal summaries that can be considered. All examples of longitudinal summaries examined here (Examples 1--4) satisfy this condition (proof in Section~\ref{sec:more_theory}). A key object in the study of $m_{n,p,\tau}$ is its influence function $\varphi_{0,p,\tau}$, which is a function of the matrix of the time point-specific influence functions of $v_{n,p,t}$, denoted $\begin{bmatrix}\phi_{0,p,t},(\mathbf{L}_{t,i})\end{bmatrix}_{t \in \tau},$ and $\dot{m}(v_{0,p,\tau})$, the gradient of $m$ evaluated at $v_{0,p,\tau}$.

The main theorem presented below involves generalizing the theory in \citet{williamson2021} to allow the outcomes to be correlated over time (both unconditionally and conditionally on covariates); we prove the generalized result in Section~\ref{sec:proofs}. 

\begin{theorem}\label{thm:summary}
  If conditions (A1), (A2), (A5), and (B1)--(B3) provided in the Supplementary Material hold, then for $\tau = [t_0, t_1]$, $m_{n,p,\tau}$ is an asymptotically linear estimator of $m_{0,p,\tau}$ with influence function equal to $\varphi_{0,p,\tau} : (\ell_{t_0}, \ldots, \ell_{t_1}) \mapsto \dot{m}(v_{0,p,\tau}) \left[\phi_{0,p,t}(\ell_t)\right]_{t \in \tau}$, that is, 
  \begin{align*}
    m_{n,p,\tau} - m_{0,p,\tau} = \frac{1}{n}\sum_{i=1}^n \dot{m}(v_{0,p,\tau})\left[\phi_{0,p,t}(\mathbf{L}_{t,i})\right]_{t \in \tau} + o_P(n^{-1/2})
  \end{align*}
under sampling from $P_0$. If conditions (A3) and (A4) also hold, then $\varphi_{0,p,\tau}$ coincides with the nonparametric efficient influence function (EIF) of $P_0 \mapsto m\{V(f_{P_{t_0}}, P_{t_0}), \ldots, V(f_{P_{t_1}}, P_{t_1})\}$ at $P_0$, and so, $m_{n,p,\tau}$ is nonparametric efficient.
\end{theorem}

We provide a proof of this result and describe the technical conditions in Section~\ref{sec:proofs}. Conditions (A1)--(A4) have been verified previously to hold for timepoint-specific VIM values based on $R^2$, classification accuracy, and AUC at each time point \citep{williamson2021}. Condition (A5) relates to the smoothness of $m$, and holds for each longitudinal summary measure proposed in Section~\ref{sec:examples}. Together, the conditions guarantee the existence of the influence function $\varphi_{0,p,\tau}$. 
Conditions (B1) and (B2) describe the required convergence rate of $f_{n,p,t}$ to $f_{0,p,t}$; these conditions are satisfied for generalized additive models \citep{hastie1990} and hold approximately for more flexible estimators, including random forests \citep{breiman2001}, with ignorable first-order bias in simulations \citep{williamson2020a,williamson2021}. A similar theorem applies to summaries of each type of predictiveness, and thus to summaries of VIM values. 

We can remove the need for condition (B3) by using a $K$-fold cross-fitting procedure. This relaxation allows more flexible (e.g., machine learning-based) algorithms to be used to estimate $f_{0,p,t}$. Briefly, this involves splitting the data into $K$ folds. Then, for each $k \in \{1, \ldots, K\}$, the data in fold $k$ are used as a held-out test set for estimating predictiveness, while an estimator $f_{n,p,t,k}$ of $f_{0,p,t}$ is obtained on the remaining data. The cross-fitted predictiveness estimator is $v_{n,p,t}^* := K^{-1}\sum_{k=1}^K V(f_{n,p,t,k}, P_{n,t,k})$, resulting in cross-fitted summary measure $m_{n,p,\tau}^* := m(v_{n,p,t}^*)$. Our second result demonstrates that our cross-fitted estimator $m_{n,p,\tau}^*$ is a consistent and efficient estimator of $m_{0,p,\tau}$ under these relaxed conditions. 

\begin{theorem}\label{thm:cv_summary}
  If conditions (A1), (A2), (A5), and (B1') and (B2') hold, then for $\tau = [t_0, t_1]$, $m_{n,p,\tau}^*$ is an asymptotically linear estimator of $m_{0,p,\tau}$ with influence function equal to $\varphi_{0,p,\tau} : (\ell_{t_0}, \ldots, \ell_{t_1}) \mapsto \dot{m}(v_{0,p,\tau}) \left[\phi_{0,p,t}(\ell_t)\right]_{t \in \tau}$, that is, under sampling from $P_0$,
  \begin{align*}
    m_{n,p,\tau}^* - m_{0,p,\tau} = \frac{1}{n}\sum_{i=1}^n \dot{m}(v_{0,p,\tau})\left[\phi_{0,p,t}(\mathbf{L}_{t,i})\right]_{t \in \tau} + o_P(n^{-1/2}).
  \end{align*}
If conditions (A3) and (A4) also hold, then $\varphi_{0,p,\tau}$ coincides with the nonparametric EIF of $P_0 \mapsto m\{V(f_{P_{t_0}}, P_{t_0}), \ldots, V(f_{P_{t_1}}, P_{t_1})\}$ at $P_0$, and so, $m_{n,p,\tau}^*$ is nonparametric efficient.
\end{theorem}

The proof of this result and discussion of the modified technical conditions are provided in Section~\ref{sec:proofs}. The same cross-fitting procedure can be used to obtain a summary of each other predictiveness measure (residual, marginal, and irreducible), and a similar theorem applies to these cross-fitted values, along with the summary of the VIM values. Under the collection of all such conditions, then the cross-fitted add-in VIM summary estimator $m(\theta^*_{n,s,\tau})$ is an asymptotically linear estimator of $m(\theta_{0,s,\tau})$ with influence function $\varphi_{0,s} : \ell \mapsto \dot{m}(v_{0,s\cup w,\tau}) \left[\phi_{0,s\cup w,t}(\ell_t)\right]_{t \in \tau} - \dot{m}(v_{0,w,\tau}) \left[\phi_{0,w,t}(\ell_t)\right]_{t \in \tau}$. As before, under a simple modification, the leave-out VIM estimator $m(\psi^*_{n,s,\tau})$ of $m(\psi_{0,s,\tau})$ is also asymptotically linear. Cross-fitting may result in negative VIM estimates; in these cases, the true VIM is likely near zero (or equal to zero).

If the add-in VIM $\theta_{0,s,t} > 0$ for at least one $t \in \tau$ and $0 < \sigma_{0,s}^2 := E_0\{\varphi_{0,s}^2(L)\} < \infty$, then the asymptotic variance of $n^{-1/2}\{m(\theta_{n,s,\tau}) - m(\theta_{0,s,\tau})\}$ can be estimated by 
\begin{align*}
    \sigma^2_{n,s} := \frac{1}{n}\sum_{i=1}^n\left\{\dot{m}(v_{n,s\cup w,\tau}) \left[\phi_{n,s\cup w,t}(\mathbf{L}_{i,t})\right]_{t \in \tau} - \dot{m}(v_{n,w,\tau}) \left[\phi_{n,w,t}(\mathbf{L}_{i,t})\right]_{t \in \tau}\right\}^2,
\end{align*}
where the appropriate components of the EIF are replaced with their estimators, implying that $(m(\theta_{n,s,\tau}) - z_{1 - \alpha/2}\sigma_{n,s}n^{-1/2}, m(\theta_{n,s,\tau}) + z_{1 - \alpha/2}\sigma_{n,s}n^{-1/2})$ is a valid $(1-\alpha)$ confidence interval for $m(\theta_{0,s,\tau})$, where $z_{1 - \alpha/2}$ denotes the $(1 - \alpha/2)$ quantile of the standard normal distribution. However, just as cross-fitting can improve estimation of $m(\theta_{0,s,\tau})$, it can also improve estimation of $\sigma^2_{0,s}$. The same calculations apply to the leave-out VIM $\psi_{0,s,t}$. We recommend the use of these cross-fitted estimators whenever machine learning tools are used to estimate $f_{0,p,t}$. 
We provide the EIFs for the summary measures described above in Section~\ref{sec:more_theory} of the Supplementary Materials.
Our R package \texttt{lvimp} \citep{lvimppkg} implements the point and variance estimators described above.

When $\theta_{0,s,t} = 0$ for some $t \in \tau$, in which case the variable group has zero importance at the given time point, the influence function of $\theta_{n,s,t}$ (or $\psi_{n,s,t}$) is zero. Since our interest is in a summary over the time series, this only presents an issue if $\theta_{0,s,t} = 0$ (or $\psi_{0,s,t} = 0$) for \textit{all} $t \in \tau$. To guard against this case, we adopt an approach to hypothesis testing based on sample-splitting as advocated for in \citet{williamson2021} that is valid under the case of zero VIM. The same considerations apply to the leave-out VIM $\psi_{0,s,\tau}$. This sample-splitting procedure involves estimating the total predictiveness $v_{n,p,t}$ on independent data from the residual predictiveness $v_{n,p\setminus s,t}$, or equivalently estimating the marginal predictiveness $v_{n,s\cup w,t}$ on independent data from the irreducible predictiveness $v_{n,w,t}$. 
In small-sample settings or settings with small effective sample size (e.g., rare-outcome settings with moderate sample sizes), the use of sample-splitting may necessitate simpler estimators of the prediction functions $f$, e.g., generalized linear models, or selecting tuning parameters that induce substantial regularization. These modifications can mitigate the impact of the reduced information available to estimate the prediction functions.

Finally, in some cases --- particularly in settings with small samples or rare events, where between-sample variability is expected to be large, or with true VIM near zero --- the VIM point estimate may be negative. In these cases, a sensible strategy is to report an estimated VIM of zero; this approach is also common in the random forest variable importance literature \citep[see, e.g.,][]{ishwaran2008}. We adopt this strategy in all results reported below.

\section{Numerical experiments}\label{sec:sims}

\subsection{Experimental setup}
We now present empirical results describing the performance of our proposed VIM longitudinal summary estimators. We consider $T=4$ timepoints, $p=10$ variables, and binary outcome $Y_t \in \{0,1\}$. In the data generating process $(X_{t,1},\ldots,X_{t,7})$ are predictive of the outcome at all time points (i.e., have VIM value greater than zero), while $(X_{t,8},X_{t,9},X_{t,10})$ are not predictive of the outcome (i.e., have VIM value equal to zero) at all time points. 
The outcome is generated with a probit model: 
  $P(Y_t = 1 \mid X_t = x_t) =  \Phi\left(\sum_{j=1}^p \beta_{t,j}x_{t,j}\right).$
We set $\beta_{t,1} = 0.1$ for all $t$; $\beta_{t,2} = 0.1 + t / 15$; $\beta_{t,3} = (-1/4) \{1 + \exp(-t)\}^{-1} + 0.1$, and $\beta_{t,j} = 0.01$ for $j \in \{4, \ldots, 7\}$. The remaining variables have zero VIM for each $t$: $\beta_{t,8:10} = \mathbf{0}_{3}$. 

We consider an autocorrelation structure between features over time, which induces correlation in the outcomes. At timepoint $t = 1$, we generate $(X_{t,4}, X_{t,5}, X_{t,6}, X_{t,7})$ following a multivariate normal distribution with mean zero and identity covariance matrix. To specify the distribution of $(X_{t,1},X_{t,2},X_{t,3})$, we use
\begin{align*}
  E\{X_{t,j} \mid (X_{t,4}, X_{t,5}, X_{t,6}, X_{t,7}) = x_t\} = 0.05x_{t}\mathbf{1}_4^\top
\end{align*}
and add multivariate normal random errors with mean zero and identity covariance matrix. This specification induces a small amount of correlation between variables $X_{t,1}$, $X_{t,2}$, and $X_{t,3}$. For $t \in \{2, 3, 4\}$, we generate each feature using an autocorrelation structure with correlation parameter $\rho = 0.5$. For $j \in \{4, 5, 6, 7\}$,
\begin{align*}
    X_{t,j} =& \ X_{t-1,j} \rho + \epsilon,
\end{align*}
where $\epsilon \stackrel{iid}{\sim} N(0,1)$. Then for $j \in \{1, 2, 3\}$,
\begin{align*}
    X_{t,j} =& \ X_{t-1,j}\rho + \epsilon_j \\
    E\{\epsilon_j \mid (X_{t,4}, X_{t,5}, X_{t,6}, X_{t,7}) = x_t\}\ =& \ 0.05x_t\mathbf{1}_4^\top,
\end{align*}
and the conditional variance of $\epsilon$ is 1. Variables $X_{t,8}$, $X_{t,9}$, and $X_{t,10}$ are always generated from a multivariate normal distribution with mean zero and identity covariance matrix. Based on this mechanism, the outcome proportion is approximately 16\% at each timepoint; the correlation between outcomes at contiguous timepoints is approximately 0.03, decaying to 0.01 one timepoint apart and near zero two timepoints apart.

\begin{figure}
    \centering
    \includegraphics[width=1\textwidth]{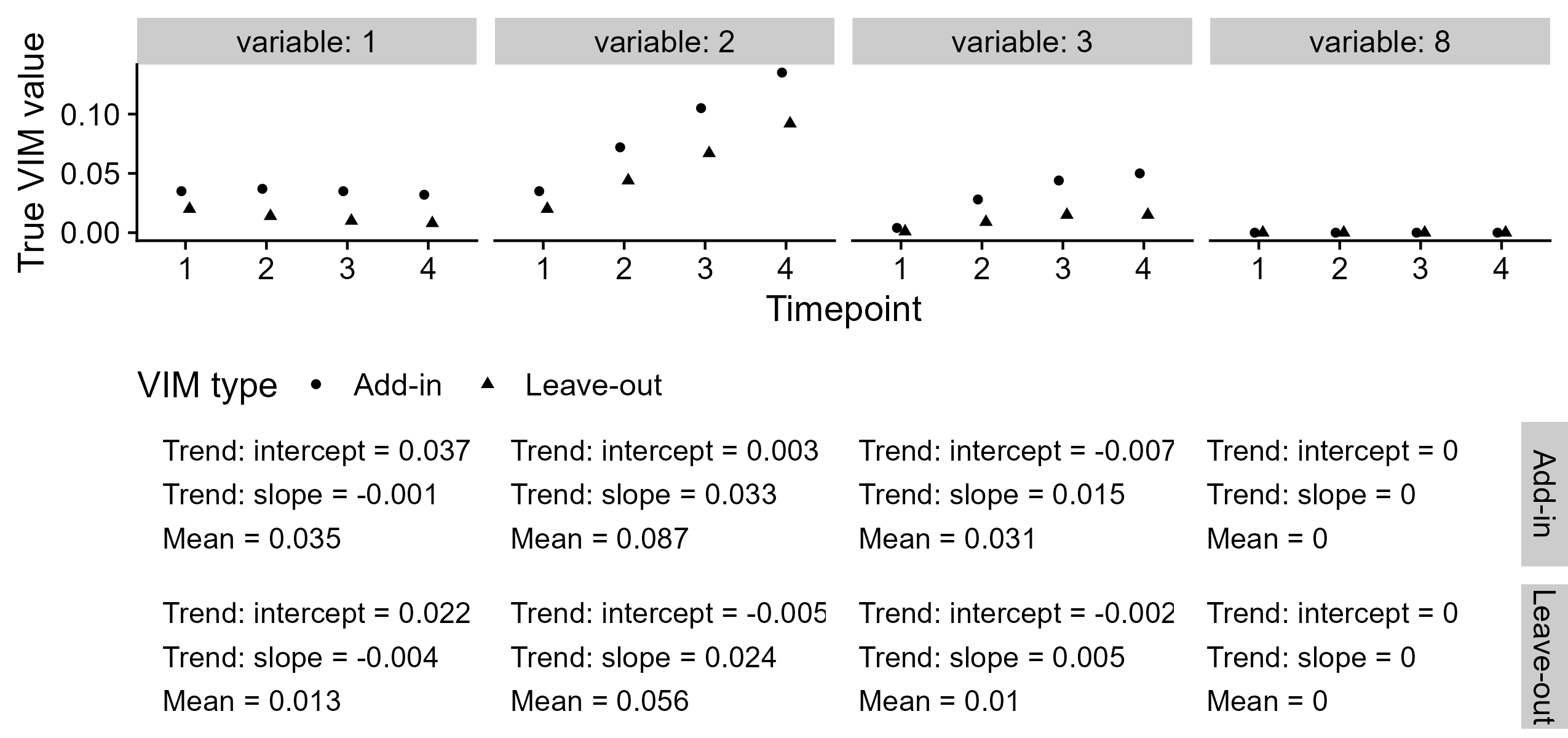}
    \caption{True variable importance values defined using AUC (both add-in and leave-out) at each timepoint and summarized over time for variables 1, 2, 3, and 8 based on the data-generating mechanism described in the simulation setup.}
    \label{fig:sim_true_vims_0.5_dgm_2}
\end{figure}

For each $n \in \{250, 1000, 5000\}$ (corresponding to expected number of events at each timepoint $\{40, 160, 800\}$, respectively), we generate 1000 random datasets according to this data-generating mechanism and consider both the add-in and leave-out importance of each variable in $S = \{1, 2, 3, 8, 9, 10\}$, the variable set of interest, fixing predictors $(X_{t,4},\ldots,X_{t,7})$ as a base set. As a reminder, in our set-up $(X_{t,1},X_{t,2},X_{t,3})$ have a VIM value greater than zero at all time points, while $(X_{t,8},X_{t,9},X_{t,10})$ have VIM value equal to zero for all $t$. 
We used AUC and positive predictive value (PPV) and sensitivity at the 95$^\text{th}$ percentile of risk to define predictiveness and considered the mean and linear trend to summarize the VIMs across the time series. The true values of the VIMs at each time point and the mean and linear trend over the time series are provided in Figure~\ref{fig:sim_true_vims_0.5_dgm_2} (Tables~\ref{tab:sim_true_vims_0.5_auc_dgm_2}--\ref{tab:sim_true_vims_0.5_sensitivity_dgm_2}; Tables~\ref{tab:sim_true_pred_0.5_auc_dgm_2}--\ref{tab:sim_true_pred_0.5_sensitivity_dgm_2} display the true predictiveness values). The leave-out VIM tends to be smaller than the add-in VIM, because it is defined relative to all other variables, including those in $S$, while add-in VIMs are compared to predictive models using only those variables in the base set. While the VIM trajectory for each variable is monotone, they are not all in the same direction: $X_1$ has a slight decrease in importance over time due to the correlation between variables $X_1$, $X_2$, and $X_3$; $X_2$ has increasing importance over the time series; and $X_3$ has nonlinearly increasing importance over time. The number of events at each timepoint in our observed data ranged from 170--400 (Section~\ref{sec:data}); we chose the data-generating mechanism above to show performance with both fewer events (40 when $n = 250$), equivalent numbers of events (for $n = 1000$), and more events (800 when $n = 5000$). For AUC in a rare-event setting, the number of events drives large-sample behavior, not the event rate \citep{minus2025behavior}. 

We used several methods to estimate the required prediction functions: logistic regression; logistic regression with lasso for variable selection \citep[][R package \texttt{glmnet}]{tibshirani1996}; random forests \citep[RF;][R package \texttt{ranger}]{breiman2001}; gradient boosted trees \citep[XGB;][R package \texttt{xgboost}]{friedman2001}; and a super learner \citep[SL;][R package \texttt{SuperLearner}]{vanderlaan2007}. The latter estimator is an ensemble of the former estimators, using cross-validation to determine the convex combination of the individual algorithms that minimizes the cross-validated negative log-likelihood. Several possible tuning parameters values were considered for each algorithm (details in Supplementary Material Table~\ref{tab:sl-algs}). In the simulated settings considered here, each of these regression methods can estimate $f_{0,p,t}$ consistently (provided that the more flexible methods are properly tuned); therefore, each provides a consistent estimator of the true VIM value. If a procedure is misspecified, then it will not provide a consistent estimator of $f_{0,p,t}$, leading to biased VIM estimates. Using a large library of candidate algorithms in the super learner can help protect against this bias. At a sample size of 10000, we removed RF and XGB from the super learner to reduce computation time.

All analyses were performed using cross-fitting to relax the technical conditions and sample-splitting to guarantee valid type I error rates under the case where the VIM is always zero. The analyses use code implemented in our R package \texttt{lvimp} \citep{lvimppkg} and may be reproduced using code available online (Section~\ref{sec:more_sims}). 

\subsection{Primary empirical results}

\begin{figure}
    \centering
    \includegraphics[width=1\textwidth]{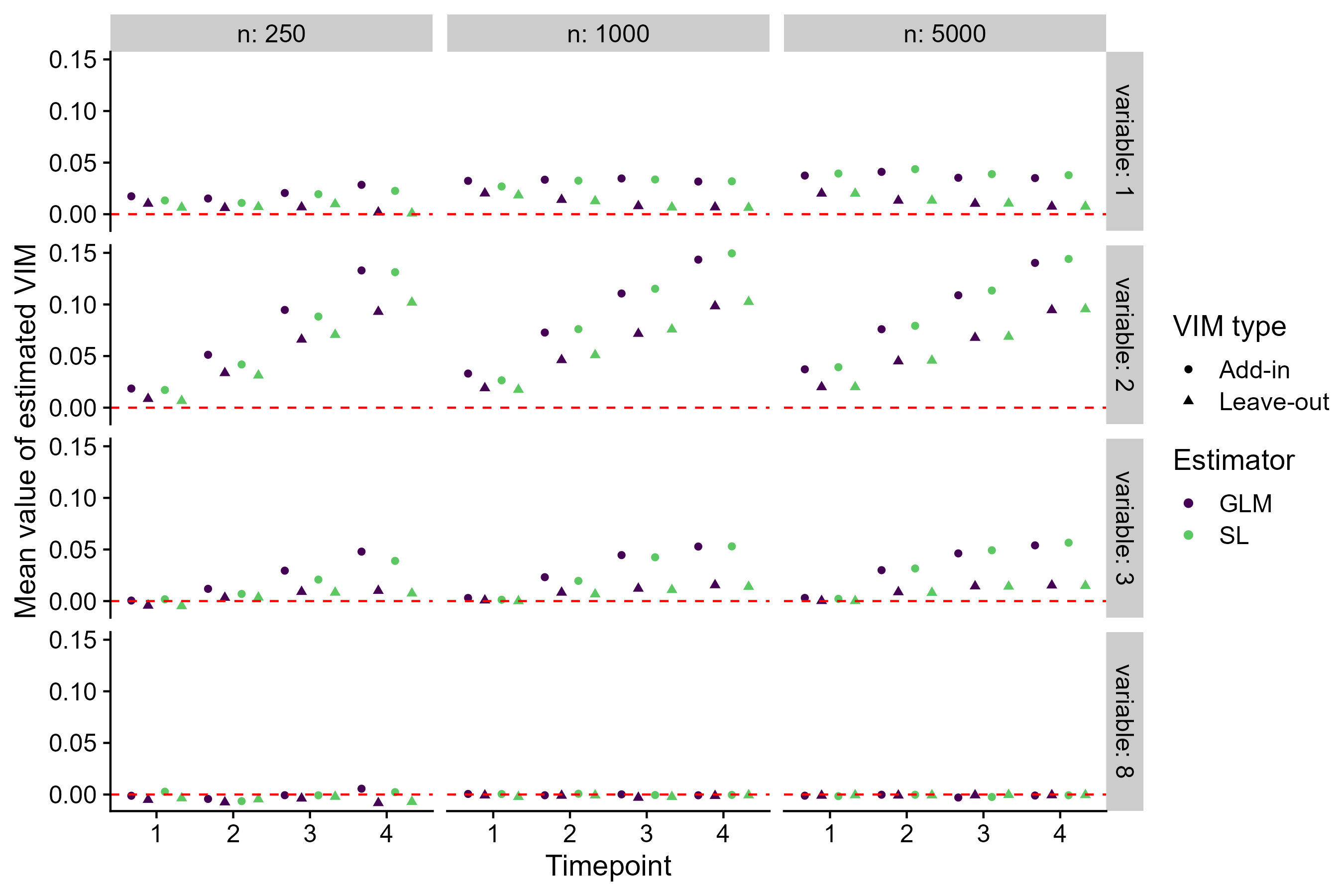}
    \caption{Mean estimated variable importance defined using AUC over 1000 simulated replicates versus time for four variables: three with true VIM value greater than zero (variables 1, 2, and 3) and one with true VIM value equal to zero (variable 8). Columns: sample size. Rows: variable. Shapes (circles and triangles) denote VIM type, while colors denote the estimator: logistic regression (GLM) or super learner (SL).}
    \label{fig:sim_vim_over_time}
\end{figure}

We display the estimated VIM defined using AUC for all sample sizes at each time point using logistic regression and a super learner in Figure~\ref{fig:sim_vim_over_time} (for PPV and sensitivity, see Figures~\ref{fig:sim_vim_over_time_0.5_ppv_dgm_2} and \ref{fig:sim_vim_over_time_0.5_sensitivity_dgm_2}). Estimation and inferential performance summaries for a sample of size 5,000 for the super learner with respect to AUC, PPV and sensitivity at the 95$\text{th}$ percentile are provided in Table~\ref{tab:sim_bign_all_measures_0.5_dgm_2}. We see that for both the mean and linear trajectory summary measures, the mean bias is small ($<$0.001). Mean bias is less than 0.05 for all $n$ and converges to zero with increasing $n$ (Tables~\ref{tab:sim_all_avg_1_0.5_auc_dgm_2}--\ref{tab:sim_all_slope_8_0.5_sensitivity_dgm_2}). Coverage of nominal 95\% confidence intervals is near 95\% for both add-in and leave-out importance of all variables (both average and slope of the linear trend) for AUC and PPV at the 95$\text{th}$ percentile, and for leave-out importance (both average and slope of the linear trend) for all variables for sensitivity at the 95$\text{th}$ percentile. However, for add-in importance with respect to sensitivity, coverage is lower than 95\%. This is likely due to the fact that the true variable importance values are near zero (i.e., adding the variable does not substantially improve prediction performance) and the small effective sample size (the number of events at $n = 5000$ is 800, with 400 per data split). In these cases, we can see negative point estimates of additive variable importance. Negative add-in VIMs are a signal that both there may be residual finite-sample bias in our estimate and also that coverage may not be at the nominal level, because adding a variable to a prediction model should never decrease prediction performance unless there is insufficient data to estimate either the model or its performance well enough. Sample size issues are mitigated somewhat when using less flexible models; coverage when using logistic regression is near 95\% (Tables~\ref{tab:sim_all_avg_1_0.5_sensitivity_dgm_2}--\ref{tab:sim_all_slope_8_0.5_sensitivity_dgm_2}). 

For the mean VIM over the time series, we have high power for detecting that the add-in AUC and PPV average is greater than zero for variables 1, 2, and 3; for variable 8, which truly has zero importance, the type I error rate is controlled at the 0.05 level (the same is true for variables 9 and 10, results not shown). For the slope of the linear VIM trend over the time series, power is lower for variable 1 (with true slope near zero) than for variables 2 and 3; type I error for variable 8 remains controlled at the 0.05 level. These trends hold for both the add-in and leave-out VIMs. While the VIM values differ between these formulations, the trends in the VIM values over time are similar and match the data-generating mechanism. The different VIM magnitudes for add-in and leave-out, while expected, are important: the interpretation of results from a variable importance estimation task depend greatly on the variables that are conditioned upon. As we discussed above, since the true (and estimated) VIM trajectories are monotone for each variable but are not all either increasing or decreasing, it would be difficult to compare AUTC or GMRC across variables in this context.

\begin{table}
\centering
\caption{\small Performance of estimators of the average VIM, defined using PPV and sensitivity at the 95th percentile of predicted risk and AUC, and slope of the linear trend in VIM over the time series, summarized over 1000 Monte-Carlo replications. Performance at sample size $n = 5000$ shown for the super learner. Both add-in (compared to four covariates) and leave-out (compared to nine covariates) variable importance are displayed for each variable. The true importance of variable 1 is nearly constant over time; variables 2 and 3 have increasing importance; and variable 8 has zero importance. \label{tab:sim_bign_all_measures_0.5_dgm_2}}
\centering
\resizebox{\ifdim\width>\linewidth\linewidth\else\width\fi}{!}{
\fontsize{7}{9}\selectfont
\begin{tabular}[t]{lrrrrrr}
\toprule
VIM type: variable & Mean est. & True value & Empirical SE & Coverage & Rejection prop. & CI width\\
\midrule
\addlinespace[0.3em]
\multicolumn{7}{l}{\textbf{Average (AUC)}}\\
\hspace{1em}Add-in: 1 & 0.040 & 0.035 & 0.011 & 0.956 & 0.956 & 0.049\\
\hspace{1em}Leave-out: 1 & 0.013 & 0.013 & 0.010 & 0.976 & 0.272 & 0.045\\
\hspace{1em}Add-in: 2 & 0.094 & 0.087 & 0.012 & 0.920 & 1.000 & 0.048\\
\hspace{1em}Leave-out: 2 & 0.057 & 0.056 & 0.011 & 0.956 & 0.999 & 0.046\\
\hspace{1em}Add-in: 3 & 0.035 & 0.031 & 0.011 & 0.966 & 0.910 & 0.049\\
\hspace{1em}Leave-out: 3 & 0.009 & 0.010 & 0.010 & 0.971 & 0.170 & 0.045\\
\bottomrule
\end{tabular}}
\end{table}

\subsection{Additional empirical results: other algorithms, sample sizes, and uncorrelated data}

In Supplementary Material Section~\ref{sec:more_dgm_2_results} (Tables~\ref{tab:sim_true_vims_0.5_auc_dgm_2}--\ref{tab:sim_all_slope_8_0.5_sensitivity_dgm_2}, Figures~\ref{fig:sim_vim_over_time_0.5_ppv_dgm_2}--\ref{fig:sim_vim_over_time_supp_0.5_sensitivity_dgm_2}), we provide further empirical results under the data-generating mechanism described above, for the full range of sample sizes and algorithms. 
As expected from theory, there is some bias and undercoverage at small sample sizes, but the bias decreases to zero and coverage increases to the nominal level with increasing sample size. The results do not differ across algorithms used to estimate the nuisance functions, except for sensitivity, where as we mentioned above there is better coverage when logistic regression is used.

In Supplementary Material Section~\ref{sec:common}, we investigated the performance of our proposed methods in a case with a more common outcome, both when the features (and outcomes) are uncorrelated and when they are correlated in a similar manner to the data-generating mechanism described above (though with higher magnitude of correlation). In this more common outcome setting, we considered only VIMs defined using AUC. In this setup, the true VIM values are different from the case presented above (Tables~\ref{tab:sim_true_vims_0_auc_dgm_1}--\ref{tab:sim_true_vims_0.5_auc_dgm_1}, Figures~\ref{fig:sim_true_vims_0_auc_dgm_1}--\ref{fig:sim_vim_over_time_0.5_auc_dgm_1}). The qualitative results in this scenario are identical to the results described above for AUC: bias decreases to zero and coverage increases to the nominal level as sample size increases (Tables~\ref{tab:sim_bign_all_0_auc_dgm_1}--\ref{tab:sim_all_slope_8_0.5_auc_dgm_1}, Figures~\ref{fig:sim_vim_over_time_0_auc_dgm_1}--\ref{fig:sim_vim_over_time_0.5_auc_dgm_1}). Taken together, these results suggest that our approach can be used in cases with uncorrelated or correlated features and outcomes.

\section{Evaluating VIMs for predictors of suicide risk}\label{sec:data}

We now return to our motivating example of estimating variable importance for predicting suicide attempt. We used data gathered from the electronic health records of patients seeking mental health care between Jan 1, 2009 and Sep 30, 2017 at one of two health systems, Kaiser Permanente Washington and HealthPartners in Minnesota. The PHQ is a patient-reported measure routinely used at both health systems to evaluate depressive symptoms (PHQ-8 total score) and to assess current suicidal ideation (PHQi9). Clinically, it is well-known that prior self-harm is one of the strongest predictors of the risk of future self-harm \citep{jacobs2010,mundt2013prediction}. In this example, we assess the longitudinal variable importance of the PHQi9 and prior recorded self-harm. Age and sex are known to be associated with risk of suicide attempt \citep{jacobs2010}. We define our first base set for add-in VIMs as age (11-17; 18-29; 30-44; 45-64; 65+ years), sex, and prior self-harm. At the time of data extraction, health system records included one measure of sex, which most likely represents sex assigned at birth. Our second base set for add-in VIMs was all demographic, prior self-harm, and diagnosis and utilization variables (see Section~\ref{sec:more_data}). Finally, we considered the leave-out VIM of PHQi9 versus all other variables. The add-in VIM for our second base set should be similar to the leave-out VIM.

Our goal was to predict a suicide attempt in the 90 days following a mental health visit; the 90 day interval was chosen in collaboration with health care providers and health system leaders. Longitudinal electronic health record data represent a complex visit process, particularly in mental health care, where some individuals seek care frequently (e.g., weekly) with a psychotherapist while others seek care infrequently (e.g., every three months) with a primary care provider to monitor symptoms. Greater than half of the patients in our dataset (described more fully below) made only one mental health visit over an 18-month time period. We simplified this process for our data analysis. Specifically, we sampled up to six visits per individual to mimic a prospective longitudinal study. For each individual, we defined their first mental health visit as their initial observation. We then sampled their second visit as the earliest visit that was at least 90 days after their first visit. We continued this sampling scheme to select up to four additional visits, each at least 90 days apart, leading to a maximum total followup of 18 months (6 periods of 90 days).  For people who died or disenrolled from the health system during the 18-month period following their first visit, we use all information up until the time of their death or disenrollment. Individuals who died by suicide have their information included in the prediction models up to their time of death, and their outcome of death by suicide is treated as a self-harm event for prediction modeling.

This sampling scheme anchors timepoints for variable importance estimation and inference relative to the first visit with a provider rather than calendar time, and places all study participants on the same time scale. While simplifying the visit process avoids issues of bias related to informative visit timing \citep[see, e.g.,][]{coulumbeATSvisittime}, it limits the scientific conclusions that we can draw from the results. In particular, ignoring the visit process may lead to the true (and estimated) VIMs being impacted by predictors of the visit timing process. However, the resulting VIMs are still of interest both in clinical care and clinical informatics: they describe the importance of the measured variables at subsequent visits after a patient's first visit to a provider for a given condition, and can therefore inform the types of questions a clinician may ask in subsequent health care visits or the data that are important to collect in the electronic medical record.

This longitudinal sampling scheme resulted in 343,950 visits made by 184,782 people without a diagnosis of schizophrenia or bipolar disorder to mental health specialty clinics (43\% of visits) and general medical providers (57\% of visits).
A total of 99,991 people (54.1\%) had exactly one visit during the 18-month time period; 44,046 (23.8\%) had two; 20,779 (11.2\%) had three; 10,393 (5.6\%) had four; 5,480 (2.9\%) had five; and 4,093 (2.2\%) had exactly six visits. Importantly, we use all available information at each time point to estimate VIMs; thus, all 184,782 people contribute to the analysis, regardless of their visit pattern. Confidence intervals, however, are based on those people with full information across the study; this is a conservative approach to inference in this case. The rate of suicide attempt was nearly constant over time, at approximately 0.5\%. The distribution of PHQi9 scores, PHQ-8 total score, age, and sex were also constant over time. Further details regarding the sampling process and the study population are in the Supplementary Material (Table~\ref{tab:table_1}). For this analysis, we considered the VIM of PHQi9 measured at each visit, since this information is readily available for use by the healthcare provider; however, one could also use the framework presented above to perform an analysis using historical capture of the PHQi9 (and other variables) \citep{wolock2024importance}.

To estimate VIMs at each of the six timepoints and summaries across the time series, we estimated prediction functions based on several variable sets (all variable sets in Table~\ref{tab:varsets}). The variable sets that we consider are: no variables; PHQi9 alone; age, sex, and prior self-harm (first base set); age, sex, prior self-harm, and PHQi9; demographic, prior self-harm, and diagnosis and utilization variables (second base set); demographic, prior self-harm, diagnosis and utilization variables, and Charlson comorbidity score (all variables but PHQi9); demographic, prior self-harm, diagnosis and utilization variables, and PHQi9; and all variables. These correspond to variable sets 1, 2, 5, 6, 9, 12, 13, and 15 of Table~\ref{tab:varsets}. The add-in VIM for PHQi9 can be computed by comparing variable set 2 to variable set 1 (importance of PHQi9 alone); variable set 6 to variable set 5 (importance of PHQi9 relative to age, sex, and prior self-harm); and variable set 13 to variable set 6 (importance of PHQi9 relative to demographic, prior self-harm, and diagnosis and utilization variables). The leave-out VIM of PHQi9 is computed by comparing variable set 15 to variable set 12. We used AUC to define the VIM at each timepoint, because AUC is often used to describe the overall fit of suicide risk prediction models \citep[see, e.g.,][]{simon2016risk,kessler2017predicting,walsh2018,sanderson2019predicting,shortreed2023,minus2025behavior}. We also considered PPV and sensitivity at the 95$^\text{th}$ percentile of predicted risk to complement this approach \citep{wolock2024importance}.

As in the numerical experiments described above, we used logistic regression, logistic regression with the lasso for variable selection, random forests, gradient boosted trees, and a super learner to estimate these prediction functions. In this data analysis, we also considered the discrete super learner, which uses cross-validation to select the single best-performing algorithm among the candidate library. As in the numerical simulations, several tuning parameter values were considered for each algorithm (Table~\ref{tab:sl-algs_data}). Unlike in the simulations presented above, here we do not know the true data-generating mechanism, so some of these procedures may be misspecified. Additionally, there may be finite-sample differences between the results based on different procedures: the effective sample size in this case is approximately 400 (due to the extremely low event rate). To guard against bias resulting from prediction model misspecification, we recommend using the super learner with a large library of candidate learners. We used multiple estimation procedures here to further emphasize the possible finite-sample differences between approaches. However, in practice, we recommend choosing an estimation procedure prior to analysis, and using cross-validation to assess performance of this procedure. To guard against false discovery inflation due to multiple testing for the VIM summaries (10 tests per estimation procedure, two estimation procedures), we used a conservative Bonferroni correction to control the family-wise error rate at 0.05, implying a corrected p-value threshold of 0.0025.

In Tables~\ref{tab:auc_data_analysis_pred_perf_all_supp}--\ref{tab:sensitivity_data_analysis_pred_perf_all_supp}, we present the mean, minimum, and maximum estimated value for each variable set across time and all algorithms for AUC, PPV, and sensitivity, respectively. The results are in line with prior expectations: prediction performance tends to increase with an increasing number of variables; and the AUC for all variables ranges from 0.73 to 0.85, similar to other suicide risk prediction models in this population. 

\begin{figure}
    \centering
    \includegraphics[width=1\textwidth]{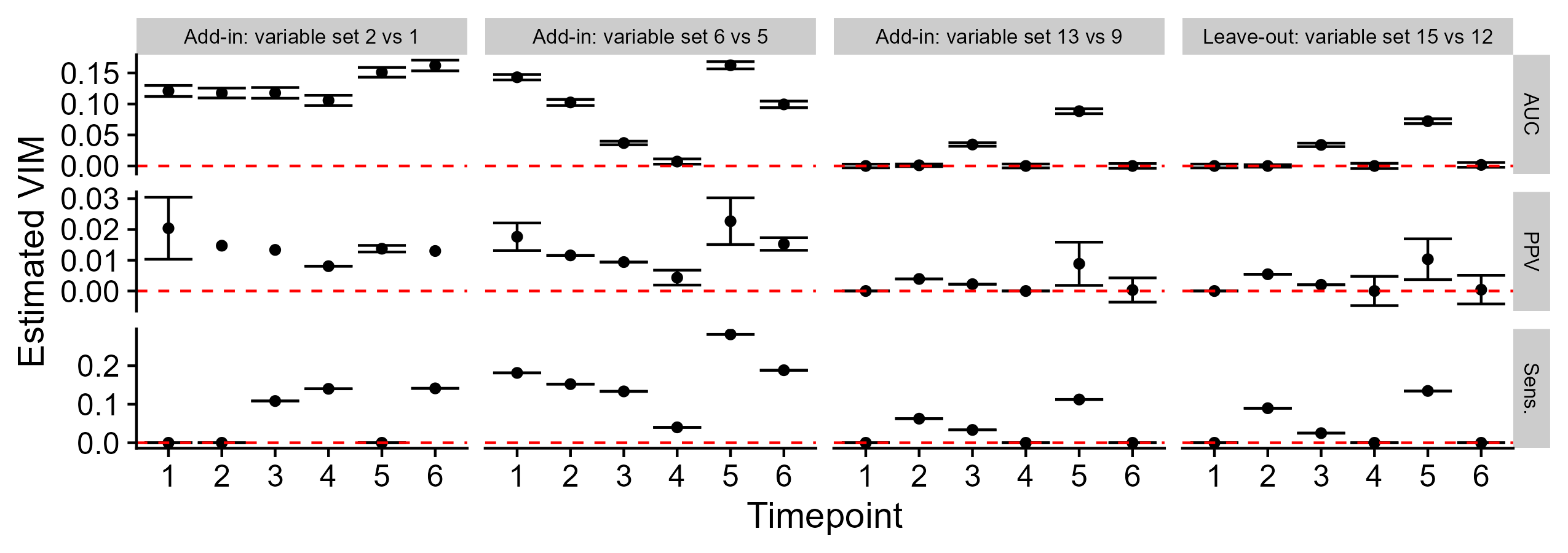}
    \caption{Variable importance estimates for PHQi9, considering the importance of PHQi9 when compared to other variables. Comparisons are in the columns: PHQi9 versus no variables (add-in: variable set 2 to 1); PHQi9 versus age, sex, and prior self-harm (add-in: variable set 6 versus 5); PHQi9 versus demographic, prior self-harm, and diagnosis \& utilization variables (add-in: variable set 13 versus 9); and PHQi9 versus all other variables (leave-out: variable set 15 versus 12). Rows denote the prediction performance metric: AUC, PPV, and sensitivity (Sens.). VIM values were truncated at zero; y-axis limits vary by prediction performance metric.}
    \label{fig:data_analysis_vim_over_time}
\end{figure}

We present the estimated add-in VIM for PHQi9 relative to various sets of variables and leave-out VIM of PHQi9 relative to all other variables in Figure~\ref{fig:data_analysis_vim_over_time} and Table~\ref{tab:data_analysis_vim_summaries}. We present results obtained using the super learner, with VIMs measured using AUC, PPV, and sensitivity; the remaining results (other variable sets and algorithms) are provided in Figures~\ref{fig:auc_data_analysis_vim_over_time_all}--\ref{fig:sensitivity_data_analysis_vim_over_time_all} and Tables~\ref{tab:auc_data_analysis_vim_summaries_all}--\ref{tab:sensitivity_data_analysis_vim_summaries_all}. Comparing the importance of PHQi9 across variable sets, we see that the add-in importance of PHQi9 compared to no variables (column 1) tends to be largest for all measures. Add-in importance compared to other variable sets tends to be lower but remains nonzero at many timepoints; comparing PHQi9 to age, sex, and prior self-harm (column 2) had some point estimates in a similar range to PHQi9 versus no variables (column 1). Importance of PHQi9 is smaller compared to the larger groups of variables (columns 3 and 4), but is nonzero at several timepoints. These trends hold across both metrics (AUC, PPV, sensitivity) and estimation procedures (Section~\ref{sec:more_data}). For a given metric and estimation procedure (e.g., super learner vs random forests) the magnitude of importance can differ, reflecting either possible bias due to model misspecification or finite-sample differences. 

Table~\ref{tab:data_analysis_vim_summaries} contains the average VIM and slope of the linear trend in VIM over time for PHQi9 compared against the same variable sets (remaining results are in Section~\ref{sec:more_data}). We focus on the super learner results since logistic regression may be misspecified. The average AUC VIM of PHQi9 is quite large for its add-in importance compared to no variables (variable set 2 versus 1; p-value $< 0.0001$ for a test that this average is zero, smaller than the Bonferroni threshold of 0.0025), is smaller when compared to age, sex, and prior self-harm (variable set 6 versus 5; p-value $< 0.0001$, smaller than the Bonferroni threshold of 0.0025), and is smaller still when compared with the full set of variables (p-values smaller than 0.05 but above the Bonferroni threshold); this is consistent with Figure~\ref{fig:data_analysis_vim_over_time}. However, the PHQi9 has the same estimated AUC VIM as all prior self-harm variables (Tables~\ref{tab:auc_data_analysis_vim_summaries_all_supp}--\ref{tab:sensitivity_data_analysis_vim_summaries_all_supp}). The slope of the linear trend in AUC VIM for PHQi9 tends to be near zero. The results suggest a slight increasing trend in AUC VIM for PHQi9 alone and a decreasing trend in VIM for PHQi9 compared to age, sex, and prior-self harm variables (variable set 6 versus 5). VIMs based on PPV had smaller magnitude than those based on AUC, but showed similar trends. Those based on sensitivity showed different trends, with a smaller average VIM for PHQi9 versus no variables compared to the other comparisons, and a large positive trend in VIM for PHQi9 versus no variables. In the Supplementary Material, we describe the add-in and leave-out importance of PHQi9 relative to several other sets of variables (including the PHQ-8 total score; see Tables~\ref{tab:auc_data_analysis_vim_summaries_all_supp}--\ref{tab:sensitivity_data_analysis_vim_summaries_all_supp}), which followed similar trends to those reported here. 

Taken together, as in the simulations, these results highlight that variable importance can differ greatly based on the chosen comparator set. By considering the importance of PHQi9 relative to different sets of variables, we gain complementary information that may be helpful towards a deeper understanding of this variable and its role in predicting suicide risk. Our results are in line with other work using simpler approaches to estimation and variable importance that similarly found that model performance does not change much over time and that PHQi9 is important over time \citep{simon2016risk,simon2017between,simon2024stability}. This implies that clinicians should continue to focus on obtaining and discussing the PHQi9 in follow-up visits with patients as part of care.

\begin{table}
\centering
\caption{Estimates of the average VIM and slope of the linear trend in VIM over the time series,  considering the importance of PHQi9 when compared to other variables. Comparisons are: PHQi9 vs no variables (add-in: comparing variable set 2 to 1), PHQi9 vs age, sex, and prior self-harm (add-in: comparing variable set 6 to 5), PHQi9 vs demographic, prior self-harm, diagnosis \& utilization variables (add-in: comparing variable set 13 to 9), PHQi9 vs all other variables (leave-out: comparing variable set 15 to 12). Estimates are shown for AUC, PPV, and sensitivity (Sens.). \label{tab:data_analysis_vim_summaries}}
\centering
\resizebox{\ifdim\width>\linewidth\linewidth\else\width\fi}{!}{
\fontsize{9}{11}\selectfont
\begin{tabular}[t]{lrrrr}
\toprule
VIM type: comparison & Estimate & SE & 95\% CI & p-value\\
\midrule
\addlinespace[0.3em]
\multicolumn{5}{l}{\textbf{Average (AUC)}}\\
\hspace{1em}PHQi9 vs no variables & 0.129 & 0.019 & {}[0.093, 0.166] & < 0.001\\
\hspace{1em}PHQi9 vs age, sex, and prior self-harm & 0.092 & 0.012 & {}[0.069, 0.115] & < 0.001\\
\hspace{1em}PHQi9 vs demographic, prior self-harm, and diagnosis \& utilization variables & 0.018 & 0.009 & {}[-0.000, 0.036] & 0.026\\
\hspace{1em}PHQi9 vs all other variables & 0.016 & 0.009 & {}[-0.003, 0.034] & 0.047\\
\addlinespace[0.3em]
\multicolumn{5}{l}{\textbf{Trend - slope (AUC)}}\\
\hspace{1em}PHQi9 vs no variables & 0.008 & 0.011 & {}[-0.013, 0.029] & 0.432\\
\hspace{1em}PHQi9 vs age, sex, and prior self-harm & -0.002 & 0.007 & {}[-0.015, 0.011] & 0.771\\
\hspace{1em}PHQi9 vs demographic, prior self-harm, and diagnosis \& utilization variables & 0.006 & 0.005 & {}[-0.004, 0.016] & 0.254\\
\hspace{1em}PHQi9 vs all other variables & 0.006 & 0.005 & {}[-0.004, 0.016] & 0.254\\
\addlinespace[0.3em]
\multicolumn{5}{l}{\textbf{Average (PPV)}}\\
\hspace{1em}PHQi9 vs no variables & 0.014 & 0.002 & {}[0.010, 0.018] & < 0.001\\
\hspace{1em}PHQi9 vs age, sex, and prior self-harm & 0.013 & 0.002 & {}[0.010, 0.017] & < 0.001\\
\hspace{1em}PHQi9 vs demographic, prior self-harm, and diagnosis \& utilization variables & 0.002 & 0.002 & {}[-0.003, 0.006] & 0.213\\
\hspace{1em}PHQi9 vs all other variables & 0.002 & 0.002 & {}[-0.002, 0.007] & 0.181\\
\addlinespace[0.3em]
\multicolumn{5}{l}{\textbf{Trend - slope (PPV)}}\\
\hspace{1em}PHQi9 vs no variables & -0.001 & 0.001 & {}[-0.004, 0.001] & 0.288\\
\hspace{1em}PHQi9 vs age, sex, and prior self-harm & 0.000 & 0.001 & {}[-0.002, 0.002] & 0.647\\
\hspace{1em}PHQi9 vs demographic, prior self-harm, and diagnosis \& utilization variables & 0.000 & 0.001 & {}[-0.002, 0.003] & 0.631\\
\hspace{1em}PHQi9 vs all other variables & 0.000 & 0.001 & {}[-0.002, 0.003] & 0.615\\
\addlinespace[0.3em]
\multicolumn{5}{l}{\textbf{Average (Sens.)}}\\
\hspace{1em}PHQi9 vs no variables & 0.002 & 0.010 & {}[-0.018, 0.022] & 0.408\\
\hspace{1em}PHQi9 vs age, sex, and prior self-harm & 0.163 & 0.018 & {}[0.127, 0.198] & < 0.001\\
\hspace{1em}PHQi9 vs demographic, prior self-harm, and diagnosis \& utilization variables & 0.029 & 0.020 & {}[-0.011, 0.069] & 0.079\\
\hspace{1em}PHQi9 vs all other variables & 0.029 & 0.020 & {}[-0.011, 0.068] & 0.078\\
\addlinespace[0.3em]
\multicolumn{5}{l}{\textbf{Trend - slope (Sens.)}}\\
\hspace{1em}PHQi9 vs no variables & 0.051 & 0.006 & {}[0.040, 0.063] & < 0.001\\
\hspace{1em}PHQi9 vs age, sex, and prior self-harm & 0.009 & 0.010 & {}[-0.011, 0.029] & 0.355\\
\hspace{1em}PHQi9 vs demographic, prior self-harm, and diagnosis \& utilization variables & 0.008 & 0.012 & {}[-0.015, 0.031] & 0.477\\
\hspace{1em}PHQi9 vs all other variables & 0.006 & 0.012 & {}[-0.017, 0.029] & 0.631\\
\bottomrule
\end{tabular}}
\end{table}

\section{Discussion}\label{sec:discussion}

We developed methods to estimate and summarize longitudinal variable importance to determine the longitudinal importance of the PHQ ninth item (PHQi9), a measure of suicidal ideation that is often asked during visits to mental health providers. 
To do this we proposed a framework for estimating and making inference on these summaries of population-level variable importance trajectories over time. As with the VIMs themselves, these trajectory-summarizing measures are population quantities, and the interpretation of these summaries depends on the VIM used (e.g., $R^2$ or AUC). We proposed several different summary measures of longitudinal VIMs. We recommend that scientific motivation guide the selection of the VIM summary statistic. It is likely that multiple summary measures will be needed to describe complex time-varying relationships.

Our longitudinal VIMs summarize over a contiguous set of timepoints. Thus, the interpretation of the final VIM summaries depends critically on how time is measured and the visit patterns that give rise to the data. In some cases, it may be of interest to consider calendar time; however, this renders estimation and inference difficult because there may be many timepoints and few observations per timepoint. For this reason, and because prediction models in suicide prevention research are often intended for prospective use in clinical care at health care visits \citep{coley2023empirical}, visit number is often used in place of calendar time. This allows estimation of, for example, the importance of a variable at the third health care visit, or a longitudinal VIM describing the importance of the variable across the first six care visits. The interpretation of the VIM further depends on both the definition of the variable itself (e.g., PHQi9 measured at the visit or the number of PHQi9 scores of 3 in the past 6 months) and the other variables included in the comparator set; these must be chosen with care with a clear focus on the scientific question at hand. VIM interpretation must always be made in the context of the possible predictors considered during model building. In this suicide risk prediction setting, the PHQ-9 was routinely collected in the two health systems. While risk of suicidal behavior has been shown to increase with increasing responses to item 9 of the PHQ \citep{SimonPHQi92019}, there are a number of other measures [e.g., Columbia-Suicide Severity Rating Scale \citep[C-SSRS;][]{posner2011columbia}, Ask Suicide-Screening Questions \citep[ASQ;][]{horowitz2012ask}] that are also likely predictive of suicidal behavior, and should be included in suicide risk prediction models when available. We also acknowledge differing views regarding when predictions of risk are accurate enough to be actionable or how tools to identify risk (either self-report questionnaires or risk prediction models) should be deployed in different healthcare settings. The results presented here do not address those questions. We use risk prediction models to illustrate new methods for evaluating and communicating longitudinal importance of specific risk factors or predictors.

The goal of a variable importance analysis is often to identify variables for further study. These inherently exploratory analyses should still be pre-specified to the extent possible, as we did in our analysis in Section~\ref{sec:data}. Further, it may be appropriate to account for multiple comparisons, by controlling the family-wise error rate, false discovery rate, or some other measure at an appropriate level. The optimal tradeoff between multiple comparison control and power for identifying variables depends on the setting at hand.

As in traditional intrinsic VIM estimation, correlation among predictors presents challenges in interpreting summaries of VIM trajectories. Correlation between variables at a given time point affects the value of the VIM itself; considering both add-in and leave-out variable importance, or ultimately considering variable importance relative to all possible sets of variables \citep[e.g.,][]{williamson2020c}, can lead to a fuller view of the importance of a variable than considering a single approach in isolation. While our point and interval estimators are valid in the presence of correlation, other approaches, including a cluster bootstrap \citep{field2007}, could yield improvements in small sample settings. Further, interactions between variables can either be handled explicitly (e.g., by a regression model which includes some pre-specified interactions or which considers all two-way interactions) or implicitly (e.g., by random forests or gradient boosted trees with sufficient depth). When calculating variable importance, all interactions (either explicit or implicit) are taken into account by either leaving out or adding in the variable of interest and resulting interactions. Finally, while the theoretical conditions (Section~\ref{sec:proofs}) might at first appear to restrict the algorithms that can be used, flexible algorithms that only approximately satisfy conditions (B1)--(B3) can often be used, as evidenced by good performance of random forests and boosted trees in our simulations.

In our analysis of electronic health records data from two large healthcare systems, we found that the PHQ ninth item (PHQi9) had large average variable importance over time, in line with previous results \citep{simon2016risk,simon2024stability}. Further, we found that this importance did not change much over time; the slope of the linear trend in importance was near zero when comparing the PHQi9 to both no other variables and a simple model with age and sex. This can help clinicians continue to focus on obtaining and discussing the PHQi9 in all visits, which may improve screening for suicide risk and other preventive care. 

In binary prediction settings the large-sample behavior of many prediction performance metrics (including AUC and sensitivity) depends on the number of events in the rarer class \citep{minus2025behavior}. In cases with a small number of events, the flexibility of the prediction algorithm should be reduced for stable estimation. The self-harm rate in our analysis was rarer than the outcome rate in our simulations. However, we designed the simulations to have similar numbers of events to the self-harm analysis; since this is the effective sample size for AUC and sensitivity, our simulation results cover the case observed in the analysis.

The simulations we conducted assumed a non-informative longitudinal visit time process. We applied our approach to data collected from electronic health care records where the visit time process is often complex and may be informative, i.e., the number and timing of visits may be related to health outcomes \citep{coulumbeATSvisittime}.  Our current approach is a necessary foundation to begin to study the problem of longitudinal variable importance in these more complex settings. One possible extension is to incorporate the time since last visit as a covariate; this could reasonably capture patient engagement with the healthcare system from our prediction standpoint.
Future work should explore the impact of informative visit time processes on longitudinal VIMs and develop approaches to separate out measures of variable importance of predictors of the health outcome of interest from predictors of the visit time process.
Importantly, in the suicide prevention context we studied here, the outcome of interest is fully observable on all individuals at any given time, while predictors may not be updated between visits. It is likely that accounting for the informative visit time process will involve a fundamentally different estimand than those we considered here. 

Finally, in many causal inference settings there is a treatment or exposure variable, or a treatment strategy, that should be handled differently from other predictors or confounding variables. In our framework, the average value under the optimal treatment strategy \citep[e.g.,][]{vanderlaan2015} could be used to define the VIM at each time point, and inference could be made on summaries of the trajectory of this VIM. This is an example of a myopic strategy, where the average value at each timepoint is maximized. Other non-myopic strategies for both overall variable importance and the importance of tailoring variables are being pursued in ongoing research, and could be used to improve individualized treatment regimes in suicide prevention \citep{kessler2020suicide}.


\section*{Acknowledgments}

The first, second, and fifth author were supported in part by the National Institute of Mental Health of the National Institutes of Health under Award Number R01 MH114873. The content is solely the responsibility of the authors and does not necessarily represent the official views of the National Institutes of Health. 

The second author acknowledges support from a Discovery Grant from the Natural Sciences and Engineering Research Council of Canada. The second author holds a Canada Research Chair (Tier 1, Canadian Institutes of Health Research) in Statistical Methods for Precision Medicine and acknowledges the support of a chercheur de m\'{e}rite career award from the Fonds de recherche du Qu\'{e}bec – Sant\'{e}.

\section*{Supplementary material}

The supplementary material contains technical details, including proofs of theorems; results from additional numerical experiments; and additional data analysis results.

{\small
\bibliographystyle{chicago}
\bibliography{brian-papers}
}

\newpage

\section*{SUPPLEMENTARY MATERIAL}
\renewcommand{\thefigure}{S\arabic{figure}}
\renewcommand{\theequation}{S\arabic{equation}}
\renewcommand{\thetable}{S\arabic{table}}
\renewcommand{\thetheorem}{S\arabic{theorm}}
\renewcommand{\thelemma}{S\arabic{lemma}}
\renewcommand{\thesection}{S\arabic{section}}
\setcounter{figure}{0}
\setcounter{table}{0}
\setcounter{section}{0}

\doublespacing

\section{Predictiveness values under a simple outcome model}\label{sec:simple_example}

Suppose that we have observations at a single time point of data unit $(W, U, Y)$, where $U = (U_1, U_2)$. In this simple example, suppose further that $(W, U) \sim N_{3}(0, I)$ are continuous variables and that $Y$ is a continuous outcome, where $Y \mid (W, U) = (w, u) \sim N(0.5(w_1 + u_1) + 2u_2, 1)$. Finally, suppose that we use the R-squared predictiveness measure, i.e., $V(f, P) = 1 - \frac{E_P\{Y - f(W, U)\}^2}{E_P\{Y - E_P(Y)\}^2}$.

In this example, $t = 1$ and $s = 3$, i.e., the variable of interest is $U_2$. Then 
\begin{align*}
  f_{0,p,t} =& \ E_0(Y \mid W = w, U = u) = 0.5(w + u_1) + 2u_2 \\
  f_{0,p\setminus s, t} =& \ E_0(Y \mid W = w, U_1 = u_1) = 0.5(w + u_1) \\
  f_{0,s\cup w, t} =& \ E_0(Y \mid W = w, U_2 = u_2) = 0.5w + 2u_2 \\
  f_{0,w,t} =& \ E_0(Y \mid W = w) = 0.5w.
\end{align*}
The marginal outcome variance is 5.5. The total predictiveness is $v_{0,p,t} = V(f_{0,p,t},P_0) \approx 0.82$. The residual predictiveness is $v_{0,p\setminus s, t} = V(f_{0,p\setminus s, t}, P_0) \approx 0.09$. The marginal predictiveness is $v_{0,s \cup w, t} = V(f_{0,s\cup w, t}, P_0) \approx 0.77$. The irreducible predictiveness is $v_{0,w, t} = V(f_{0,w, t}, P_0) \approx 0.05$. So in this example, $\psi_{0,s,t} \approx 0.73$ and $\theta_{0,s,t} \approx 0.73$ (in fact, the two are equal in this case because $W$, $U_1$, and $U_2$ are mutually independent).

\section{Proofs of theorems}\label{sec:proofs}

\subsection{Notation and regularity conditions}

We assume that $(L_0, L_1, \ldots, L_T)$ follow a distribution $\mathcal{P}_0$, where each time point has a unique marginal data-generating mechanism $P_{0,t}$ that lies in a nonparametric class of distributions $\mathcal{M}$. Note that this setup allows for data to be correlated across time, and for covariates to be correlated within a given timepoint. We further assume that the set $s$ provides the indices of a group of variables (or single variable) for which it is of scientific interest to do inference on variable importance. For a given prediction function $f_t$ defined at time point $t$ and a data-generating mechanism $P_t$ corresponding to the same time point, we define $V(f_t, P_t)$ as a measure of the predictiveness of $f_t$ under distribution $P_t$. For a vector $x \in \mathbb{R}^p$, we use the notation $x_{p,t} \in \mathbb{R}^p$ to refer to the entire vector; $x_{w,t} \in \mathbb{R}^{\lvert W_t \rvert}$ refers to the confounders alone; $x_{s \cup w, t} \in \mathbb{R}^{\lvert s \rvert + \lvert W_t \rvert}$ refers to the confounders and the variables with index in $s$; and $x_{p \setminus s, t} \in \mathbb{R}^{p - \lvert s \rvert}$ refers to the variables with index not in $s$. Note that in all cases, $\lvert \cdot \rvert$ may refer to the absolute value or the length of a vector depending on context. We refer to a contiguous subset of the time series $\tau = [t_0, t_1]$.

For rich classes of functions $\mathcal{F}_{p,t}$ that makes use of all variables at timepoint $t$; a subset $\mathcal{F}_{p \setminus s, t}$ that does not make use of the variables with index in $s$ at timepoint $t$; a subset $\mathcal{F}_{s \cup w, t}$ that makes use of the variables with index in $s$ and the potential confounders; and a subset $\mathcal{F}_{w,t}$ that makes use of the confounders alone, we define the predictiveness maximizing functions
\begin{align*}
  f_{0,p,t} \in& \ \argmax_{f \in \mathcal{F}_{p,t}} V(f, P_{0,t}) \\
  f_{0,p\setminus s, t} \in& \ \argmax_{f \in \mathcal{F}_{p \setminus s, t}} V(f, P_{0,t}) \\
  f_{0, s \cup w, t} \in& \ \argmax_{f \in \mathcal{F}_{s \cup w, t}} V(f, P_{0,t}) \\
  f_{0, w, t} \in& \ \argmax_{f \in \mathcal{F}_{w, t}} V(f, P_{0,t}),
\end{align*}
where the middle subscript refers to the set of variables that the function may make use of. These predictiveness-maximizing functions can also be referred to as the oracle prediction functions at each time point with respect to the variables that are used. Suppose that at each time point, we obtain estimators $f_{n,p,t}$, $f_{n,p\setminus s, t}$, $f_{n,s\cup w, t}$, and $f_{n,w,t}$ of $f_{0,p,t}$, $f_{0,p\setminus s, t}$, $f_{0,s\cup w, t}$, and $f_{0,w,t}$, respectively; here, the subscript $n$ refers to the fact that these estimators are obtained on a sample of size $n$. Below, we will focus on inference for the total predictiveness $v_{0,p,t}:= V(f_{0,p,t}, P_{0,t})$, since results can be readily extended to the residual predictiveness $v_{0,p\setminus s, t}:= V(f_{0,p\setminus s, t}, P_{0,t})$, the marginal predictiveness $v_{0,s\cup w, t} := V(f_{0, s\cup w, t}, P_{0,t})$, and the irreducible predictiveness $v_{0,w,t} := V(f_{0, w, t}, P_{0,t})$. The predictiveness values are the `oracle' values in the sense that they describe the best possible predictiveness of a given set of variables under sampling from $P_{0,t}$, because their definition involves the function that maximizes predictiveness over a large function class.

The results can also be easily extended to the variable importance measure (VIM) values $\psi_{0,s,t} := v_{0,p,t} - v_{0,p\setminus s, t}$ and $\theta_{0,s, t} := v_{0,s\cup w, t} - v_{0,w,t}$. The two VIM parameters differ in the maximal set of variables under consideration, which defines the type of VIM: $\psi_{0,s,t}$ is a \textit{leave-out}, or \textit{conditional}, VIM, comparing the predictiveness using all $p$ features to the predictiveness removing the features in $s$; and $\theta_{0,s,t}$ is an \textit{add-in}, or \textit{marginal}, VIM, comparing the predictiveness of the variables in $s$ plus the confounders to the predictiveness of the confounders alone. Finally, we define estimators $v_{n,p,t}:=V(f_{n,p,t},P_{n,t})$, $v_{n,p\setminus s, t}:=V(f_{n,p\setminus s, t},P_{n,t})$, $v_{n,s\cup w, t} := V(f_{n,s\cup w, t}, P_{n,t})$, $v_{n,w,t}:= V(f_{n,w,t}, P_{n,t})$, $\psi_{n,s,t}:= v_{n,p,t} - v_{n,p\setminus s, t}$ and $\theta_{n,s\cup w, t}:=v_{n,s\cup w, t} - v_{n,w,t}$ of $v_{0,p,t}$, $v_{0,p\setminus s, t}$, $v_{0,s\cup w, t}$, $v_{0,w,t}$, $\psi_{0,s,t}$, and $\theta_{0,s,t}$, respectively.

We now generalize notation from \citet{williamson2021} and several conditions that are required for our results to hold across multiple time points. We define the linear space $\mathcal{R} := \{c(P_1 - P_2): c \in [0, \infty), P_1, P_2 \in \mathcal{M}\}$ of finite signed measures generated by $\mathcal{M}$. For any $R \in \mathcal{R}$, say $R = c(P_1 - P_2)$, we refer to the supremum norm $\lVert R \rVert_{\infty} := c\sup_z\lvert F_1(z) - F_2(z) \rvert$, where $F_1$ and $F_2$ are the distribution functions corresponding to $P_1$ and $P_2$, respectively. Furthermore, we denote by $\dot{V}(f, P_{0,t}; h)$ the G\^ateaux derivative of $P \mapsto V(f, P)$ at $P_{0,t}$ in the direction $h \in \mathcal{R}$, and define the random function $g_{n,r,t}: \ell \mapsto \dot{V}(f_{n,r,t}, P_{0,t}; \delta_\ell - P_{0,t}) - \dot{V}(f_{0,r,t}, P_{0,t}; \delta_\ell - P_{0,t})$, where for a data realization $\ell$, $\delta_\ell$ is the degenerate distribution on $\{\ell\}$ and $r$ refers to a collection of variables (e.g., $r = p$ denotes all variables). For any $P_t \in \mathcal{M}$ and index set $r \subseteq \{1, \ldots, p\}$, we also denote by $f_{t,r,P_t}$ any $P_t$-population maximizer of $f \mapsto V(f, P_t)$ over $\mathcal{F}_{r,t}$. Again, we focus first on inference for $v_{0,p,t}$, since we can extend the necessary conditions (and results) to $v_{0,p\setminus s, t}$, $v_{0,s\cup w , t}$, $v_{0,w,t}$, $\psi_{0,s,t}$, and $\theta_{0,s,t}$. We define the following set of conditions at each time point $t \in \{0, \ldots, T\}$, classified as being either deterministic (A) or stochastic (B) in nature:
\begin{itemize}
   \item[(A1)] there is some $C > 0$ such that for each sequence $f_1, f_2, \cdots \in \mathcal{F}_{p,t}$ with $\lVert f_j - f_{0,p,t}\rVert_{\mathcal{F}_{p,t}} \to 0$, there is a $J$ such that for all $j > J$, $\lvert V(f_j, P_{0,t}) - V(f_{0,p,t}, P_{0,t})\rvert \leq C \lVert f_j - f_{0,p,t}\rVert_{\mathcal{F}}^2$;
   \item[(A2)] there is some $\delta > 0$ such that for each sequence $\epsilon_1, \epsilon_2, \ldots \in \mathbb{R}$ and $h, h_1, h_2, \ldots \in \mathcal{R}$ satisfying that $\epsilon_j \to 0$ and $\lVert h_j - h \rVert_\infty \to 0$, it holds that
   \begin{align*}
     \sup_{f \in \mathcal{F}: \lVert f - f_{0,t} \rVert_{\mathcal{F}} < \delta} \big\lvert \frac{V(f, P_{0,t} + \epsilon_j h_j) - V(f, P_{0,t})}{\epsilon_j} - \dot{V}(f, P_{0,t}; h_j)\big\rvert \to 0;
   \end{align*}
   \item[(A3)] $\lVert f_{P_{0,t} + \epsilon h,p,t} - f_{0,p,t} \rVert_{\mathcal{F}} = o(\epsilon)$ for each $h \in \mathcal{R}$;
   \item[(A4)] $f \mapsto \dot{V}(f, P_{0,t}; h)$ is continuous at $f_{0,p,t}$ relative to $\mathcal{F}_{p,t}$ for each $h \in \mathcal{R}$;
   \item[(A5)] there exists $\dot{m}(v_{0,p,\tau})$ such that $m(v) = m(v_{0,p,\tau}) + \dot{m}(v_{0,p,\tau})(v - v_{0,p,\tau}) + R(v, v_{0,p,\tau})$, where $R(v, v_{0,p,\tau}) = O(\lVert v - v_{0,p,\tau} \rVert^2)$; 
   \item[(B1)] $\lVert f_{n,p,t} - f_{0,p,t} \rVert_{\mathcal{F}_{p,t}} = o_P(n^{-1/4})$;
   \item[(B2)] $E_{P_{0,t}}[\int \{g_{n,p,t}(\ell)\}^2 dP_{0,t}(\ell)] = o_P(1)$;
   \item[(B3)] there exists some $P_{0,t}$-Donsker class $\mathcal{G}_{0,t}$ such that $P_{0,t}(g_{n,p,t} \in \mathcal{G}_{0,t})\to 1$.
\end{itemize} 
Conditions (A1)--(A4), as mentioned in the main manuscript, are deterministic conditions on the data-generating process and function class $\mathcal{F}_{p,t}$ necessary for consistency and the existence of the efficient influence function of $v_{0,p,t}$. Condition (A1) states that there is no first-order contribution from estimation of $f_{0,p,t}$. While this follows as a consequence of the optimality of $f_{0,p,t}$, this optimality in turn requires appropriate regularity conditions on $P_{0,t}$ and $\mathcal{F}_{p,t}$. Condition (A2) states that $P_t \mapsto V(f_{0,p,t}, P_t)$ must be locally uniformly Hadamard differentiable at $P_{0,t}$ in a neighborhood of $f_{0,p,t}$. Condition (A3) requires that $f_{P_t,p,t}$ vary smoothly in $P_t$ around $P_{0,t}$. Condition (A4) requires that the Hadamard derivative of $P_t \mapsto V(f, P_t)$ at $P_{0,t}$ vary smoothly in $f$ around $f_{0,p,t}$. Each of these conditions has been verified to hold for the $R^2$, classification accuracy, AUC, and deviance measures of predictiveness in \citet{williamson2021}. Condition (A5) restricts the class of summary measures over the longitudinal variable importance trajectory to functions that are smooth enough for the delta method for influence functions \citep{bkrw} to apply. 
Each of the examples discussed in Section~2.3 (main manuscript) satisfies condition (A5); we verify this explicitly below.

The stochastic conditions (B1)--(B3) describe the required behavior of estimators of the prediction function $f_{0,p,t}$, specifically their required convergence rate to $f_{0,p,t}$ and their complexity. Condition (B1) specifies the convergence rate of an estimator $f_{n,p,t}$ to its estimand $f_{0,p,t}$ necessary for second-order terms to be asymptotically negligible. Condition (B2) states that a functional of $f_{n,p,t}$ must converge to the corresponding functional evaluated at $f_{0,p,t}$, and is often implied by consistency of $f_{n,p,t}$. Finally, condition (B3) restricts the complexity of $f_{n,p,t}$. We can modify conditions (B1) and (B2) into $K$-fold cross-fitted versions, which will remove the need for condition (B3) to hold, and allows flexible (e.g., machine learning-based) algorithms that are not known to satisfy (B3) to be used to estimate $f_{0,p,t}$. Briefly, this involves splitting the data into $K$ folds; for each $k \in \{1, \ldots, K\}$, using the data in fold $k$ as a held-out test set and obtaining estimators $f_{n,p,t,k}$ of $f_{0,p,t}$ on the remaining data; and updating (B1) and (B2) to 
\begin{itemize}
  \item[(B1')] $\lVert f_{n,p,t,k} - f_{0,p,t} \rVert_{\mathcal{F}_{p,t}} = o_P(n^{-1/4})$ for each $k \in \{1, \ldots, K\}$; and
  \item[(B2')] $E_{P_{0,t}}[\int \{g_{n,p,t,k}(\ell)\}^2 dP_{0,t}(\ell)] = o_P(1)$ for each $k \in \{1, \ldots, K\}$,
\end{itemize}
where $g_{n,p,t,k} := \ell \mapsto \dot{V}(f_{n,p,t,k}, P_{0,t}; \delta_\ell - P_{0,t}) - \dot{V}(f_{0,p,t}, P_{0,t}; \delta_\ell - P_{0,t})$. In practice, we have found cross-fitting to be crucial whenever complex algorithms are used \citep{williamson2021}.

We define a contiguous subset of the time series, $\tau := [t_0,t_1] \subseteq \{0, \ldots, T\}$, that will play a role in several of the summaries below. Set $v_{0,p} := (v_{0,p,0}, \ldots, v_{0,p,T})$ and $v_{n,p} := (v_{n,p,0}, \ldots, v_{n,p,T})$.
 
\subsection{Proof of Theorem 1}

Under the collection of conditions (A1)--(A4) and (B1)--(B3), Theorem 1 of \citet{williamson2021} applies to each time-specific predictiveness estimator and corresponding VIM estimator. Thus, for each time point $t = 1,\ldots,T$, $v_{n,p,t}$ is an asymptotically linear estimator of $v_{0,p,t}$ with influence function equal to $\phi_{0,p,t}$, so
\begin{align*}
    v_{n,p,t} - v_{0,p,t} = \frac{1}{n}\sum_{i=1}^n \phi_{0,p,t}(L_{t,i}),
\end{align*}
implying that $\sqrt{n}(v_{n,p,t} - v_{0,p,t}) \to_d Z \sim N(0, \sigma^2)$, where $\sigma^2 = E_0 \{\phi_{0,p,t}(L_t)^2\}$. Further, the multivariate central limit theorem applies to the vector $\boldsymbol{v}_{n,p} = [v_{n,p,1},\ldots,v_{n,p,T}]^\top$ and its limit $\boldsymbol{v}_{0,p} = [v_{0,p,1},\ldots,v_{0,p,T}]^\top$:
\begin{align*}
    \sqrt{n}(\boldsymbol{v}_{n,p} - \boldsymbol{v}_{0,p}) \to_d \boldsymbol{Z} \sim N_T(0, \Sigma), \text{ where} \\
    \Sigma = E_0 \left(\begin{bmatrix}
    \phi_{0,p,1}(L_1) \\ \vdots \\ \phi_{0,p,T}(L_T)
\end{bmatrix}\begin{bmatrix}
    \phi_{0,p,1}(L_1) \\ \vdots \\ \phi_{0,p,T}(L_T)
\end{bmatrix}^\top\right),
\end{align*}
which holds even in the case of correlated data.

Under condition (A5), the functional delta method may be applied \citep{neugebauer2006causal,benkeser2019estimating} and yields that 
\begin{align*}
  m_{n,p,\tau} - m_{0,p,\tau} = \frac{1}{n}\sum_{i=1}^n \dot{m}(v_{0,p,\tau})\left[\phi_{0,p,t}(L_{t,i})\right]_{t \in \tau} + o_P(n^{-1/2}),
\end{align*}
precisely as claimed.

\subsection{Proof of Theorem 2}

Under the collection of conditions (A1)--(A4) and (B1'), (B2'), Theorem 2 of \citet{williamson2021} applies to each time-specific predictiveness estimator and corresponding VIM estimator.  

As in the proof of Theorem 1 above, under condition (A5), the functional delta method and the multivariate central limit theorem may be applied, yielding the desired result.

\section{Additional theoretical results}\label{sec:more_theory}

\subsection{Verification of condition (A5) for examples in Section~2.3 (main manuscript)}

First, we note that if $P_t \mapsto V(f, P_t)$ is linear in $f$, then $V(f, P_t)$ is Hadamard differentiable uniformly in $f$. It has been shown that several common measures $V$, including $R^2$, classification accuracy, AUC, and cross-entropy, are linear in this sense \citep[][supplementary materials section 3.3]{williamson2021}. For these measures, then, it remains to show that the summary measure is similarly linear in $f$.

\noindent\textit{Example 1: mean over a contiguous subset of the time series.} In this example, $m(v) := \lVert \tau \rVert^{-1}\sum_{t \in \tau} V(f, P_t)$. If each $V(f, P_t)$ is linear in $f$, then the mean is clearly also a linear functional of $f$, so condition (A5) holds with $R(v, v_{0,p,\tau}) = 0$. 

\noindent\textit{Example 2:  linear trend over a contiguous subset of the time series.} In this example, $m(v) = (U^\top U)^{-1}U^\top v_{\tau}$. Since matrix multiplication is a linear transformation, then if each $V(f, P_t)$ is linear in $f$, condition (A5) holds with $R(v, v_{0,p,\tau}) = 0$.

\noindent\textit{Example 3: area under a subset of the trajectory curve.} In this example, $m(v) = \int_{t_0}^{t_1} h(t, v) dt$. Integrals are a linear transformation, so if $h$ is also linear (e.g., a piecewise linear function or spline interpolator) and each $V(f, P_t)$ is linear in $f$, then condition (A5) holds with $R(v, v_{0,p,\tau}) = 0$.

\textit{Example 4: geometric mean rate of change over a subset of the trajectory curve.} In this example, $m(v) = \exp\left[\lVert \tau \rVert^{-1}\sum_{t \in \tau}\log \{\lvert h'(t, v) \rvert\}\right]$. We have defined $h$ as a differentiable linear function, so $m$ is continuous. Thus, by the mean value theorem, $\dot{m}$ exists, and a Taylor series expansion yields that
\begin{align*}
    m(v) = m(v_{0,p,\tau}) + \dot{m}(v_{0,p,\tau})(v - v_{0,p,\tau}) + g(v)(v - v_{0,p,\tau})
\end{align*}
where $\lim_{v \to v_{0,p,\tau}} g(v) = 0$. Thus, $R(v, v_{0,p,\tau}) = g(v)(v - v_{0,p,\tau}) = O(\lVert v - v_{0,p,\tau}\rVert^2)$.

\subsection{Efficient influence functions for Examples 1--4}

The efficient influence function for the mean total predictiveness over the time period $\tau$ is
\begin{align*}
    \phi_{0,p,\tau}: \ell \mapsto \lvert \tau \rvert^{-1} \sum_{t \in \tau} \phi_{0,p,t}(\ell_t),
\end{align*}
which results from the functional delta method, since $m(x) = \lvert \tau \rvert^{-1} \sum_{t \in \tau} x_t$.

The efficient influence function for the linear trend in the total predictiveness over the time period $\tau$ is
\begin{align*}
    \phi_{0,p,\tau}: \ell \mapsto (U^\top U)^{-1}U^\top \begin{bmatrix} \phi_{0,p,t}(\ell_t) \end{bmatrix}_{t \in \tau},
\end{align*}
which results from the functional delta method, since $m(x) = (U^\top U)^{-1}U^\top x$.

The efficient influence function for the AUTC of the total predictiveness over the time period $\tau$ is
\begin{align*}
    \phi_{0,p,\tau}: \ell \mapsto \sum_{t \in \tau} \int_{x \in \tau_{0,c}} h'(x, v_{0,p})dx\phi_{0,p,t}(\ell),
\end{align*}
where for the piecewise linear trajectory function $h'(x, v_{0,p}) = \frac{1}{2}(v_{0,p,t} + v_{0,p,t+1})$ for $t \in [\tau + 1, \tau_1 - 1]$ and $h'(x, v_{0,p}) = \frac{1}{2}v_{0,p,t}$ for $t \in \{\tau, \tau_1\}$. For the spline trajectory function, $h'$ depends on the type of spline fit.

The efficient influence function for the GMRC of the total predictiveness over the time period $\tau$ is
\begin{align*}
    \phi_{0,p,\tau}: \ell \mapsto \gamma_{0,p}\lVert \tau \rVert^{-1}\sum_{t \in \tau}\lvert h'(t, v_{0,p})\rvert^{-1}\sign\{h'(t, v_{0,p})\}h''(t, v_{0,p})\phi_{0,p,t}(\ell) ,
\end{align*}
where $\gamma_{0,p}$ is the true GMRC and $h''$ is the derivative of $h'$ with respect to its second argument. This derivative exists for the spline interpolator, but is zero for the piecewise linear interpolator; thus, using the efficient influence function for variance estimation for the GMRC is not possible using the piecewise linear interpolator.

\section{Additional numerical results}\label{sec:more_sims}

\subsection{Replicating all numerical experiments}

All numerical experiments presented here and in the main manuscript can be replicated using code available on GitHub at \url{https://github.com/bdwilliamson/lvimp_supplementary}.

\subsection{Super learner specification}

The specific candidate learners and their corresponding tuning parameters for our super learner library \citep{slpkg} are provided in Table~\ref{tab:sl-algs}. In most cases, we used the default tuning parameters, since our goal was to showcase the performance of our proposed variable importance estimator under different scenarios rather than to use a library flexible enough to be robust in many situations. It is possible that with a different library of learners, different results could be obtained. We used 5-fold cross-validation to select the optimal convex combination of candidate learners when $n > 1000$, and 10-fold cross-validation otherwise. This is similar to recent guidelines proposed by \citet{phillips2023}. Recall that at the largest sample size ($n$ = 10000) we removed random forests (RF) and boosted trees (XGB) from the super learner to reduce computation time, so we do not obtain results from these algorithms at that sample size.

\begin{table}
    \centering
    \caption{Candidate learners in the super learner ensemble used in the simulations, along with their R implementation, tuning parameter values, and description of the tuning parameters. All tuning parameters besides those listed here are set to their default values. In particular, the random forests are grown with a minimum node size of 5 for continuous outcomes and 1 for binary outcomes and a subsampling fraction of 1; the boosted trees are grown with shrinkage rate of 0.1, and a minimum of 10 observations per node. \\
    ${}^{\dagger}$: $p$ denotes the total number of predictors. }
    \label{tab:sl-algs}
    \begin{tabular}{c|ccc}
       Candidate Learner & R & Tuning Parameter & Tuning parameter  \\
       & Implementation & and possible values & description\\ \hline
        Random forests & \texttt{ranger} & \texttt{mtry} $= 1/2\sqrt{p}$ ${}^{\dagger}$ & Number of variables \\
        & \citep{rangerpkg} & & to possibly split \\
        & & & at in each node \\ 
        & & \texttt{num.trees} $= 500$ & Number of trees \\
        \hline
        Gradient boosted & \texttt{xgboost} & \texttt{max.depth} $ = 1$ &  Maximum tree depth\\
        trees & \citep{xgboostpkg} & \texttt{ntree} $=500$ & Number of iterations \\ \hline
        Lasso & \texttt{glmnet} & $\lambda$ & $\ell_1$ regularization  \\
        & \citep{glmnetpkg} & chosen via 10-fold CV & parameter \\ \hline
    \end{tabular}
\end{table}

\clearpage 

\subsection{Results for sensitivity, PPV, and other algorithms}\label{sec:more_dgm_2_results}

In Tables~\ref{tab:sim_true_vims_0.5_auc_dgm_2}--\ref{tab:sim_true_vims_0.5_sensitivity_dgm_2}, we present the true VIM values with respect to AUC, PPV at the 95$^\text{th}$ percentile, and sensitivity at the 95$^\text{th}$ percentile, respectively.

\begin{table}
\centering
\caption{True variable importance values (defined using AUC) at each time point and summarized over the time series. Each feature has correlation 0.5 with the previous timepoint. \label{tab:sim_true_vims_0.5_auc_dgm_2}}
\centering
\fontsize{9}{11}\selectfont
\begin{tabular}[t]{llllllll}
\toprule
VIM type: variable & Timepoint 1 & Timepoint 2 & Timepoint 3 & Timepoint 4 & Mean & Trend intercept & Trend slope\\
\midrule
Add-in: 1 & 0.035 & 0.037 & 0.035 & 0.032 & 0.035 & 0.037 & -0.001\\
Leave-out: 1 & 0.02 & 0.014 & 0.01 & 0.008 & 0.013 & 0.022 & -0.004\\
Add-in: 2 & 0.035 & 0.072 & 0.105 & 0.135 & 0.087 & 0.003 & 0.033\\
Leave-out: 2 & 0.02 & 0.044 & 0.067 & 0.092 & 0.056 & -0.005 & 0.024\\
Add-in: 3 & 0.004 & 0.028 & 0.044 & 0.05 & 0.031 & -0.007 & 0.015\\
Leave-out: 3 & 0.001 & 0.009 & 0.015 & 0.015 & 0.01 & -0.002 & 0.005\\
Add-in: 8 & 0 & 0 & 0 & 0 & 0 & 0 & 0\\
Leave-out: 8 & 0 & 0 & 0 & 0 & 0 & 0 & 0\\
\bottomrule
\end{tabular}
\end{table}
\begin{table}
\centering
\caption{True variable importance values (defined using PPV at the 95th percentile of predicted risk) at each time point and summarized over the time series. Each feature has correlation 0.5 with the previous timepoint. \label{tab:sim_true_vims_0.5_ppv_dgm_2}}
\centering
\fontsize{9}{11}\selectfont
\begin{tabular}[t]{llllllll}
\toprule
VIM type: variable & Timepoint 1 & Timepoint 2 & Timepoint 3 & Timepoint 4 & Mean & Trend intercept & Trend slope\\
\midrule
Add-in: 1 & 0.039 & 0.042 & 0.041 & 0.039 & 0.04 & 0.04 & 0\\
Leave-out: 1 & 0.023 & 0.019 & 0.015 & 0.013 & 0.018 & 0.026 & -0.003\\
Add-in: 2 & 0.039 & 0.086 & 0.135 & 0.185 & 0.111 & -0.011 & 0.049\\
Leave-out: 2 & 0.023 & 0.058 & 0.096 & 0.14 & 0.079 & -0.018 & 0.039\\
Add-in: 3 & 0.004 & 0.031 & 0.053 & 0.062 & 0.037 & -0.011 & 0.019\\
Leave-out: 3 & 0.001 & 0.013 & 0.022 & 0.025 & 0.015 & -0.005 & 0.008\\
Add-in: 8 & 0 & 0 & 0 & 0 & 0 & 0 & 0\\
Leave-out: 8 & 0 & 0 & 0 & 0 & 0 & 0 & 0\\
\bottomrule
\end{tabular}
\end{table}
\begin{table}
\centering
\caption{True variable importance values (defined using sensitivity at the 95th percentile of predicted risk) at each time point and summarized over the time series. Each feature has correlation 0.5 with the previous timepoint. \label{tab:sim_true_vims_0.5_sensitivity_dgm_2}}
\centering
\fontsize{9}{11}\selectfont
\begin{tabular}[t]{llllllll}
\toprule
VIM type: variable & Timepoint 1 & Timepoint 2 & Timepoint 3 & Timepoint 4 & Mean & Trend intercept & Trend slope\\
\midrule
Add-in: 1 & 0.012 & 0.013 & 0.012 & 0.011 & 0.012 & 0.013 & 0\\
Leave-out: 1 & 0.007 & 0.006 & 0.004 & 0.004 & 0.005 & 0.008 & -0.001\\
Add-in: 2 & 0.012 & 0.026 & 0.04 & 0.052 & 0.032 & -0.001 & 0.014\\
Leave-out: 2 & 0.007 & 0.017 & 0.028 & 0.04 & 0.023 & -0.004 & 0.011\\
Add-in: 3 & 0.001 & 0.009 & 0.015 & 0.017 & 0.011 & -0.003 & 0.005\\
Leave-out: 3 & 0 & 0.004 & 0.007 & 0.007 & 0.004 & -0.001 & 0.002\\
Add-in: 8 & 0 & 0 & 0 & 0 & 0 & 0 & 0\\
Leave-out: 8 & 0 & 0 & 0 & 0 & 0 & 0 & 0\\
\bottomrule
\end{tabular}
\end{table}

In Tables~\ref{tab:sim_true_pred_0.5_auc_dgm_2}--\ref{tab:sim_true_pred_0.5_sensitivity_dgm_2}, we present the true predictiveness values measured using AUC, PPV at the 95$^\text{th}$ percentile, and sensitivity at the 95$^\text{th}$ percentile, respectively.

\begin{table}
\centering
\caption{True predictiveness values (defined using AUC) at each time point and summarized over the time series. Each feature has correlation 0.5 with the previous timepoint. \label{tab:sim_true_pred_0.5_auc_dgm_2}}
\centering
\fontsize{9}{11}\selectfont
\begin{tabular}[t]{llllllll}
\toprule
Pred. type: variable & Timepoint 1 & Timepoint 2 & Timepoint 3 & Timepoint 4 & Mean & Trend intercept & Trend slope\\
\midrule
Add-in: 1 & 0.554 & 0.56 & 0.561 & 0.562 & 0.559 & 0.553 & 0.002\\
Leave-out: 1 & 0.555 & 0.605 & 0.647 & 0.68 & 0.622 & 0.518 & 0.042\\
Add-in: 2 & 0.554 & 0.595 & 0.631 & 0.664 & 0.611 & 0.519 & 0.037\\
Leave-out: 2 & 0.555 & 0.575 & 0.59 & 0.596 & 0.579 & 0.545 & 0.014\\
Add-in: 3 & 0.523 & 0.551 & 0.57 & 0.579 & 0.556 & 0.509 & 0.019\\
Leave-out: 3 & 0.574 & 0.61 & 0.642 & 0.673 & 0.625 & 0.542 & 0.033\\
Add-in: 8 & 0.519 & 0.523 & 0.526 & 0.529 & 0.524 & 0.516 & 0.003\\
Leave-out: 8 & 0.575 & 0.619 & 0.657 & 0.688 & 0.635 & 0.54 & 0.038\\
Baseline & 0.519 & 0.523 & 0.526 & 0.529 & 0.524 & 0.516 & 0.003\\
All & 0.575 & 0.619 & 0.657 & 0.688 & 0.635 & 0.54 & 0.038\\
\bottomrule
\end{tabular}
\end{table}
\begin{table}
\centering
\caption{True predictiveness values (defined using PPV at the 95th percentile of predicted risk) at each time point and summarized over the time series. Each feature has correlation 0.5 with the previous timepoint. \label{tab:sim_true_pred_0.5_ppv_dgm_2}}
\centering
\fontsize{9}{11}\selectfont
\begin{tabular}[t]{llllllll}
\toprule
Pred. type: variable & Timepoint 1 & Timepoint 2 & Timepoint 3 & Timepoint 4 & Mean & Trend intercept & Trend slope\\
\midrule
Add-in: 1 & 0.219 & 0.232 & 0.24 & 0.248 & 0.235 & 0.211 & 0.01\\
Leave-out: 1 & 0.221 & 0.29 & 0.358 & 0.42 & 0.322 & 0.155 & 0.067\\
Add-in: 2 & 0.219 & 0.276 & 0.334 & 0.395 & 0.306 & 0.16 & 0.059\\
Leave-out: 2 & 0.221 & 0.251 & 0.277 & 0.294 & 0.26 & 0.199 & 0.025\\
Add-in: 3 & 0.185 & 0.221 & 0.252 & 0.271 & 0.232 & 0.16 & 0.029\\
Leave-out: 3 & 0.243 & 0.296 & 0.351 & 0.408 & 0.324 & 0.187 & 0.055\\
Add-in: 8 & 0.181 & 0.19 & 0.199 & 0.209 & 0.195 & 0.171 & 0.01\\
Leave-out: 8 & 0.244 & 0.308 & 0.373 & 0.434 & 0.34 & 0.181 & 0.063\\
Baseline & 0.181 & 0.19 & 0.199 & 0.209 & 0.195 & 0.171 & 0.01\\
All & 0.244 & 0.308 & 0.373 & 0.434 & 0.34 & 0.181 & 0.063\\
\bottomrule
\end{tabular}
\end{table}
\begin{table}
\centering
\caption{True predictiveness values (defined using sensitivity at the 95th percentile of predicted risk) at each time point and summarized over the time series. Each feature has correlation 0.5 with the previous timepoint. \label{tab:sim_true_pred_0.5_sensitivity_dgm_2}}
\centering
\fontsize{9}{11}\selectfont
\begin{tabular}[t]{llllllll}
\toprule
Pred. type: variable & Timepoint 1 & Timepoint 2 & Timepoint 3 & Timepoint 4 & Mean & Trend intercept & Trend slope\\
\midrule
Add-in: 1 & 0.068 & 0.07 & 0.07 & 0.07 & 0.07 & 0.068 & 0.001\\
Leave-out: 1 & 0.068 & 0.087 & 0.105 & 0.119 & 0.095 & 0.053 & 0.017\\
Add-in: 2 & 0.068 & 0.083 & 0.098 & 0.112 & 0.09 & 0.054 & 0.015\\
Leave-out: 2 & 0.068 & 0.076 & 0.081 & 0.083 & 0.077 & 0.065 & 0.005\\
Add-in: 3 & 0.057 & 0.067 & 0.074 & 0.077 & 0.069 & 0.052 & 0.006\\
Leave-out: 3 & 0.075 & 0.089 & 0.103 & 0.115 & 0.096 & 0.062 & 0.013\\
Add-in: 8 & 0.056 & 0.057 & 0.058 & 0.059 & 0.058 & 0.055 & 0.001\\
Leave-out: 8 & 0.076 & 0.093 & 0.109 & 0.123 & 0.1 & 0.061 & 0.016\\
Baseline & 0.056 & 0.057 & 0.058 & 0.059 & 0.058 & 0.055 & 0.001\\
All & 0.076 & 0.093 & 0.109 & 0.123 & 0.1 & 0.061 & 0.016\\
\bottomrule
\end{tabular}
\end{table}

\begin{figure}
    \centering
    \includegraphics[width=1\textwidth]{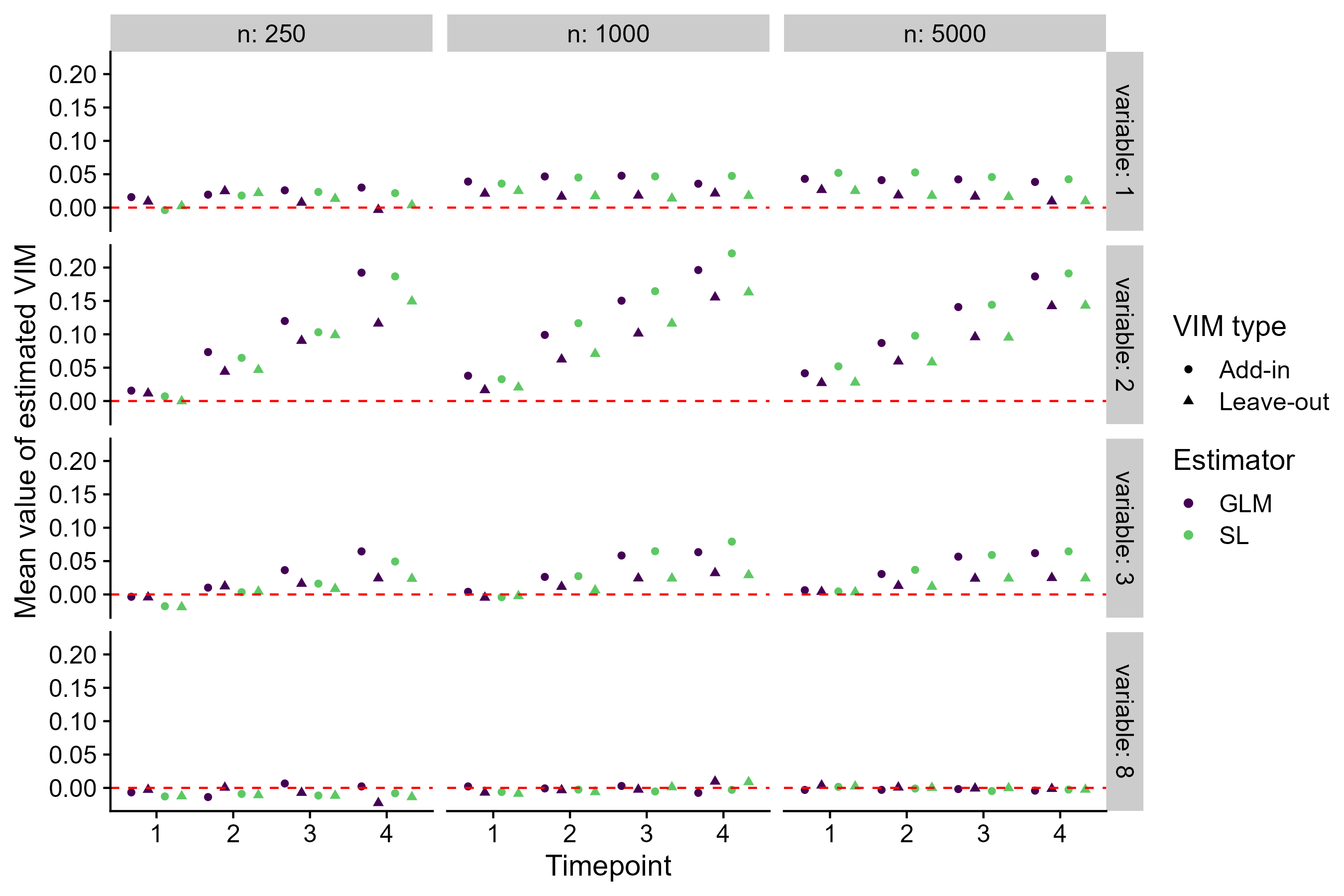}
    \caption{Mean estimated variable importance defined using PPV at the 95$^\text{th}$ percentile over 1000 simulated replicates versus time for four variables: three with true VIM value greater than zero (variables 1, 2, and 3) and one with true VIM value equal to zero (variable 8). Columns: sample size. Rows: variable. Shapes (circles and triangles) denote VIM type, while colors denote the estimator: logistic regression (GLM) or super learner (SL).}
    \label{fig:sim_vim_over_time_0.5_ppv_dgm_2}
\end{figure}

\begin{figure}
    \centering
    \includegraphics[width=1\textwidth]{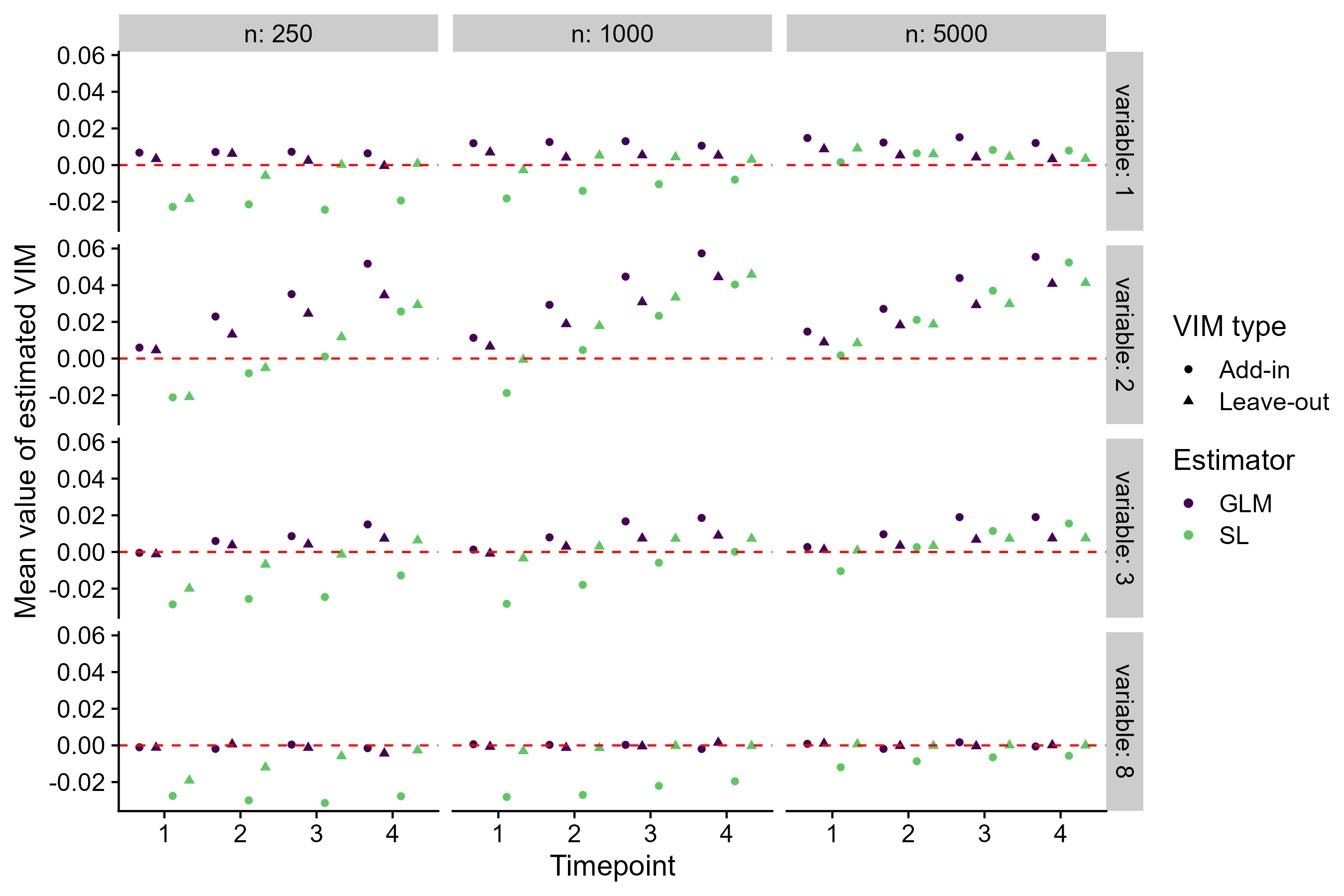}
    \caption{Mean estimated variable importance defined using sensitivity at the 95$^\text{th}$ percentile over 1000 simulated replicates versus time for four variables: three with true VIM value greater than zero (variables 1, 2, and 3) and one with true VIM value equal to zero (variable 8). Columns: sample size. Rows: variable. Shapes (circles and triangles) denote VIM type, while colors denote the estimator: logistic regression (GLM) or super learner (SL).}
    \label{fig:sim_vim_over_time_0.5_sensitivity_dgm_2}
\end{figure}

\begin{figure}
    \centering
    \includegraphics[width=1\textwidth]{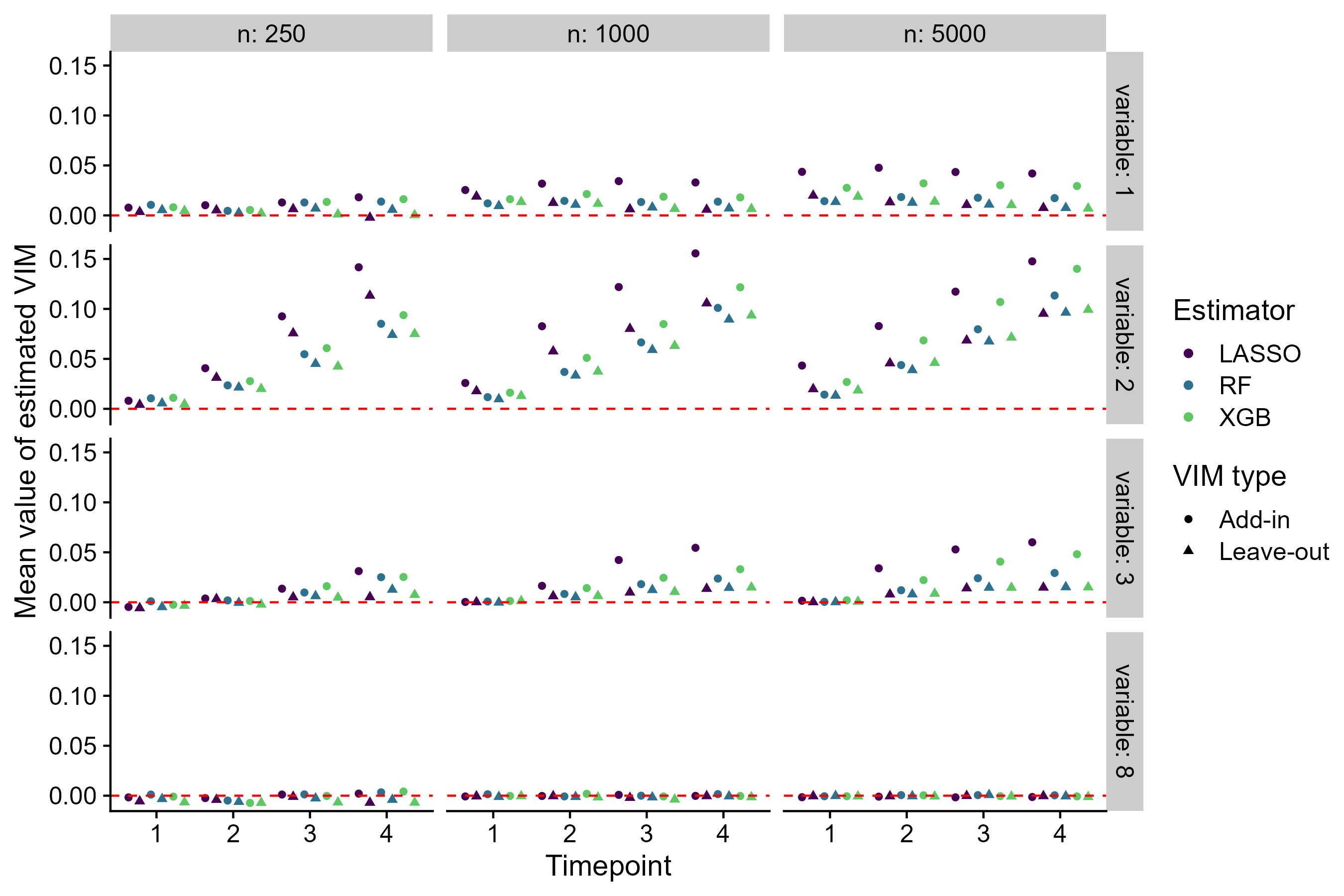}
    \caption{Mean estimated variable importance defined using AUC over 1000 simulated replicates versus time for four variables: three with true VIM value greater than zero (variables 1, 2, and 3) and one with true VIM value equal to zero (variable 8). Columns: sample size. Rows: variable. Shapes (circles and triangles) denote VIM type, while colors denote the estimator: lasso, random forests (RF), or boosted trees (XGB).}
    \label{fig:sim_vim_over_time_supp_0.5_auc_dgm_2}
\end{figure}

\begin{figure}
    \centering
    \includegraphics[width=1\textwidth]{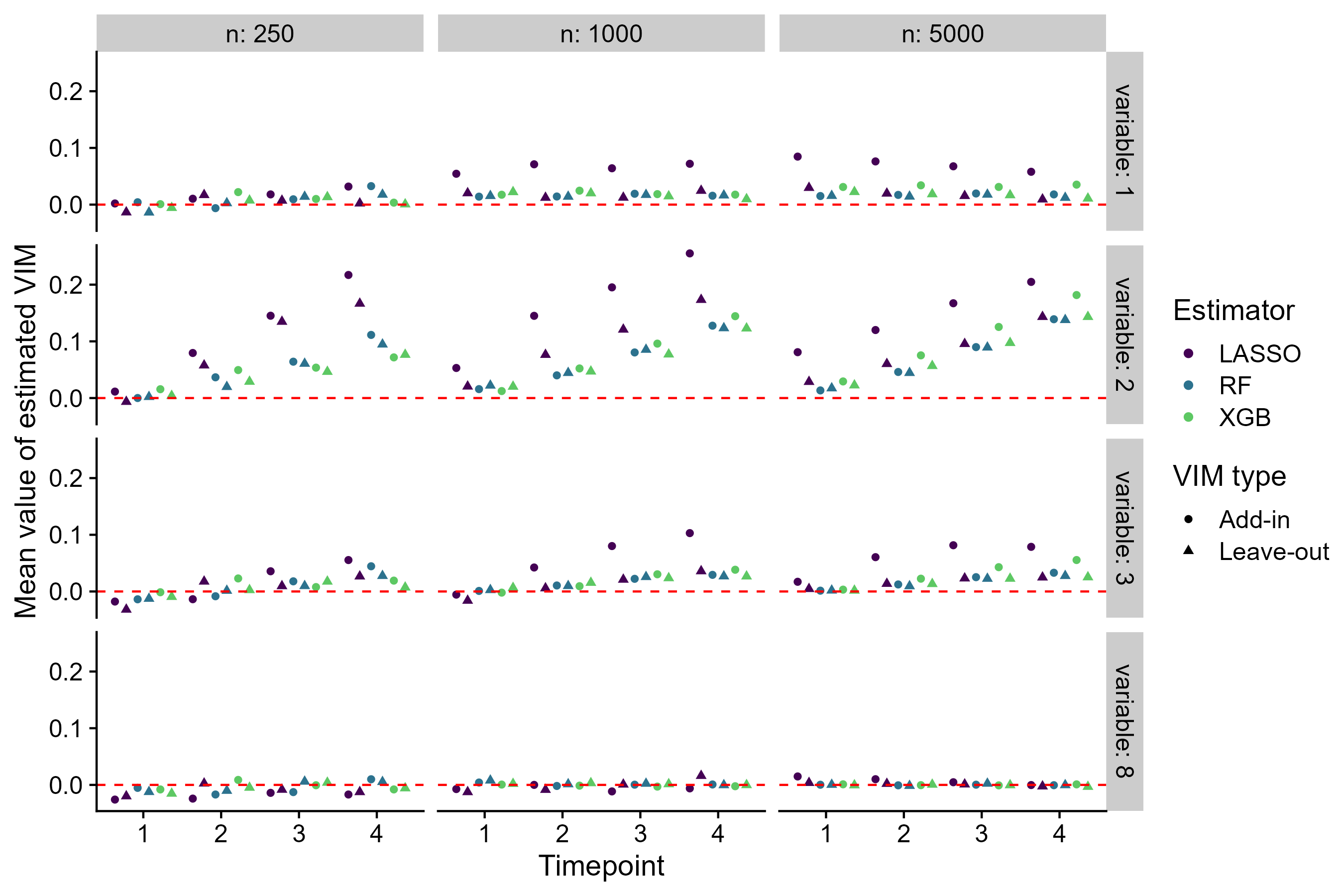}
    \caption{Mean estimated variable importance defined using PPV at the 95$^\text{th}$ percentile over 1000 simulated replicates versus time for four variables: three with true VIM value greater than zero (variables 1, 2, and 3) and one with true VIM value equal to zero (variable 8). Columns: sample size. Rows: variable. Shapes (circles and triangles) denote VIM type, while colors denote the estimator: lasso, random forests (RF), or boosted trees (XGB).}
    \label{fig:sim_vim_over_time_supp_0.5_ppv_dgm_2}
\end{figure}

\begin{figure}
    \centering
    \includegraphics[width=1\textwidth]{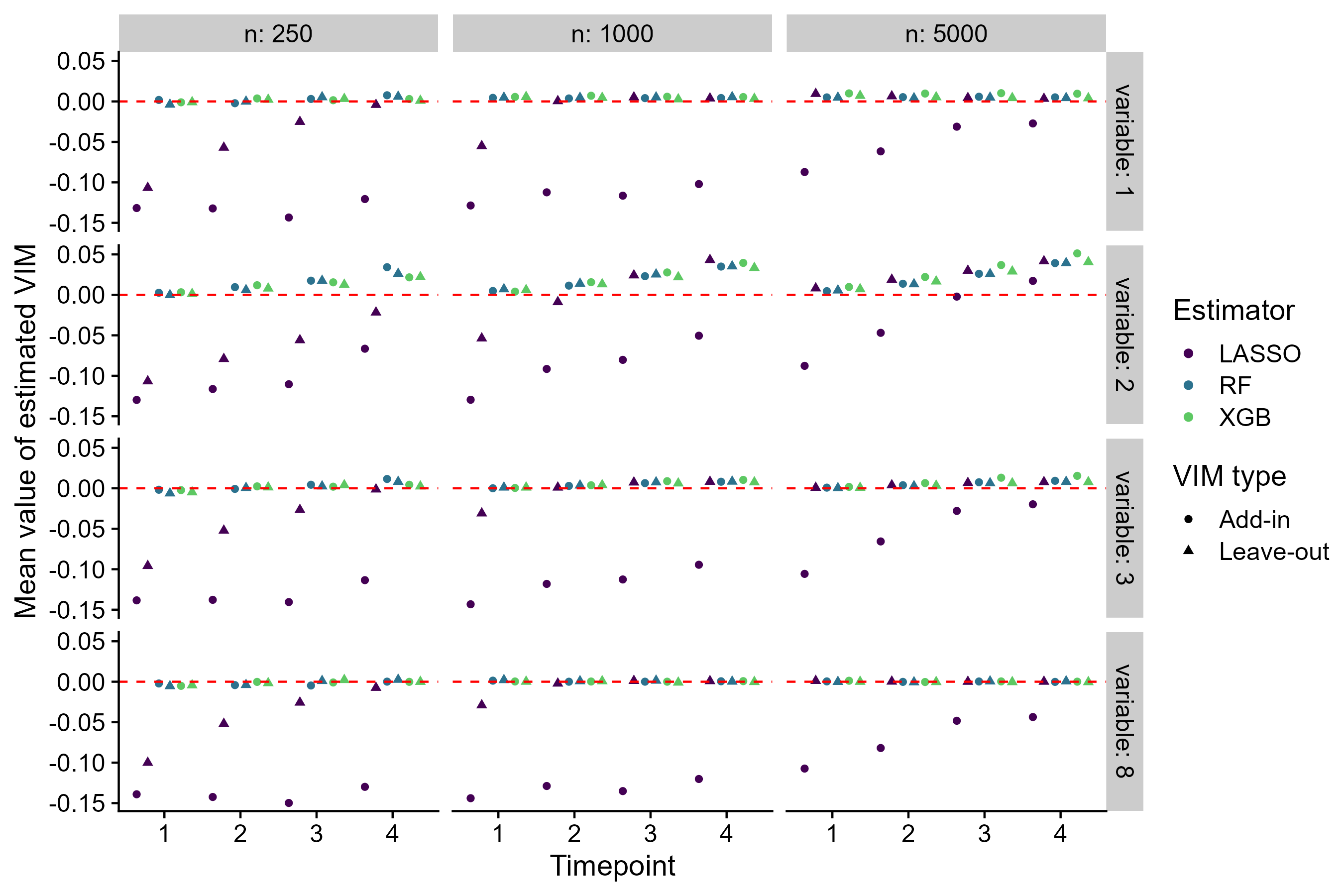}
    \caption{Mean estimated variable importance defined using sensitivity at the 95$^\text{th}$ percentile over 1000 simulated replicates versus time for four variables: three with true VIM value greater than zero (variables 1, 2, and 3) and one with true VIM value equal to zero (variable 8). Columns: sample size. Rows: variable. Shapes (circles and triangles) denote VIM type, while colors denote the estimator: lasso, random forests (RF), or boosted trees (XGB).}
    \label{fig:sim_vim_over_time_supp_0.5_sensitivity_dgm_2}
\end{figure}

In Figures~\ref{fig:sim_vim_over_time_0.5_ppv_dgm_2} and \ref{fig:sim_vim_over_time_0.5_sensitivity_dgm_2}, we display the estimated VIMs defined using PPV and sensitivity at the 95$^\text{th}$ percentile at all sample sizes using logistic regression and a super learner. In Figures~\ref{fig:sim_vim_over_time_supp_0.5_auc_dgm_2}--
\ref{fig:sim_vim_over_time_supp_0.5_sensitivity_dgm_2}, we show the mean estimated variable importance over time using lasso, random forests (RF), and boosted trees (XGB) for AUC, PPV and sensitivity at the 95$^\text{th}$ percentile.

\clearpage

In Tables~\ref{tab:sim_all_avg_1_0.5_auc_dgm_2}--\ref{tab:sim_all_slope_8_0.5_sensitivity_dgm_2}, we display the estimated average VIM and slope of the linear VIM trend for each variable, including all estimators and all sample sizes. In most cases, we observe bias decreasing to zero as $n$ grows and coverage increasing to the nominal level. As noted in the main manuscript, for sensitivity at the 95$^\text{th}$ percentile of predicted risk, we observe good coverage for add-in importance when using logistic regression (GLM). The low coverage observed in some cases for estimators based on random forests and boosted trees (including the super learner, which includes these in its library) may have arisen because we did not use cross-validation to tune these estimators, which is necessary in practical applications; we included simple versions of these algorithms here to show the ease with which one can compare results across algorithms.

\begingroup\fontsize{9}{11}\selectfont

\begin{longtable}[t]{llrrrrrrr}
\caption{Performance of estimators of the average VIM, defined using AUC, for variable 1 over the time series, summarized over 1000 Monte-Carlo replications. Performance shown for each sample size $n \in \{250, 1000, 5000\}$, and all estimators: logistic regression (GLM), lasso, random forests (RF), super learner (SL), and boosted trees (XGB). Both add-in (compared to four covariates) and leave-out (compared to nine covariates) variable importance are displayed. Each feature has correlation 0.5 with the previous timepoint. \label{tab:sim_all_avg_1_0.5_auc_dgm_2}}\\
\toprule
VIM type: variable & Estimator & $n$ & Mean est. & True value & Empirical SE & Coverage & Rejection prop. & CI width\\
\midrule
\endfirsthead
\caption[]{Performance of estimators of the average VIM, defined using AUC, for variable 1 over the time series, summarized over 1000 Monte-Carlo replications. \textit{(continued)}}\\
\toprule
VIM type: variable & Estimator & $n$ & Mean est. & True value & Empirical SE & Coverage & Rejection prop. & CI width\\
\midrule
\endhead

\endfoot
\bottomrule
\endlastfoot
Add-in: 1 & GLM & 250 & 0.020 & 0.035 & 0.052 & 0.905 & 0.146 & 0.179\\
Add-in: 1 & GLM & 1000 & 0.033 & 0.035 & 0.026 & 0.928 & 0.395 & 0.094\\
Add-in: 1 & GLM & 5000 & 0.037 & 0.035 & 0.011 & 0.966 & 0.931 & 0.048\\
Leave-out: 1 & GLM & 250 & 0.006 & 0.013 & 0.048 & 0.930 & 0.084 & 0.174\\
Leave-out: 1 & GLM & 1000 & 0.012 & 0.013 & 0.024 & 0.940 & 0.143 & 0.091\\
Leave-out: 1 & GLM & 5000 & 0.013 & 0.013 & 0.010 & 0.979 & 0.274 & 0.045\\
Add-in: 1 & LASSO & 250 & 0.012 & 0.035 & 0.035 & 0.993 & 0.017 & 0.233\\
Add-in: 1 & LASSO & 1000 & 0.031 & 0.035 & 0.023 & 0.992 & 0.158 & 0.128\\
\end{longtable}
\endgroup{}
\begingroup\fontsize{9}{11}\selectfont

\begin{longtable}[t]{llrrrrrrr}
\caption{Performance of estimators of the average VIM, defined using AUC, for variable 2 over the time series, summarized over 1000 Monte-Carlo replications. Performance shown for each sample size $n \in \{250, 1000, 5000\}$, and all estimators: logistic regression (GLM), lasso, random forests (RF), super learner (SL), and boosted trees (XGB). Both add-in (compared to four covariates) and leave-out (compared to nine covariates) variable importance are displayed. Each feature has correlation 0.5 with the previous timepoint. \label{tab:sim_all_avg_2_0.5_auc_dgm_2}}\\
\toprule
VIM type: variable & Estimator & $n$ & Mean est. & True value & Empirical SE & Coverage & Rejection prop. & CI width\\
\midrule
\endfirsthead
\caption[]{Performance of estimators of the average VIM, defined using AUC, for variable 2 over the time series, summarized over 1000 Monte-Carlo replications. \textit{(continued)}}\\
\toprule
VIM type: variable & Estimator & $n$ & Mean est. & True value & Empirical SE & Coverage & Rejection prop. & CI width\\
\midrule
\endhead

\endfoot
\bottomrule
\endlastfoot
Add-in: 2 & GLM & 250 & 0.074 & 0.087 & 0.054 & 0.886 & 0.497 & 0.177\\
Add-in: 2 & GLM & 1000 & 0.090 & 0.087 & 0.027 & 0.918 & 0.971 & 0.093\\
Add-in: 2 & GLM & 5000 & 0.091 & 0.087 & 0.012 & 0.947 & 1.000 & 0.047\\
Leave-out: 2 & GLM & 250 & 0.050 & 0.056 & 0.051 & 0.902 & 0.322 & 0.176\\
Leave-out: 2 & GLM & 1000 & 0.059 & 0.056 & 0.025 & 0.927 & 0.796 & 0.092\\
Leave-out: 2 & GLM & 5000 & 0.057 & 0.056 & 0.011 & 0.961 & 0.998 & 0.046\\
Add-in: 2 & LASSO & 250 & 0.071 & 0.087 & 0.044 & 0.991 & 0.177 & 0.257\\
Add-in: 2 & LASSO & 1000 & 0.096 & 0.087 & 0.022 & 0.996 & 0.962 & 0.138\\
\end{longtable}
\endgroup{}
\begingroup\fontsize{9}{11}\selectfont

\begin{longtable}[t]{llrrrrrrr}
\caption{Performance of estimators of the average VIM, defined using AUC, for variable 3 over the time series, summarized over 1000 Monte-Carlo replications. Performance shown for each sample size $n \in \{250, 1000, 5000\}$, and all estimators: logistic regression (GLM), lasso, random forests (RF), super learner (SL), and boosted trees (XGB). Both add-in (compared to four covariates) and leave-out (compared to nine covariates) variable importance are displayed. Each feature has correlation 0.5 with the previous timepoint. \label{tab:sim_all_avg_3_0.5_auc_dgm_2}}\\
\toprule
VIM type: variable & Estimator & $n$ & Mean est. & True value & Empirical SE & Coverage & Rejection prop. & CI width\\
\midrule
\endfirsthead
\caption[]{Performance of estimators of the average VIM, defined using AUC, for variable 3 over the time series, summarized over 1000 Monte-Carlo replications. \textit{(continued)}}\\
\toprule
VIM type: variable & Estimator & $n$ & Mean est. & True value & Empirical SE & Coverage & Rejection prop. & CI width\\
\midrule
\endhead

\endfoot
\bottomrule
\endlastfoot
Add-in: 3 & GLM & 250 & 0.022 & 0.031 & 0.052 & 0.918 & 0.152 & 0.179\\
Add-in: 3 & GLM & 1000 & 0.031 & 0.031 & 0.026 & 0.919 & 0.369 & 0.094\\
Add-in: 3 & GLM & 5000 & 0.033 & 0.031 & 0.011 & 0.965 & 0.885 & 0.048\\
Leave-out: 3 & GLM & 250 & 0.005 & 0.010 & 0.047 & 0.930 & 0.073 & 0.174\\
Leave-out: 3 & GLM & 1000 & 0.009 & 0.010 & 0.024 & 0.934 & 0.110 & 0.091\\
Leave-out: 3 & GLM & 5000 & 0.010 & 0.010 & 0.010 & 0.974 & 0.180 & 0.045\\
Add-in: 3 & LASSO & 250 & 0.011 & 0.031 & 0.034 & 0.996 & 0.008 & 0.234\\
Add-in: 3 & LASSO & 1000 & 0.028 & 0.031 & 0.022 & 0.991 & 0.133 & 0.122\\
\end{longtable}
\endgroup{}
\begingroup\fontsize{9}{11}\selectfont

\begin{longtable}[t]{llrrrrrrr}
\caption{Performance of estimators of the average VIM, defined using AUC, for variable 8 over the time series, summarized over 1000 Monte-Carlo replications. Performance shown for each sample size $n \in \{250, 1000, 5000\}$, and all estimators: logistic regression (GLM), lasso, random forests (RF), super learner (SL), and boosted trees (XGB). Both add-in (compared to four covariates) and leave-out (compared to nine covariates) variable importance are displayed. Each feature has correlation 0.5 with the previous timepoint. \label{tab:sim_all_avg_8_0.5_auc_dgm_2}}\\
\toprule
VIM type: variable & Estimator & $n$ & Mean est. & True value & Empirical SE & Coverage & Rejection prop. & CI width\\
\midrule
\endfirsthead
\caption[]{Performance of estimators of the average VIM, defined using AUC, for variable 8 over the time series, summarized over 1000 Monte-Carlo replications. \textit{(continued)}}\\
\toprule
VIM type: variable & Estimator & $n$ & Mean est. & True value & Empirical SE & Coverage & Rejection prop. & CI width\\
\midrule
\endhead

\endfoot
\bottomrule
\endlastfoot
Add-in: 8 & GLM & 250 & 0.000 & 0 & 0.048 & 0.931 & 0.064 & 0.179\\
Add-in: 8 & GLM & 1000 & 0.000 & 0 & 0.024 & 0.951 & 0.052 & 0.094\\
Add-in: 8 & GLM & 5000 & -0.001 & 0 & 0.010 & 0.973 & 0.021 & 0.048\\
Leave-out: 8 & GLM & 250 & -0.006 & 0 & 0.047 & 0.925 & 0.047 & 0.174\\
Leave-out: 8 & GLM & 1000 & -0.002 & 0 & 0.024 & 0.936 & 0.050 & 0.090\\
Leave-out: 8 & GLM & 5000 & -0.001 & 0 & 0.010 & 0.975 & 0.021 & 0.045\\
Add-in: 8 & LASSO & 250 & 0.000 & 0 & 0.029 & 0.996 & 0.003 & 0.229\\
Add-in: 8 & LASSO & 1000 & 0.000 & 0 & 0.015 & 0.999 & 0.003 & 0.114\\
\end{longtable}
\endgroup{}
\begingroup\fontsize{9}{11}\selectfont

\begin{longtable}[t]{llrrrrrrr}
\caption{Performance of estimators of the linear VIM trend, defined using AUC, for variable 1 over the time series, summarized over 1000 Monte-Carlo replications. Performance shown for each sample size $n \in \{250, 1000, 5000\}$, and all estimators: logistic regression (GLM), lasso, random forests (RF), super learner (SL), and boosted trees (XGB). Both add-in (compared to four covariates) and leave-out (compared to nine covariates) variable importance are displayed. Each feature has correlation 0.5 with the previous timepoint. \label{tab:sim_all_slope_1_0.5_auc_dgm_2}}\\
\toprule
VIM type: variable & Estimator & $n$ & Mean est. & True value & Empirical SE & Coverage & Rejection prop. & CI width\\
\midrule
\endfirsthead
\caption[]{Performance of estimators of the linear VIM trend, defined using AUC, for variable 1 over the time series, summarized over 1000 Monte-Carlo replications. \textit{(continued)}}\\
\toprule
VIM type: variable & Estimator & $n$ & Mean est. & True value & Empirical SE & Coverage & Rejection prop. & CI width\\
\midrule
\endhead

\endfoot
\bottomrule
\endlastfoot
Add-in: 1 & GLM & 250 & 0.004 & -0.001 & 0.046 & 0.918 & 0.082 & 0.161\\
Add-in: 1 & GLM & 1000 & 0.000 & -0.001 & 0.024 & 0.925 & 0.080 & 0.084\\
Add-in: 1 & GLM & 5000 & -0.001 & -0.001 & 0.010 & 0.961 & 0.042 & 0.042\\
Leave-out: 1 & GLM & 250 & -0.002 & -0.004 & 0.042 & 0.923 & 0.072 & 0.156\\
Leave-out: 1 & GLM & 1000 & -0.005 & -0.004 & 0.021 & 0.946 & 0.060 & 0.081\\
Leave-out: 1 & GLM & 5000 & -0.004 & -0.004 & 0.009 & 0.975 & 0.045 & 0.041\\
Add-in: 1 & LASSO & 250 & 0.003 & -0.001 & 0.032 & 0.993 & 0.007 & 0.207\\
Add-in: 1 & LASSO & 1000 & 0.003 & -0.001 & 0.020 & 0.984 & 0.015 & 0.102\\
\end{longtable}
\endgroup{}
\begingroup\fontsize{9}{11}\selectfont

\begin{longtable}[t]{llrrrrrrr}
\caption{Performance of estimators of the linear VIM trend, defined using AUC, for variable 2 over the time series, summarized over 1000 Monte-Carlo replications. Performance shown for each sample size $n \in \{250, 1000, 5000\}$, and all estimators: logistic regression (GLM), lasso, random forests (RF), super learner (SL), and boosted trees (XGB). Both add-in (compared to four covariates) and leave-out (compared to nine covariates) variable importance are displayed. Each feature has correlation 0.5 with the previous timepoint. \label{tab:sim_all_slope_2_0.5_auc_dgm_2}}\\
\toprule
VIM type: variable & Estimator & $n$ & Mean est. & True value & Empirical SE & Coverage & Rejection prop. & CI width\\
\midrule
\endfirsthead
\caption[]{Performance of estimators of the linear VIM trend, defined using AUC, for variable 2 over the time series, summarized over 1000 Monte-Carlo replications. \textit{(continued)}}\\
\toprule
VIM type: variable & Estimator & $n$ & Mean est. & True value & Empirical SE & Coverage & Rejection prop. & CI width\\
\midrule
\endhead

\endfoot
\bottomrule
\endlastfoot
Add-in: 2 & GLM & 250 & 0.039 & 0.033 & 0.048 & 0.898 & 0.223 & 0.159\\
Add-in: 2 & GLM & 1000 & 0.037 & 0.033 & 0.024 & 0.915 & 0.454 & 0.083\\
Add-in: 2 & GLM & 5000 & 0.034 & 0.033 & 0.010 & 0.961 & 0.901 & 0.042\\
Leave-out: 2 & GLM & 250 & 0.029 & 0.024 & 0.044 & 0.918 & 0.136 & 0.157\\
Leave-out: 2 & GLM & 1000 & 0.026 & 0.024 & 0.022 & 0.935 & 0.258 & 0.082\\
Leave-out: 2 & GLM & 5000 & 0.025 & 0.024 & 0.010 & 0.965 & 0.664 & 0.041\\
Add-in: 2 & LASSO & 250 & 0.045 & 0.033 & 0.036 & 0.989 & 0.059 & 0.206\\
Add-in: 2 & LASSO & 1000 & 0.043 & 0.033 & 0.020 & 0.986 & 0.381 & 0.099\\
\end{longtable}
\endgroup{}
\begingroup\fontsize{9}{11}\selectfont

\begin{longtable}[t]{llrrrrrrr}
\caption{Performance of estimators of the linear VIM trend, defined using AUC, for variable 3 over the time series, summarized over 1000 Monte-Carlo replications. Performance shown for each sample size $n \in \{250, 1000, 5000\}$, and all estimators: logistic regression (GLM), lasso, random forests (RF), super learner (SL), and boosted trees (XGB). Both add-in (compared to four covariates) and leave-out (compared to nine covariates) variable importance are displayed. Each feature has correlation 0.5 with the previous timepoint. \label{tab:sim_all_slope_3_0.5_auc_dgm_2}}\\
\toprule
VIM type: variable & Estimator & $n$ & Mean est. & True value & Empirical SE & Coverage & Rejection prop. & CI width\\
\midrule
\endfirsthead
\caption[]{Performance of estimators of the linear VIM trend, defined using AUC, for variable 3 over the time series, summarized over 1000 Monte-Carlo replications. \textit{(continued)}}\\
\toprule
VIM type: variable & Estimator & $n$ & Mean est. & True value & Empirical SE & Coverage & Rejection prop. & CI width\\
\midrule
\endhead

\endfoot
\bottomrule
\endlastfoot
Add-in: 3 & GLM & 250 & 0.016 & 0.015 & 0.045 & 0.927 & 0.095 & 0.160\\
Add-in: 3 & GLM & 1000 & 0.017 & 0.015 & 0.022 & 0.939 & 0.142 & 0.084\\
Add-in: 3 & GLM & 5000 & 0.017 & 0.015 & 0.010 & 0.967 & 0.336 & 0.042\\
Leave-out: 3 & GLM & 250 & 0.005 & 0.005 & 0.042 & 0.932 & 0.067 & 0.156\\
Leave-out: 3 & GLM & 1000 & 0.005 & 0.005 & 0.020 & 0.951 & 0.047 & 0.081\\
Leave-out: 3 & GLM & 5000 & 0.005 & 0.005 & 0.009 & 0.975 & 0.046 & 0.041\\
Add-in: 3 & LASSO & 250 & 0.012 & 0.015 & 0.031 & 0.999 & 0.004 & 0.208\\
Add-in: 3 & LASSO & 1000 & 0.019 & 0.015 & 0.018 & 0.996 & 0.035 & 0.104\\
\end{longtable}
\endgroup{}
\begingroup\fontsize{9}{11}\selectfont

\begin{longtable}[t]{llrrrrrrr}
\caption{Performance of estimators of the linear VIM trend, defined using AUC, for variable 8 over the time series, summarized over 1000 Monte-Carlo replications. Performance shown for each sample size $n \in \{250, 1000, 5000\}$, and all estimators: logistic regression (GLM), lasso, random forests (RF), super learner (SL), and boosted trees (XGB). Both add-in (compared to four covariates) and leave-out (compared to nine covariates) variable importance are displayed. Each feature has correlation 0.5 with the previous timepoint. \label{tab:sim_all_slope_8_0.5_auc_dgm_2}}\\
\toprule
VIM type: variable & Estimator & $n$ & Mean est. & True value & Empirical SE & Coverage & Rejection prop. & CI width\\
\midrule
\endfirsthead
\caption[]{Performance of estimators of the linear VIM trend, defined using AUC, for variable 8 over the time series, summarized over 1000 Monte-Carlo replications. \textit{(continued)}}\\
\toprule
VIM type: variable & Estimator & $n$ & Mean est. & True value & Empirical SE & Coverage & Rejection prop. & CI width\\
\midrule
\endhead

\endfoot
\bottomrule
\endlastfoot
Add-in: 8 & GLM & 250 & 0.002 & 0 & 0.043 & 0.936 & 0.064 & 0.160\\
Add-in: 8 & GLM & 1000 & 0.000 & 0 & 0.021 & 0.951 & 0.049 & 0.084\\
Add-in: 8 & GLM & 5000 & 0.000 & 0 & 0.009 & 0.982 & 0.018 & 0.043\\
Leave-out: 8 & GLM & 250 & -0.001 & 0 & 0.041 & 0.940 & 0.060 & 0.155\\
Leave-out: 8 & GLM & 1000 & 0.000 & 0 & 0.020 & 0.958 & 0.042 & 0.080\\
Leave-out: 8 & GLM & 5000 & 0.000 & 0 & 0.009 & 0.973 & 0.027 & 0.040\\
Add-in: 8 & LASSO & 250 & 0.001 & 0 & 0.025 & 0.999 & 0.001 & 0.205\\
Add-in: 8 & LASSO & 1000 & 0.000 & 0 & 0.013 & 0.999 & 0.001 & 0.102\\
\end{longtable}
\endgroup{}

\begingroup\fontsize{9}{11}\selectfont

\begin{longtable}[t]{llrrrrrrr}
\caption{Performance of estimators of the average VIM, defined using PPV at the 95th percentile of predicted risk, for variable 1 over the time series, summarized over 1000 Monte-Carlo replications. Performance shown for each sample size $n \in \{250, 1000, 5000\}$, and all estimators: logistic regression (GLM), lasso, random forests (RF), super learner (SL), and boosted trees (XGB). Both add-in (compared to four covariates) and leave-out (compared to nine covariates) variable importance are displayed. Each feature has correlation 0.5 with the previous timepoint. \label{tab:sim_all_avg_1_0.5_ppv_dgm_2}}\\
\toprule
VIM type: variable & Estimator & $n$ & Mean est. & True value & Empirical SE & Coverage & Rejection prop. & CI width\\
\midrule
\endfirsthead
\caption[]{Performance of estimators of the average VIM, defined using PPV at the 95th percentile of predicted risk, for variable 1 over the time series, summarized over 1000 Monte-Carlo replications. \textit{(continued)}}\\
\toprule
VIM type: variable & Estimator & $n$ & Mean est. & True value & Empirical SE & Coverage & Rejection prop. & CI width\\
\midrule
\endhead

\endfoot
\bottomrule
\endlastfoot
Add-in: 1 & GLM & 250 & 0.023 & 0.040 & 0.124 & 0.472 & 0.273 & 0.542\\
Add-in: 1 & GLM & 1000 & 0.042 & 0.040 & 0.072 & 0.877 & 0.226 & 0.236\\
Add-in: 1 & GLM & 5000 & 0.041 & 0.040 & 0.034 & 0.955 & 0.385 & 0.132\\
Leave-out: 1 & GLM & 250 & 0.009 & 0.018 & 0.142 & 0.643 & 0.133 & 0.631\\
Leave-out: 1 & GLM & 1000 & 0.019 & 0.018 & 0.074 & 0.884 & 0.125 & 0.238\\
Leave-out: 1 & GLM & 5000 & 0.018 & 0.018 & 0.029 & 0.971 & 0.117 & 0.130\\
Add-in: 1 & LASSO & 250 & 0.016 & 0.040 & 0.099 & 0.618 & 0.000 & 0.857\\
Add-in: 1 & LASSO & 1000 & 0.066 & 0.040 & 0.070 & 0.952 & 0.047 & 0.785\\
\end{longtable}
\endgroup{}
\begingroup\fontsize{9}{11}\selectfont

\begin{longtable}[t]{llrrrrrrr}
\caption{Performance of estimators of the average VIM, defined using PPV at the 95th percentile of predicted risk, for variable 2 over the time series, summarized over 1000 Monte-Carlo replications. Performance shown for each sample size $n \in \{250, 1000, 5000\}$, and all estimators: logistic regression (GLM), lasso, random forests (RF), super learner (SL), and boosted trees (XGB). Both add-in (compared to four covariates) and leave-out (compared to nine covariates) variable importance are displayed. Each feature has correlation 0.5 with the previous timepoint. \label{tab:sim_all_avg_2_0.5_ppv_dgm_2}}\\
\toprule
VIM type: variable & Estimator & $n$ & Mean est. & True value & Empirical SE & Coverage & Rejection prop. & CI width\\
\midrule
\endfirsthead
\caption[]{Performance of estimators of the average VIM, defined using PPV at the 95th percentile of predicted risk, for variable 2 over the time series, summarized over 1000 Monte-Carlo replications. \textit{(continued)}}\\
\toprule
VIM type: variable & Estimator & $n$ & Mean est. & True value & Empirical SE & Coverage & Rejection prop. & CI width\\
\midrule
\endhead

\endfoot
\bottomrule
\endlastfoot
Add-in: 2 & GLM & 250 & 0.100 & 0.111 & 0.134 & 0.537 & 0.556 & 0.567\\
Add-in: 2 & GLM & 1000 & 0.121 & 0.111 & 0.072 & 0.895 & 0.607 & 0.243\\
Add-in: 2 & GLM & 5000 & 0.114 & 0.111 & 0.034 & 0.946 & 0.939 & 0.135\\
Leave-out: 2 & GLM & 250 & 0.066 & 0.079 & 0.134 & 0.626 & 0.375 & 0.585\\
Leave-out: 2 & GLM & 1000 & 0.084 & 0.079 & 0.071 & 0.893 & 0.423 & 0.232\\
Leave-out: 2 & GLM & 5000 & 0.081 & 0.079 & 0.029 & 0.975 & 0.825 & 0.127\\
Add-in: 2 & LASSO & 250 & 0.113 & 0.111 & 0.119 & 0.627 & NaN & 0.731\\
Add-in: 2 & LASSO & 1000 & 0.162 & 0.111 & 0.073 & 0.965 & 0.365 & 0.656\\
\end{longtable}
\endgroup{}
\begingroup\fontsize{9}{11}\selectfont

\begin{longtable}[t]{llrrrrrrr}
\caption{Performance of estimators of the average VIM, defined using PPV at the 95th percentile of predicted risk, for variable 3 over the time series, summarized over 1000 Monte-Carlo replications. Performance shown for each sample size $n \in \{250, 1000, 5000\}$, and all estimators: logistic regression (GLM), lasso, random forests (RF), super learner (SL), and boosted trees (XGB). Both add-in (compared to four covariates) and leave-out (compared to nine covariates) variable importance are displayed. Each feature has correlation 0.5 with the previous timepoint. \label{tab:sim_all_avg_3_0.5_ppv_dgm_2}}\\
\toprule
VIM type: variable & Estimator & $n$ & Mean est. & True value & Empirical SE & Coverage & Rejection prop. & CI width\\
\midrule
\endfirsthead
\caption[]{Performance of estimators of the average VIM, defined using PPV at the 95th percentile of predicted risk, for variable 3 over the time series, summarized over 1000 Monte-Carlo replications. \textit{(continued)}}\\
\toprule
VIM type: variable & Estimator & $n$ & Mean est. & True value & Empirical SE & Coverage & Rejection prop. & CI width\\
\midrule
\endhead

\endfoot
\bottomrule
\endlastfoot
Add-in: 3 & GLM & 250 & 0.027 & 0.037 & 0.126 & 0.481 & 0.500 & 0.524\\
Add-in: 3 & GLM & 1000 & 0.038 & 0.037 & 0.071 & 0.878 & 0.214 & 0.237\\
Add-in: 3 & GLM & 5000 & 0.039 & 0.037 & 0.034 & 0.951 & 0.332 & 0.134\\
Leave-out: 3 & GLM & 250 & 0.012 & 0.015 & 0.136 & 0.627 & 0.000 & 0.629\\
Leave-out: 3 & GLM & 1000 & 0.016 & 0.015 & 0.074 & 0.880 & 0.134 & 0.239\\
Leave-out: 3 & GLM & 5000 & 0.016 & 0.015 & 0.029 & 0.974 & 0.100 & 0.130\\
Add-in: 3 & LASSO & 250 & 0.015 & 0.037 & 0.098 & 0.613 & NaN & 0.648\\
Add-in: 3 & LASSO & 1000 & 0.055 & 0.037 & 0.070 & 0.919 & 0.000 & 0.797\\
\end{longtable}
\endgroup{}
\begingroup\fontsize{9}{11}\selectfont

\begin{longtable}[t]{llrrrrrrr}
\caption{Performance of estimators of the average VIM, defined using PPV at the 95th percentile of predicted risk, for variable 8 over the time series, summarized over 1000 Monte-Carlo replications. Performance shown for each sample size $n \in \{250, 1000, 5000\}$, and all estimators: logistic regression (GLM), lasso, random forests (RF), super learner (SL), and boosted trees (XGB). Both add-in (compared to four covariates) and leave-out (compared to nine covariates) variable importance are displayed. Each feature has correlation 0.5 with the previous timepoint. \label{tab:sim_all_avg_8_0.5_ppv_dgm_2}}\\
\toprule
VIM type: variable & Estimator & $n$ & Mean est. & True value & Empirical SE & Coverage & Rejection prop. & CI width\\
\midrule
\endfirsthead
\caption[]{Performance of estimators of the average VIM, defined using PPV at the 95th percentile of predicted risk, for variable 8 over the time series, summarized over 1000 Monte-Carlo replications. \textit{(continued)}}\\
\toprule
VIM type: variable & Estimator & $n$ & Mean est. & True value & Empirical SE & Coverage & Rejection prop. & CI width\\
\midrule
\endhead

\endfoot
\bottomrule
\endlastfoot
Add-in: 8 & GLM & 250 & -0.003 & 0 & 0.121 & 0.429 & 0.500 & 0.513\\
Add-in: 8 & GLM & 1000 & -0.001 & 0 & 0.069 & 0.893 & 0.105 & 0.234\\
Add-in: 8 & GLM & 5000 & -0.003 & 0 & 0.036 & 0.942 & 0.048 & 0.138\\
Leave-out: 8 & GLM & 250 & -0.008 & 0 & 0.140 & 0.626 & 0.182 & 0.595\\
Leave-out: 8 & GLM & 1000 & -0.001 & 0 & 0.075 & 0.879 & 0.088 & 0.240\\
Leave-out: 8 & GLM & 5000 & 0.001 & 0 & 0.029 & 0.980 & 0.033 & 0.131\\
Add-in: 8 & LASSO & 250 & -0.023 & 0 & 0.086 & 0.522 & NaN & 0.644\\
Add-in: 8 & LASSO & 1000 & -0.003 & 0 & 0.062 & 0.891 & 0.077 & 0.869\\
\end{longtable}
\endgroup{}
\begingroup\fontsize{9}{11}\selectfont

\begin{longtable}[t]{llrrrrrrr}
\caption{Performance of estimators of the linear VIM trend, defined using PPV at the 95th percentile of predicted risk, for variable 1 over the time series, summarized over 1000 Monte-Carlo replications. Performance shown for each sample size $n \in \{250, 1000, 5000\}$, and all estimators: logistic regression (GLM), lasso, random forests (RF), super learner (SL), and boosted trees (XGB). Both add-in (compared to four covariates) and leave-out (compared to nine covariates) variable importance are displayed. Each feature has correlation 0.5 with the previous timepoint. \label{tab:sim_all_slope_1_0.5_ppv_dgm_2}}\\
\toprule
VIM type: variable & Estimator & $n$ & Mean est. & True value & Empirical SE & Coverage & Rejection prop. & CI width\\
\midrule
\endfirsthead
\caption[]{Performance of estimators of the linear VIM trend, defined using PPV at the 95th percentile of predicted risk, for variable 1 over the time series, summarized over 1000 Monte-Carlo replications. \textit{(continued)}}\\
\toprule
VIM type: variable & Estimator & $n$ & Mean est. & True value & Empirical SE & Coverage & Rejection prop. & CI width\\
\midrule
\endhead

\endfoot
\bottomrule
\endlastfoot
Add-in: 1 & GLM & 250 & 0.005 & 0.000 & 0.108 & 0.447 & 0.364 & 0.440\\
Add-in: 1 & GLM & 1000 & -0.001 & 0.000 & 0.064 & 0.881 & 0.131 & 0.209\\
Add-in: 1 & GLM & 5000 & -0.001 & 0.000 & 0.030 & 0.970 & 0.031 & 0.118\\
Leave-out: 1 & GLM & 250 & -0.006 & -0.003 & 0.119 & 0.613 & 0.267 & 0.517\\
Leave-out: 1 & GLM & 1000 & 0.000 & -0.003 & 0.066 & 0.878 & 0.123 & 0.210\\
Leave-out: 1 & GLM & 5000 & -0.005 & -0.003 & 0.027 & 0.957 & 0.044 & 0.116\\
Add-in: 1 & LASSO & 250 & 0.009 & 0.000 & 0.087 & 0.570 & 1.000 & 0.649\\
Add-in: 1 & LASSO & 1000 & 0.005 & 0.000 & 0.063 & 0.928 & 0.023 & 0.619\\
\end{longtable}
\endgroup{}
\begingroup\fontsize{9}{11}\selectfont

\begin{longtable}[t]{llrrrrrrr}
\caption{Performance of estimators of the linear VIM trend, defined using PPV at the 95th percentile of predicted risk, for variable 2 over the time series, summarized over 1000 Monte-Carlo replications. Performance shown for each sample size $n \in \{250, 1000, 5000\}$, and all estimators: logistic regression (GLM), lasso, random forests (RF), super learner (SL), and boosted trees (XGB). Both add-in (compared to four covariates) and leave-out (compared to nine covariates) variable importance are displayed. Each feature has correlation 0.5 with the previous timepoint. \label{tab:sim_all_slope_2_0.5_ppv_dgm_2}}\\
\toprule
VIM type: variable & Estimator & $n$ & Mean est. & True value & Empirical SE & Coverage & Rejection prop. & CI width\\
\midrule
\endfirsthead
\caption[]{Performance of estimators of the linear VIM trend, defined using PPV at the 95th percentile of predicted risk, for variable 2 over the time series, summarized over 1000 Monte-Carlo replications. \textit{(continued)}}\\
\toprule
VIM type: variable & Estimator & $n$ & Mean est. & True value & Empirical SE & Coverage & Rejection prop. & CI width\\
\midrule
\endhead

\endfoot
\bottomrule
\endlastfoot
Add-in: 2 & GLM & 250 & 0.058 & 0.049 & 0.113 & 0.508 & 0.556 & 0.463\\
Add-in: 2 & GLM & 1000 & 0.053 & 0.049 & 0.068 & 0.869 & 0.239 & 0.216\\
Add-in: 2 & GLM & 5000 & 0.049 & 0.049 & 0.030 & 0.961 & 0.353 & 0.121\\
Leave-out: 2 & GLM & 250 & 0.036 & 0.039 & 0.119 & 0.595 & 0.312 & 0.480\\
Leave-out: 2 & GLM & 1000 & 0.046 & 0.039 & 0.065 & 0.879 & 0.212 & 0.205\\
Leave-out: 2 & GLM & 5000 & 0.038 & 0.039 & 0.026 & 0.968 & 0.262 & 0.113\\
Add-in: 2 & LASSO & 250 & 0.068 & 0.049 & 0.101 & 0.569 & NaN & 0.590\\
Add-in: 2 & LASSO & 1000 & 0.066 & 0.049 & 0.065 & 0.956 & 0.135 & 0.538\\
\end{longtable}
\endgroup{}
\begingroup\fontsize{9}{11}\selectfont

\begin{longtable}[t]{llrrrrrrr}
\caption{Performance of estimators of the linear VIM trend, defined using PPV at the 95th percentile of predicted risk, for variable 3 over the time series, summarized over 1000 Monte-Carlo replications. Performance shown for each sample size $n \in \{250, 1000, 5000\}$, and all estimators: logistic regression (GLM), lasso, random forests (RF), super learner (SL), and boosted trees (XGB). Both add-in (compared to four covariates) and leave-out (compared to nine covariates) variable importance are displayed. Each feature has correlation 0.5 with the previous timepoint. \label{tab:sim_all_slope_3_0.5_ppv_dgm_2}}\\
\toprule
VIM type: variable & Estimator & $n$ & Mean est. & True value & Empirical SE & Coverage & Rejection prop. & CI width\\
\midrule
\endfirsthead
\caption[]{Performance of estimators of the linear VIM trend, defined using PPV at the 95th percentile of predicted risk, for variable 3 over the time series, summarized over 1000 Monte-Carlo replications. \textit{(continued)}}\\
\toprule
VIM type: variable & Estimator & $n$ & Mean est. & True value & Empirical SE & Coverage & Rejection prop. & CI width\\
\midrule
\endhead

\endfoot
\bottomrule
\endlastfoot
Add-in: 3 & GLM & 250 & 0.023 & 0.019 & 0.105 & 0.450 & 0.250 & 0.419\\
Add-in: 3 & GLM & 1000 & 0.021 & 0.019 & 0.065 & 0.871 & 0.165 & 0.210\\
Add-in: 3 & GLM & 5000 & 0.019 & 0.019 & 0.031 & 0.956 & 0.104 & 0.120\\
Leave-out: 3 & GLM & 250 & 0.009 & 0.008 & 0.119 & 0.600 & 0.333 & 0.503\\
Leave-out: 3 & GLM & 1000 & 0.012 & 0.008 & 0.066 & 0.879 & 0.138 & 0.211\\
Leave-out: 3 & GLM & 5000 & 0.007 & 0.008 & 0.027 & 0.968 & 0.040 & 0.116\\
Add-in: 3 & LASSO & 250 & 0.028 & 0.019 & 0.093 & 0.553 & NaN & 0.529\\
Add-in: 3 & LASSO & 1000 & 0.038 & 0.019 & 0.064 & 0.879 & 0.065 & 0.657\\
\end{longtable}
\endgroup{}
\begingroup\fontsize{9}{11}\selectfont

\begin{longtable}[t]{llrrrrrrr}
\caption{Performance of estimators of the linear VIM trend, defined using PPV at the 95th percentile of predicted risk, for variable 8 over the time series, summarized over 1000 Monte-Carlo replications. Performance shown for each sample size $n \in \{250, 1000, 5000\}$, and all estimators: logistic regression (GLM), lasso, random forests (RF), super learner (SL), and boosted trees (XGB). Both add-in (compared to four covariates) and leave-out (compared to nine covariates) variable importance are displayed. Each feature has correlation 0.5 with the previous timepoint. \label{tab:sim_all_slope_8_0.5_ppv_dgm_2}}\\
\toprule
VIM type: variable & Estimator & $n$ & Mean est. & True value & Empirical SE & Coverage & Rejection prop. & CI width\\
\midrule
\endfirsthead
\caption[]{Performance of estimators of the linear VIM trend, defined using PPV at the 95th percentile of predicted risk, for variable 8 over the time series, summarized over 1000 Monte-Carlo replications. \textit{(continued)}}\\
\toprule
VIM type: variable & Estimator & $n$ & Mean est. & True value & Empirical SE & Coverage & Rejection prop. & CI width\\
\midrule
\endhead

\endfoot
\bottomrule
\endlastfoot
Add-in: 8 & GLM & 250 & 0.005 & 0 & 0.103 & 0.409 & 0.667 & 0.443\\
Add-in: 8 & GLM & 1000 & -0.003 & 0 & 0.063 & 0.882 & 0.129 & 0.208\\
Add-in: 8 & GLM & 5000 & 0.000 & 0 & 0.031 & 0.959 & 0.042 & 0.122\\
Leave-out: 8 & GLM & 250 & -0.007 & 0 & 0.121 & 0.594 & 0.273 & 0.492\\
Leave-out: 8 & GLM & 1000 & 0.005 & 0 & 0.066 & 0.890 & 0.115 & 0.211\\
Leave-out: 8 & GLM & 5000 & -0.002 & 0 & 0.027 & 0.965 & 0.035 & 0.116\\
Add-in: 8 & LASSO & 250 & 0.001 & 0 & 0.071 & 0.548 & NaN & 0.422\\
Add-in: 8 & LASSO & 1000 & 0.000 & 0 & 0.057 & 0.849 & 0.231 & 0.708\\
\end{longtable}
\endgroup{}

\begingroup\fontsize{9}{11}\selectfont

\begin{longtable}[t]{llrrrrrrr}
\caption{Performance of estimators of the average VIM, defined using sensitivity at the 95th percentile of predicted risk, for variable 1 over the time series, summarized over 1000 Monte-Carlo replications. Performance shown for each sample size $n \in \{250, 1000, 5000\}$, and all estimators: logistic regression (GLM), lasso, random forests (RF), super learner (SL), and boosted trees (XGB). Both add-in (compared to four covariates) and leave-out (compared to nine covariates) variable importance are displayed. Each feature has correlation 0.5 with the previous timepoint. \label{tab:sim_all_avg_1_0.5_sensitivity_dgm_2}}\\
\toprule
VIM type: variable & Estimator & $n$ & Mean est. & True value & Empirical SE & Coverage & Rejection prop. & CI width\\
\midrule
\endfirsthead
\caption[]{Performance of estimators of the average VIM, defined using sensitivity at the 95th percentile of predicted risk, for variable 1 over the time series, summarized over 1000 Monte-Carlo replications. \textit{(continued)}}\\
\toprule
VIM type: variable & Estimator & $n$ & Mean est. & True value & Empirical SE & Coverage & Rejection prop. & CI width\\
\midrule
\endhead

\endfoot
\bottomrule
\endlastfoot
Add-in: 1 & GLM & 250 & 0.007 & 0.012 & 0.032 & 0.899 & 0.117 & 0.109\\
Add-in: 1 & GLM & 1000 & 0.012 & 0.012 & 0.016 & 0.955 & 0.162 & 0.065\\
Add-in: 1 & GLM & 5000 & 0.014 & 0.012 & 0.010 & 0.917 & 0.454 & 0.036\\
Leave-out: 1 & GLM & 250 & 0.003 & 0.005 & 0.040 & 0.916 & 0.088 & 0.139\\
Leave-out: 1 & GLM & 1000 & 0.005 & 0.005 & 0.021 & 0.951 & 0.079 & 0.083\\
Leave-out: 1 & GLM & 5000 & 0.005 & 0.005 & 0.010 & 0.980 & 0.090 & 0.044\\
Add-in: 1 & LASSO & 250 & -0.132 & 0.012 & 0.089 & 0.161 & 0.003 & 0.099\\
Add-in: 1 & LASSO & 1000 & -0.115 & 0.012 & 0.082 & 0.105 & 0.014 & 0.062\\
\end{longtable}
\endgroup{}
\begingroup\fontsize{9}{11}\selectfont

\begin{longtable}[t]{llrrrrrrr}
\caption{Performance of estimators of the average VIM, defined using sensitivity at the 95th percentile of predicted risk, for variable 2 over the time series, summarized over 1000 Monte-Carlo replications. Performance shown for each sample size $n \in \{250, 1000, 5000\}$, and all estimators: logistic regression (GLM), lasso, random forests (RF), super learner (SL), and boosted trees (XGB). Both add-in (compared to four covariates) and leave-out (compared to nine covariates) variable importance are displayed. Each feature has correlation 0.5 with the previous timepoint. \label{tab:sim_all_avg_2_0.5_sensitivity_dgm_2}}\\
\toprule
VIM type: variable & Estimator & $n$ & Mean est. & True value & Empirical SE & Coverage & Rejection prop. & CI width\\
\midrule
\endfirsthead
\caption[]{Performance of estimators of the average VIM, defined using sensitivity at the 95th percentile of predicted risk, for variable 2 over the time series, summarized over 1000 Monte-Carlo replications. \textit{(continued)}}\\
\toprule
VIM type: variable & Estimator & $n$ & Mean est. & True value & Empirical SE & Coverage & Rejection prop. & CI width\\
\midrule
\endhead

\endfoot
\bottomrule
\endlastfoot
Add-in: 2 & GLM & 250 & 0.029 & 0.032 & 0.035 & 0.909 & 0.256 & 0.122\\
Add-in: 2 & GLM & 1000 & 0.036 & 0.032 & 0.017 & 0.961 & 0.625 & 0.072\\
Add-in: 2 & GLM & 5000 & 0.035 & 0.032 & 0.010 & 0.925 & 0.963 & 0.038\\
Leave-out: 2 & GLM & 250 & 0.019 & 0.023 & 0.038 & 0.915 & 0.172 & 0.131\\
Leave-out: 2 & GLM & 1000 & 0.025 & 0.023 & 0.019 & 0.960 & 0.329 & 0.078\\
Leave-out: 2 & GLM & 5000 & 0.024 & 0.023 & 0.009 & 0.987 & 0.761 & 0.042\\
Add-in: 2 & LASSO & 250 & -0.106 & 0.032 & 0.089 & 0.203 & 0.020 & 0.116\\
Add-in: 2 & LASSO & 1000 & -0.088 & 0.032 & 0.081 & 0.150 & 0.050 & 0.070\\
\end{longtable}
\endgroup{}
\begingroup\fontsize{9}{11}\selectfont

\begin{longtable}[t]{llrrrrrrr}
\caption{Performance of estimators of the average VIM, defined using sensitivity at the 95th percentile of predicted risk, for variable 3 over the time series, summarized over 1000 Monte-Carlo replications. Performance shown for each sample size $n \in \{250, 1000, 5000\}$, and all estimators: logistic regression (GLM), lasso, random forests (RF), super learner (SL), and boosted trees (XGB). Both add-in (compared to four covariates) and leave-out (compared to nine covariates) variable importance are displayed. Each feature has correlation 0.5 with the previous timepoint. \label{tab:sim_all_avg_3_0.5_sensitivity_dgm_2}}\\
\toprule
VIM type: variable & Estimator & $n$ & Mean est. & True value & Empirical SE & Coverage & Rejection prop. & CI width\\
\midrule
\endfirsthead
\caption[]{Performance of estimators of the average VIM, defined using sensitivity at the 95th percentile of predicted risk, for variable 3 over the time series, summarized over 1000 Monte-Carlo replications. \textit{(continued)}}\\
\toprule
VIM type: variable & Estimator & $n$ & Mean est. & True value & Empirical SE & Coverage & Rejection prop. & CI width\\
\midrule
\endhead

\endfoot
\bottomrule
\endlastfoot
Add-in: 3 & GLM & 250 & 0.007 & 0.011 & 0.032 & 0.909 & 0.117 & 0.110\\
Add-in: 3 & GLM & 1000 & 0.011 & 0.011 & 0.016 & 0.959 & 0.154 & 0.065\\
Add-in: 3 & GLM & 5000 & 0.013 & 0.011 & 0.010 & 0.912 & 0.410 & 0.035\\
Leave-out: 3 & GLM & 250 & 0.004 & 0.004 & 0.039 & 0.923 & 0.088 & 0.138\\
Leave-out: 3 & GLM & 1000 & 0.005 & 0.004 & 0.021 & 0.954 & 0.072 & 0.083\\
Leave-out: 3 & GLM & 5000 & 0.005 & 0.004 & 0.009 & 0.983 & 0.084 & 0.044\\
Add-in: 3 & LASSO & 250 & -0.133 & 0.011 & 0.089 & 0.161 & 0.005 & 0.099\\
Add-in: 3 & LASSO & 1000 & -0.117 & 0.011 & 0.084 & 0.115 & 0.014 & 0.061\\
\end{longtable}
\endgroup{}
\begingroup\fontsize{9}{11}\selectfont

\begin{longtable}[t]{llrrrrrrr}
\caption{Performance of estimators of the average VIM, defined using sensitivity at the 95th percentile of predicted risk, for variable 8 over the time series, summarized over 1000 Monte-Carlo replications. Performance shown for each sample size $n \in \{250, 1000, 5000\}$, and all estimators: logistic regression (GLM), lasso, random forests (RF), super learner (SL), and boosted trees (XGB). Both add-in (compared to four covariates) and leave-out (compared to nine covariates) variable importance are displayed. Each feature has correlation 0.5 with the previous timepoint. \label{tab:sim_all_avg_8_0.5_sensitivity_dgm_2}}\\
\toprule
VIM type: variable & Estimator & $n$ & Mean est. & True value & Empirical SE & Coverage & Rejection prop. & CI width\\
\midrule
\endfirsthead
\caption[]{Performance of estimators of the average VIM, defined using sensitivity at the 95th percentile of predicted risk, for variable 8 over the time series, summarized over 1000 Monte-Carlo replications. \textit{(continued)}}\\
\toprule
VIM type: variable & Estimator & $n$ & Mean est. & True value & Empirical SE & Coverage & Rejection prop. & CI width\\
\midrule
\endhead

\endfoot
\bottomrule
\endlastfoot
Add-in: 8 & GLM & 250 & -0.001 & 0 & 0.031 & 0.920 & 0.078 & 0.104\\
Add-in: 8 & GLM & 1000 & 0.000 & 0 & 0.015 & 0.961 & 0.043 & 0.061\\
Add-in: 8 & GLM & 5000 & 0.000 & 0 & 0.012 & 0.839 & 0.122 & 0.034\\
Leave-out: 8 & GLM & 250 & -0.002 & 0 & 0.040 & 0.917 & 0.073 & 0.141\\
Leave-out: 8 & GLM & 1000 & 0.000 & 0 & 0.021 & 0.960 & 0.047 & 0.084\\
Leave-out: 8 & GLM & 5000 & 0.000 & 0 & 0.010 & 0.980 & 0.030 & 0.045\\
Add-in: 8 & LASSO & 250 & -0.140 & 0 & 0.089 & 0.155 & 0.002 & 0.093\\
Add-in: 8 & LASSO & 1000 & -0.132 & 0 & 0.085 & 0.096 & 0.006 & 0.055\\
Add-in: 8 & LASSO & 5000 & -0.070 & 0 & 0.084 & 0.204 & 0.122 & 0.032\\
Leave-out: 8 & LASSO & 250 & -0.046 & 0 & 0.061 & 0.659 & 0.019 & 0.135\\
Leave-out: 8 & LASSO & 1000 & -0.007 & 0 & 0.030 & 0.894 & 0.033 & 0.083\\
Leave-out: 8 & LASSO & 5000 & 0.000 & 0 & 0.011 & 0.952 & 0.056 & 0.045\\
Add-in: 8 & RF & 250 & -0.003 & 0 & 0.033 & 0.921 & 0.065 & 0.113\\
Add-in: 8 & RF & 1000 & 0.000 & 0 & 0.016 & 0.949 & 0.060 & 0.064\\
Add-in: 8 & RF & 5000 & 0.000 & 0 & 0.008 & 0.970 & 0.034 & 0.033\\
Leave-out: 8 & RF & 250 & -0.001 & 0 & 0.039 & 0.903 & 0.069 & 0.133\\
Leave-out: 8 & RF & 1000 & 0.001 & 0 & 0.020 & 0.944 & 0.061 & 0.078\\
Leave-out: 8 & RF & 5000 & 0.000 & 0 & 0.009 & 0.987 & 0.023 & 0.041\\
Add-in: 8 & SL & 250 & -0.029 & 0 & 0.049 & 0.720 & 0.027 & 0.104\\
Add-in: 8 & SL & 1000 & -0.024 & 0 & 0.038 & 0.659 & 0.030 & 0.061\\
Add-in: 8 & SL & 5000 & -0.008 & 0 & 0.030 & 0.547 & 0.207 & 0.033\\
Leave-out: 8 & SL & 250 & -0.010 & 0 & 0.042 & 0.889 & 0.049 & 0.136\\
Leave-out: 8 & SL & 1000 & -0.001 & 0 & 0.021 & 0.950 & 0.040 & 0.083\\
Leave-out: 8 & SL & 5000 & 0.000 & 0 & 0.011 & 0.963 & 0.046 & 0.044\\
Add-in: 8 & XGB & 250 & -0.002 & 0 & 0.033 & 0.916 & 0.070 & 0.111\\
Add-in: 8 & XGB & 1000 & 0.000 & 0 & 0.016 & 0.955 & 0.048 & 0.064\\
Add-in: 8 & XGB & 5000 & 0.000 & 0 & 0.008 & 0.970 & 0.041 & 0.033\\
Leave-out: 8 & XGB & 250 & -0.001 & 0 & 0.038 & 0.909 & 0.059 & 0.132\\
Leave-out: 8 & XGB & 1000 & 0.000 & 0 & 0.020 & 0.955 & 0.043 & 0.079\\
Leave-out: 8 & XGB & 5000 & 0.000 & 0 & 0.009 & 0.980 & 0.021 & 0.043\\*
\end{longtable}
\endgroup{}
\begingroup\fontsize{9}{11}\selectfont

\begin{longtable}[t]{llrrrrrrr}
\caption{Performance of estimators of the linear VIM trend, defined using sensitivity at the 95th percentile of predicted risk, for variable 1 over the time series, summarized over 1000 Monte-Carlo replications. Performance shown for each sample size $n \in \{250, 1000, 5000\}$, and all estimators: logistic regression (GLM), lasso, random forests (RF), super learner (SL), and boosted trees (XGB). Both add-in (compared to four covariates) and leave-out (compared to nine covariates) variable importance are displayed. Each feature has correlation 0.5 with the previous timepoint. \label{tab:sim_all_slope_1_0.5_sensitivity_dgm_2}}\\
\toprule
VIM type: variable & Estimator & $n$ & Mean est. & True value & Empirical SE & Coverage & Rejection prop. & CI width\\
\midrule
\endfirsthead
\caption[]{Performance of estimators of the linear VIM trend, defined using sensitivity at the 95th percentile of predicted risk, for variable 1 over the time series, summarized over 1000 Monte-Carlo replications. \textit{(continued)}}\\
\toprule
VIM type: variable & Estimator & $n$ & Mean est. & True value & Empirical SE & Coverage & Rejection prop. & CI width\\
\midrule
\endhead

\endfoot
\bottomrule
\endlastfoot
Add-in: 1 & GLM & 250 & 0.000 & 0.000 & 0.029 & 0.914 & 0.094 & 0.095\\
Add-in: 1 & GLM & 1000 & 0.000 & 0.000 & 0.014 & 0.951 & 0.049 & 0.058\\
Add-in: 1 & GLM & 5000 & -0.001 & 0.000 & 0.009 & 0.921 & 0.079 & 0.032\\
Leave-out: 1 & GLM & 250 & -0.002 & -0.001 & 0.036 & 0.916 & 0.088 & 0.124\\
Leave-out: 1 & GLM & 1000 & 0.000 & -0.001 & 0.018 & 0.955 & 0.048 & 0.073\\
Leave-out: 1 & GLM & 5000 & -0.002 & -0.001 & 0.008 & 0.975 & 0.030 & 0.039\\
Add-in: 1 & LASSO & 250 & 0.002 & 0.000 & 0.075 & 0.468 & 0.529 & 0.087\\
Add-in: 1 & LASSO & 1000 & 0.008 & 0.000 & 0.070 & 0.360 & 0.643 & 0.055\\
\end{longtable}
\endgroup{}
\input{plots/sims/summary_table_slope_2_0.5_sensitivity_dgm_2}
\input{plots/sims/summary_table_slope_3_0.5_sensitivity_dgm_2}
\input{plots/sims/summary_table_slope_8_0.5_sensitivity_dgm_2}

\clearpage

\subsection{Results with a more common outcome}\label{sec:common}

\subsubsection{Experimental setup}

In this section, we present empirical results describing the performance of our proposed VIM longitudinal summary estimators in a setting with a more common outcome. In these simulations, we only consider VIMs defined using AUC.

As in the main simulations, we consider $T=4$ timepoints, $p=10$ variables, and binary outcome $Y_t \in \{0,1\}$. As before, in the data generating process $(X_{t,1},\ldots,X_{t,7})$ are predictive of the outcome at all time points (i.e., have VIM value greater than zero), while $(X_{t,8},X_{t,9},X_{t,10})$ are not predictive of the outcome (i.e., have VIM value equal to zero) at all time points. 
The outcome is generated with a probit model: 
  $P(Y_t = 1 \mid X_t = x_t) =  \Phi\left(\sum_{j=1}^p \beta_{t,j}x_{t,j}\right).$
We set $\beta_{t,1} = 2$ for all $t$; $\beta_{t,2} = 2 + t / 4$; $\beta_{t,3} = (-1) \{1 + \exp(-t)\}^{-1} + 2$, and $\beta_{t,j} = 0.05$ for $j \in \{4, \ldots, 7\}$. The remaining variables have zero VIM for each $t$: $\beta_{t,8:10} = \mathbf{0}_{3}$. 

In the first set of results we present (Section~\ref{sec:common_uncorrelated}), we follow the same data-generating process for all time points. For each time point, $t$, we generate $(X_{t,4}, X_{t,5}, X_{t,6}, X_{t,7})$ following a multivariate normal distribution with mean zero and identity covariance matrix. To specify the distribution of $(X_{t,1},X_{t,2},X_{t,3})$, we use
\begin{align*}
  E\{X_{t,j} \mid (X_{t,4}, X_{t,5}, X_{t,6}, X_{t,7}) = x_t\} = 0.05x_{t}\mathbf{1}_4^\top
\end{align*}
and add multivariate normal random errors with mean zero and identity covariance matrix. This specification induces a small amount of correlation between variables $X_{t,1}$, $X_{t,2}$, and $X_{t,3}$. Variables $X_{t,8}$, $X_{t,9}$, and $X_{t,10}$ are generated from a multivariate normal distribution with mean zero and identity covariance matrix. 

In the second set of results we present (Section~\ref{sec:common_correlated}), the features are correlated across time. At $t = 1$, we generate the features and outcome as in the main simulations. For $t \in \{2, 3, 4\}$, we generate each feature using an autocorrelation structure with correlation parameter $\rho = 0.5$. For $j \in \{4, 5, 6, 7\}$,
\begin{align*}
    X_{t,j} =& \ X_{t-1,j} \rho + \epsilon,
\end{align*}
where $\epsilon \stackrel{iid}{\sim} N(0,1)$. Then for $j \in \{1, 2, 3\}$,
\begin{align*}
    X_{t,j} =& \ X_{t-1,j}\rho + \epsilon_j \\
    E\{\epsilon_j \mid (X_{t,4}, X_{t,5}, X_{t,6}, X_{t,7}) = x_t\}\ =& \ 0.05x_t\mathbf{1}_4^\top,
\end{align*}
and the conditional variance of $\epsilon$ is 1. As before, this specification induces a small amount of correlation between variables $X_{t,1}$, $X_{t,2}$, and $X_{t,3}$. Variables $X_{t,8}$, $X_{t,9}$, and $X_{t,10}$ are generated from a multivariate normal distribution with mean zero and identity covariance matrix. Importantly, by creating autocorrelation within features over time, this specification also induces correlation in the outcomes. The outcome at time 1 has correlation approximately 0.28 with the outcome at time 2; 0.13 with the outcome at time 3; and 0.07 with the outcome at time 4. Correlation values between other consecutive timepoints are similar.

\begin{figure}
    \centering
    \includegraphics[width=1\textwidth]{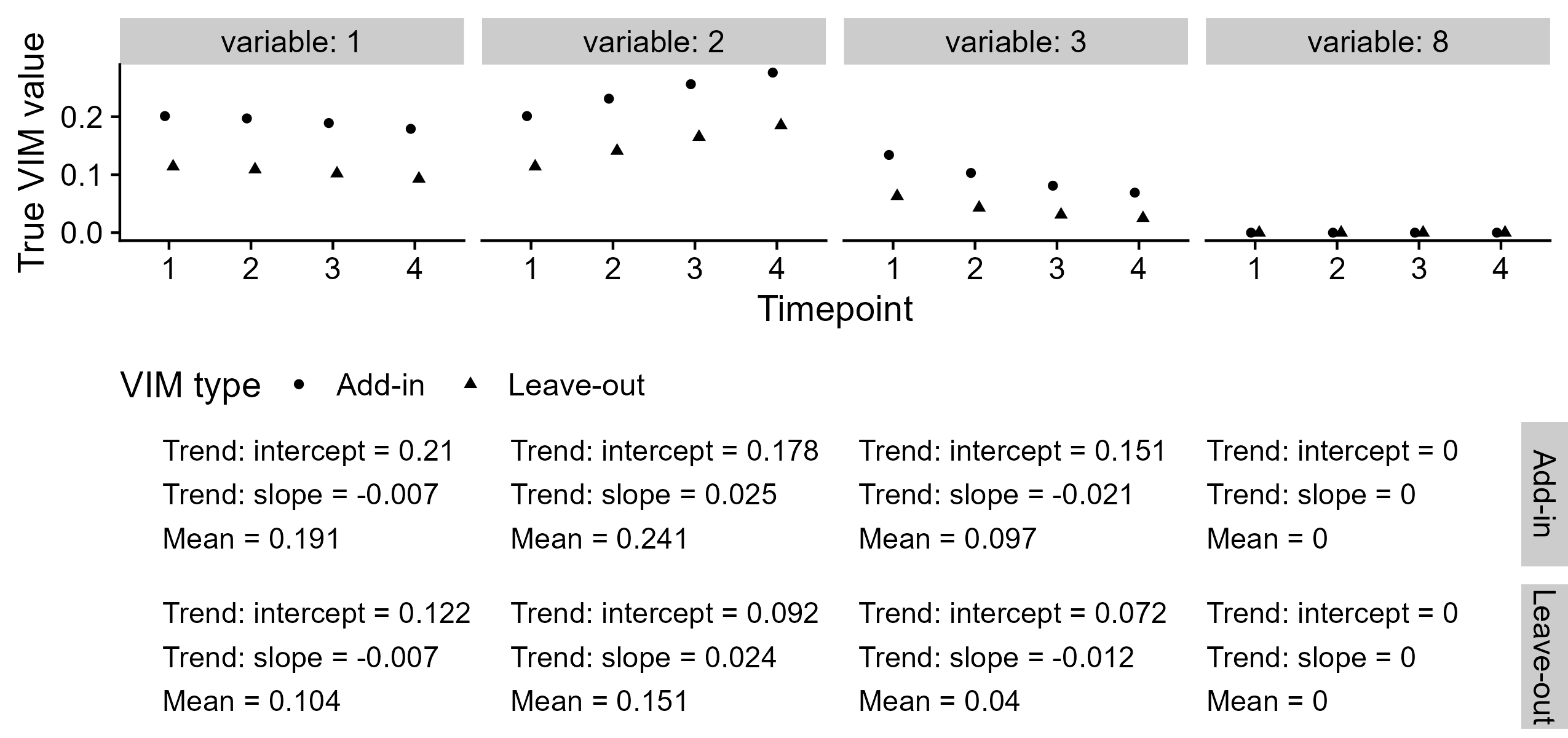}
    \caption{True variable importance values (both add-in and leave-out) defined using AUC at each timepoint and summarized over time for variables 1, 2, 3, and 8 based on the data-generating mechanism with no correlation.}
    \label{fig:sim_true_vims_0_auc_dgm_1}
\end{figure}

\begin{figure}
    \centering
    \includegraphics[width=1\textwidth]{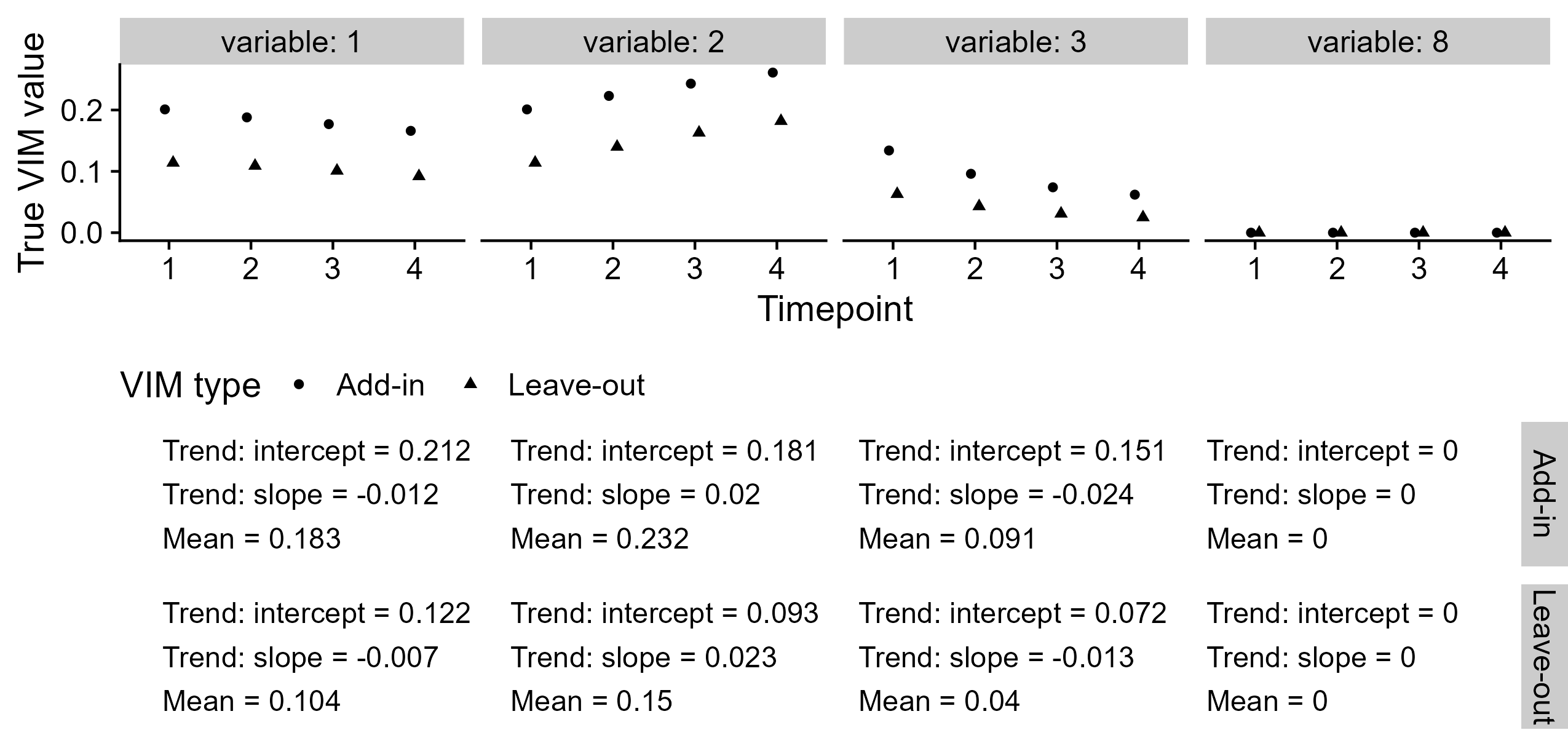}
    \caption{True variable importance values (both add-in and leave-out) defined using AUC at each timepoint and summarized over time for variables 1, 2, 3, and 8 based on the data-generating mechanism with correlation described in the simulation setup.}
    \label{fig:sim_true_vims_0.5_auc_dgm_1}
\end{figure}

In all cases, for each $n \in \{100, 250, 500, 1000, 5000, 10000\}$, we generate 1000 random datasets according to this data-generating mechanism and consider both the add-in and leave-out importance of each variable in $S = \{1, 2, 3, 8, 9, 10\}$, the variable set of interest, fixing predictors $(X_{t,4},\ldots,X_{t,7})$ as a base set. The true values of the AUC VIMs at each time point and the mean and linear trend over the time series are provided in Figures~\ref{fig:sim_true_vims_0_auc_dgm_1} and \ref{fig:sim_true_vims_0.5_auc_dgm_1}. The true VIMs in the uncorrelated and correlated cases are given in Tables~\ref{tab:sim_true_vims_0_auc_dgm_1} and \ref{tab:sim_true_vims_0.5_auc_dgm_1}, respectively. The true add-in VIMs are nearly the same as in the uncorrelated case, while the true leave-out VIMs are identical to the uncorrelated case.

\begin{table}
\centering
\caption{True variable importance values (defined using AUC) at each time point and summarized over the time series. \label{tab:sim_true_vims_0_auc_dgm_1}}
\centering
\fontsize{9}{11}\selectfont
\begin{tabular}[t]{llllllll}
\toprule
VIM type: variable & Timepoint 1 & Timepoint 2 & Timepoint 3 & Timepoint 4 & Mean & Trend intercept & Trend slope\\
\midrule
Add-in: 1 & 0.201 & 0.197 & 0.189 & 0.179 & 0.191 & 0.21 & -0.007\\
Leave-out: 1 & 0.114 & 0.109 & 0.102 & 0.093 & 0.104 & 0.122 & -0.007\\
Add-in: 2 & 0.201 & 0.231 & 0.256 & 0.276 & 0.241 & 0.178 & 0.025\\
Leave-out: 2 & 0.114 & 0.141 & 0.165 & 0.185 & 0.151 & 0.092 & 0.024\\
Add-in: 3 & 0.134 & 0.103 & 0.081 & 0.069 & 0.097 & 0.151 & -0.021\\
Leave-out: 3 & 0.063 & 0.043 & 0.031 & 0.025 & 0.04 & 0.072 & -0.012\\
Add-in: 8 & 0 & 0 & 0 & 0 & 0 & 0 & 0\\
Leave-out: 8 & 0 & 0 & 0 & 0 & 0 & 0 & 0\\
\bottomrule
\end{tabular}
\end{table}

\input{plots/sims/sim_true_vims_0.5_auc_dgm_1}

\subsubsection{Empirical results using AUC for all estimators with uncorrelated data}\label{sec:common_uncorrelated}

\begin{figure}
    \centering
    \includegraphics[width=1\textwidth]{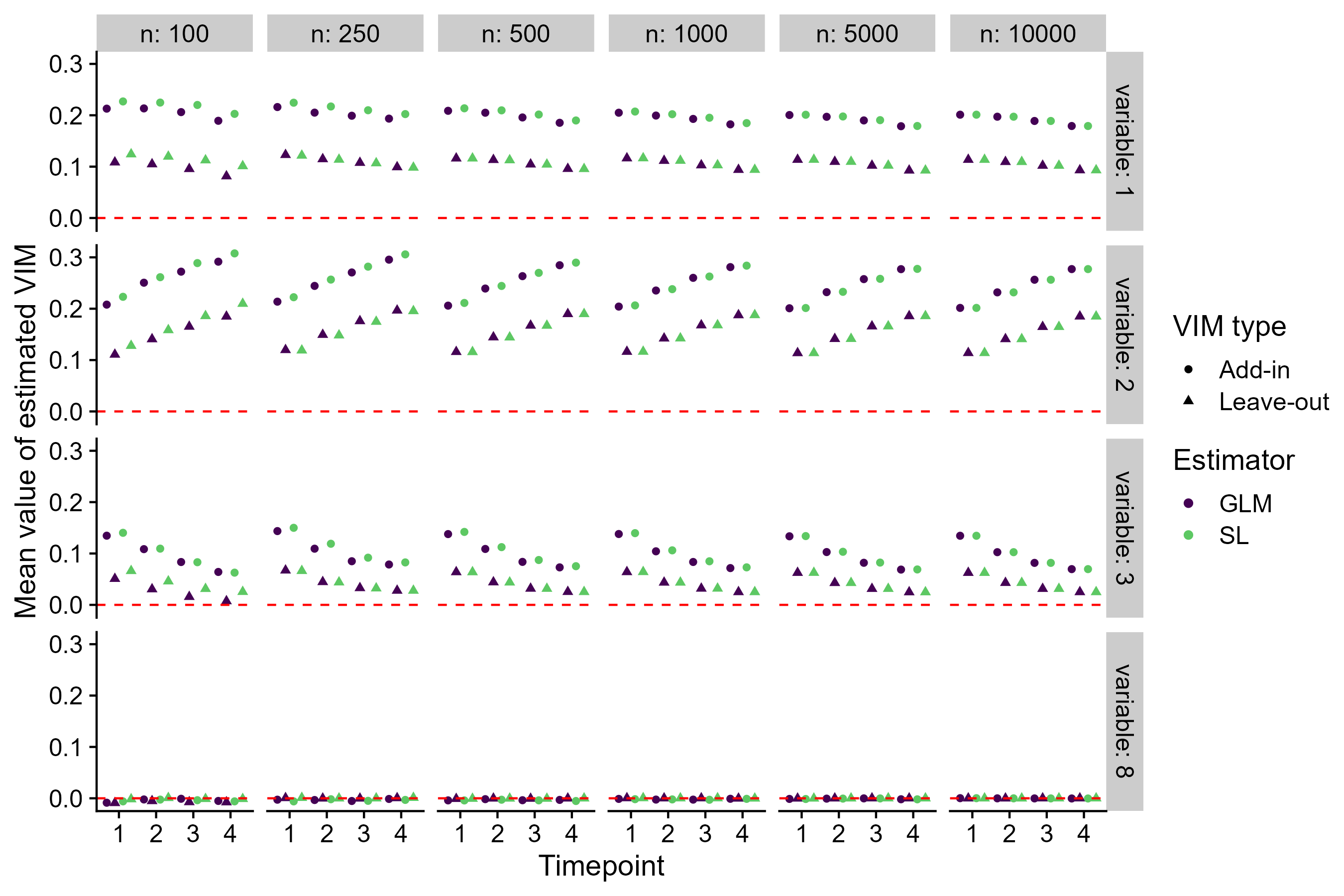}
    \caption{Mean estimated variable importance defined using AUC over 1000 simulated replicates versus time for four variables: three with true VIM value greater than zero (variables 1, 2, and 3) and one with true VIM value equal to zero (variable 8). Columns: sample size. Rows: variable. Shapes (circles and triangles) denote VIM type, while colors denote the estimator: logistic regression (GLM) or super learner (SL).}
    \label{fig:sim_vim_over_time_0_auc_dgm_1}
\end{figure}

\begin{figure}
    \centering
    \includegraphics[width=1\textwidth]{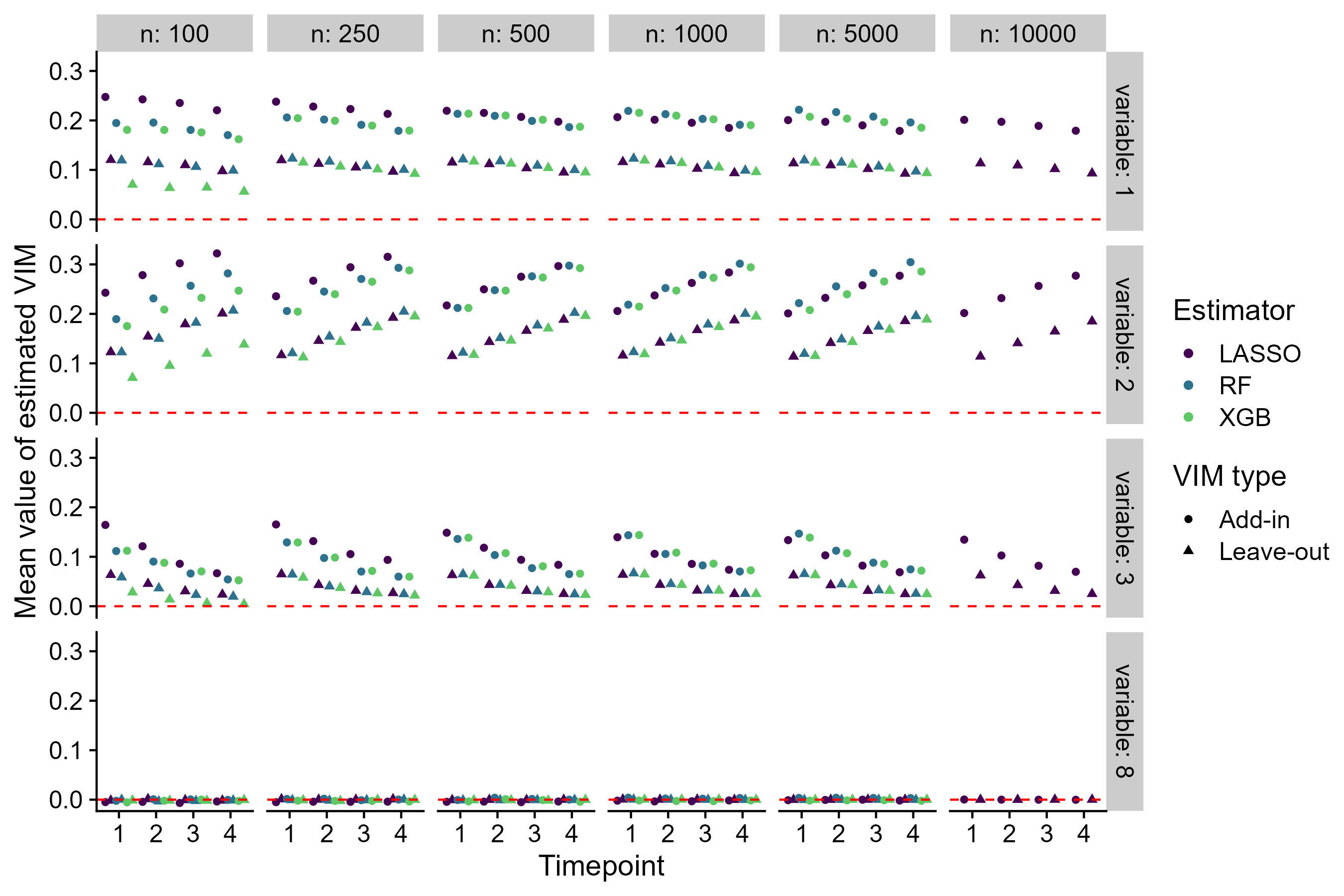}
    \caption{Mean estimated variable importance defined using AUC over 1000 simulated replicates versus time for four variables: three with true VIM value greater than zero (variables 1, 2, and 3) and one with true VIM value equal to zero (variable 8). Columns: sample size. Rows: variable. Shapes (circles and triangles) denote VIM type, while  colors denote the estimator: lasso, random forests (RF), or boosted trees (XGB). RF and XGB were not run for $n = 10000$.}
    \label{fig:sim_vim_over_time_supp_0_auc_dgm_1}
\end{figure}

In Figures~\ref{fig:sim_vim_over_time_0_auc_dgm_1} and \ref{fig:sim_vim_over_time_supp_0_auc_dgm_1}, we display the estimated VIM values at each timepoint when using the lasso, random forests (RF), and boosted trees (XGB) to estimate the underlying prediction functions. As in the main manuscript, we see little difference across estimators as sample size grows, reflecting the fact that in this example, all estimators (if properly tuned) can estimate the underlying prediction function consistently. Regardless, all procedures estimate the same VIM estimand.

We display the estimated VIM for all sample sizes at each time point using logistic regression and a super learner in Figure~\ref{fig:sim_vim_over_time_0_auc_dgm_1}. Estimation and inferential performance summaries for a sample of size 5,000 for the same prediction modelling approaches are provided in Table~\ref{tab:sim_bign_all_0_auc_dgm_1}, with full results for all sample sizes and variables provided in Tables~\ref{tab:sim_all_avg_1_0_auc_dgm_1}--\ref{tab:sim_all_slope_8_0_auc_dgm_1}.

\input{plots/sims/summary_table_bign_all_0_auc_dgm_1}

In Tables~\ref{tab:sim_all_avg_1_0_auc_dgm_1}--\ref{tab:sim_all_slope_8_0_auc_dgm_1}, we display the estimated average VIM and slope of the linear VIM trend for each variable, including all estimators and all sample sizes.
\input{plots/sims/summary_table_avg_1_0_auc_dgm_1}
\input{plots/sims/summary_table_avg_2_0_auc_dgm_1}
\input{plots/sims/summary_table_avg_3_0_auc_dgm_1}
\input{plots/sims/summary_table_avg_8_0_auc_dgm_1}
\input{plots/sims/summary_table_slope_1_0_auc_dgm_1}
\input{plots/sims/summary_table_slope_2_0_auc_dgm_1}
\input{plots/sims/summary_table_slope_3_0_auc_dgm_1}
\input{plots/sims/summary_table_slope_8_0_auc_dgm_1}

The low coverage observed in some cases for estimators based on random forests and gradient boosted trees may be due to the fact that we did not use cross-validation to tune these estimators, which is necessary in practical applications; we included simple versions of these algorithms here to show the ease with which one can compare results across algorithms and to highlight that more flexible algorithms should be tuned to achieve the required convergence rate. For variable 8, which has zero VIM, the rejection proportion is not equal to 1 - coverage in all cases (e.g., some rows in Table~\ref{tab:sim_all_avg_8_0_auc_dgm_1}) because we perform a one-sided test at level 0.05, while the confidence intervals are 95\% confidence intervals.

\subsubsection{Empirical results using AUC for all estimators with correlated data}\label{sec:common_correlated}

In this case, we only used logistic regression to estimate the required prediction functions, since our goal is to compare these results with the uncorrelated results presented above.

\begin{figure}
    \centering
    \includegraphics[width=1\textwidth]{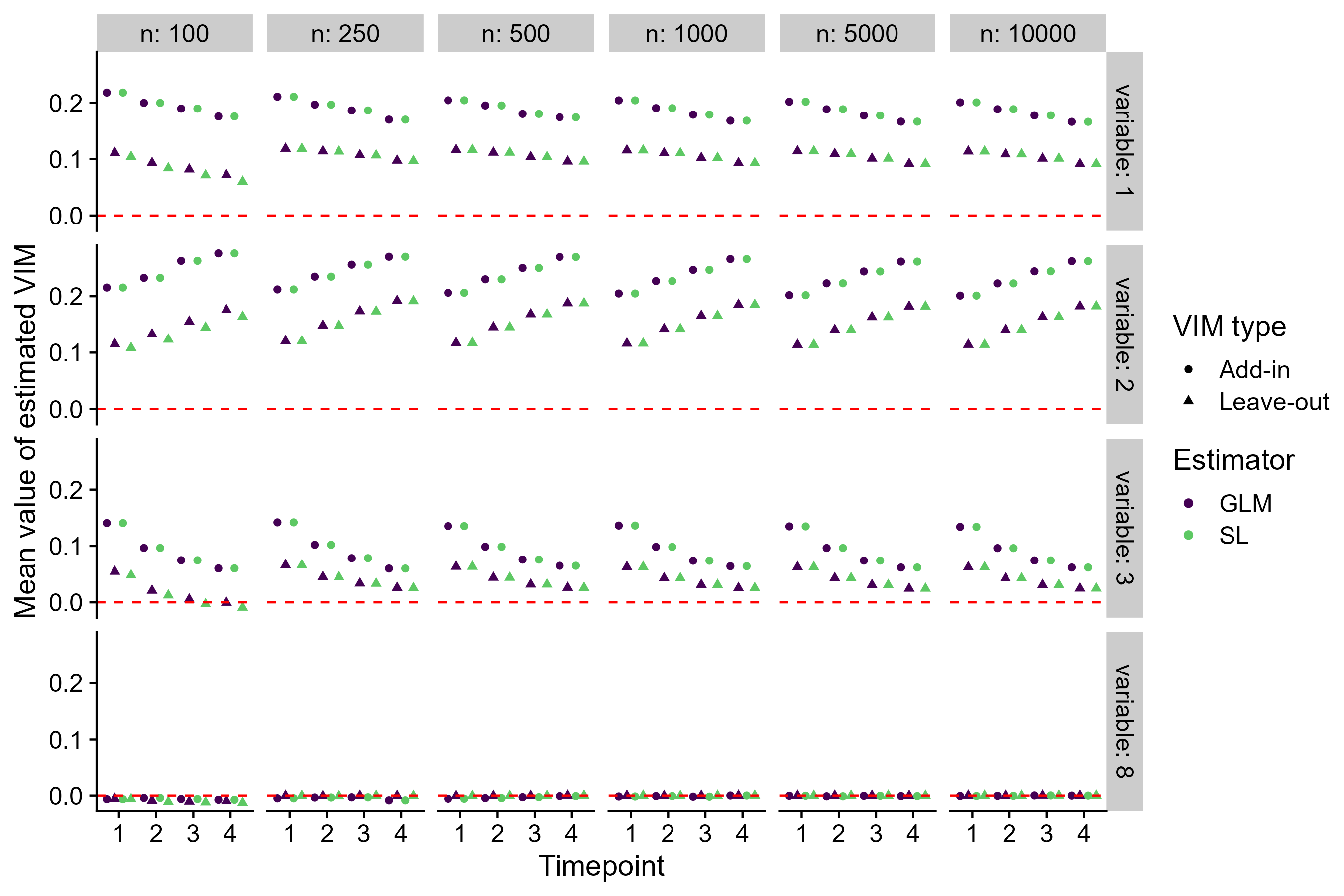}
    \caption{Mean estimated variable importance over 1000 simulated replicates in the correlated setting versus time for four variables: three with true VIM value greater than zero (variables 1, 2, and 3) and one with true VIM value equal to zero (variable 8). Columns: sample size. Rows: variable. Shapes (circles and triangles) denote VIM type.}
    \label{fig:sim_vim_over_time_0.5_auc_dgm_1}
\end{figure}

We display the estimated VIM for all sample sizes at each time point using logistic regression and a super learner in Figure~\ref{fig:sim_vim_over_time_0.5_auc_dgm_1}. Estimation and inferential performance summaries for a sample of size 5,000 for the same prediction modelling approaches are provided in Table~\ref{tab:sim_bign_all_0.5_auc_dgm_1}, with full results for all sample sizes and variables provided in Tables~\ref{tab:sim_all_avg_1_0.5_auc_dgm_1}--\ref{tab:sim_all_slope_8_0.5_auc_dgm_1}. At a large sample size (5,000), we see that our estimators are unbiased and have empirical coverage that is within Monte-Carlo error of the nominal 95\% level. Type I error (for variable 8) is controlled below the 0.05 level in all cases. In smaller-sample cases, there may be small amounts of bias and lower coverage (e.g., Table~\ref{tab:sim_all_avg_1_0.5_auc_dgm_1}); however, the bias converges to zero and coverage converges to the nominal level by approximately $n = 1000$ in all cases. This is similar to the result in the uncorrelated case, providing empirical support that our methodology works regardless of whether there is correlation between outcomes or features over time.

\input{plots/sims/summary_table_bign_all_0.5_auc_dgm_1}

\input{plots/sims/summary_table_avg_1_0.5_auc_dgm_1}
\input{plots/sims/summary_table_avg_2_0.5_auc_dgm_1}
\input{plots/sims/summary_table_avg_3_0.5_auc_dgm_1}
\input{plots/sims/summary_table_avg_8_0.5_auc_dgm_1}
\input{plots/sims/summary_table_slope_1_0.5_auc_dgm_1}
\input{plots/sims/summary_table_slope_2_0.5_auc_dgm_1}
\input{plots/sims/summary_table_slope_3_0.5_auc_dgm_1}
\input{plots/sims/summary_table_slope_8_0.5_auc_dgm_1}

\clearpage

\section{Additional results from studying predictors of suicide risk}\label{sec:more_data}

\subsection{Description of the study cohort}
We provide a full description of our sample in Table~\ref{tab:table_1}.

\clearpage
{\fontsize{7.5}{9}\selectfont
\setlength{\LTpost}{0mm}
\begin{longtable}{p{1.25in}p{0.68in}p{0.68in}p{0.68in}p{0.68in}p{0.68in}p{0.68in}p{0.68in}p{0.68in}}
\caption{Cohort description for sample used to estimate variable importance in suicide attempt risk prediction models.}\label{tab:table_1}\\
\toprule
 &  & \multicolumn{6}{c}{\textbf{Timepoint (days from initoal visit)}} \\ 
\cmidrule(lr){3-8}
\textbf{Characteristic} &  \textbf{Overall} \newline N = 343950\textsuperscript{\textit{a}} &  \textbf{90} \newline N = 87527 &  \textbf{180} \newline N = 64121 &  \textbf{270} \newline N = 52258 &  \textbf{360} \newline N = 52799 &  \textbf{450} \newline N = 45958 &  \textbf{540} \newline N = 41287 \\ 
\midrule
\multicolumn{8}{l}{Outcome} \\ 
\midrule
Suicide attempt\textsuperscript{\textit{a}} & 1,494 (0.4\%) & 398 (0.5\%) & 297 (0.5\%) & 240 (0.5\%) & 200 (0.4\%) & 189 (0.4\%) & 170 (0.4\%) \\ 
\midrule
\multicolumn{8}{l}{Demographic
 variables} \\ 
\midrule
Age in years &  &  &  &  &  &  &  \\ 
    11--17 & 48,470 (14\%) & 13,201 (15\%) & 9,716 (15\%) & 7,535 (14\%) & 7,209 (14\%) & 5,696 (12\%) & 5,113 (12\%) \\ 
    18--29 & 60,652 (18\%) & 17,591 (20\%) & 11,777 (18\%) & 9,123 (17\%) & 8,482 (16\%) & 7,246 (16\%) & 6,433 (16\%) \\ 
    30--44 & 78,186 (23\%) & 20,952 (24\%) & 14,807 (23\%) & 11,817 (23\%) & 11,544 (22\%) & 10,148 (22\%) & 8,918 (22\%) \\ 
    45--64 & 105,037 (31\%) & 24,731 (28\%) & 18,837 (29\%) & 15,889 (30\%) & 17,040 (32\%) & 15,001 (33\%) & 13,539 (33\%) \\ 
    65 and older & 51,605 (15\%) & 11,052 (13\%) & 8,984 (14\%) & 7,894 (15\%) & 8,524 (16\%) & 7,867 (17\%) & 7,284 (18\%) \\ 
Sex &  &  &  &  &  &  &  \\ 
    Female & 219,365 (64\%) & 54,868 (63\%) & 40,592 (63\%) & 33,222 (64\%) & 33,932 (64\%) & 29,833 (65\%) & 26,918 (65\%) \\ 
    Male & 124,584 (36\%) & 32,658 (37\%) & 23,529 (37\%) & 19,036 (36\%) & 18,867 (36\%) & 16,125 (35\%) & 14,369 (35\%) \\ 
    Unknown sex & 1 (<0.1\%) & 1 (<0.1\%) & 0 (0\%) & 0 (0\%) & 0 (0\%) & 0 (0\%) & 0 (0\%) \\ 
\midrule
Prior self-harm\textsuperscript{\textit{b}} & 4,486 (1.3\%) & 1,019 (1.2\%) & 893 (1.4\%) & 816 (1.6\%) & 693 (1.3\%) & 549 (1.2\%) & 516 (1.2\%) \\ 
\midrule
PHQi9 &  &  &  &  &  &  &  \\ 
    Not measured & 209,537 (61\%) & 53,213 (61\%) & 40,004 (62\%) & 32,451 (62\%) & 32,179 (61\%) & 27,347 (60\%) & 24,343 (59\%) \\ 
     0 & 110,095 (32\%) & 28,195 (32\%) & 19,683 (31\%) & 15,937 (30\%) & 17,064 (32\%) & 15,247 (33\%) & 13,969 (34\%) \\ 
     1 & 16,433 (4.8\%) & 4,158 (4.8\%) & 3,033 (4.7\%) & 2,586 (4.9\%) & 2,371 (4.5\%) & 2,247 (4.9\%) & 2,038 (4.9\%) \\ 
     2 & 5,099 (1.5\%) & 1,249 (1.4\%) & 901 (1.4\%) & 823 (1.6\%) & 782 (1.5\%) & 732 (1.6\%) & 612 (1.5\%) \\ 
     3 & 2,786 (0.8\%) & 712 (0.8\%) & 500 (0.8\%) & 461 (0.9\%) & 403 (0.8\%) & 385 (0.8\%) & 325 (0.8\%) \\ 
\midrule
PHQ-8 total score\textsuperscript{\textit{c}} &  &  &  &  &  &  &  \\ 
    Not measured & 209,602 (61\%) & 53,296 (61\%) & 40,003 (62\%) & 32,447 (62\%) & 32,172 (61\%) & 27,356 (60\%) & 24,328 (59\%) \\ 
     0--4 & 42,744 (12\%) & 10,971 (13\%) & 7,696 (12\%) & 5,993 (11\%) & 6,790 (13\%) & 5,925 (13\%) & 5,369 (13\%) \\ 
     5--10 & 44,101 (13\%) & 11,464 (13\%) & 7,929 (12\%) & 6,433 (12\%) & 6,618 (13\%) & 6,037 (13\%) & 5,620 (14\%) \\ 
     11--15 & 25,745 (7.5\%) & 6,497 (7.4\%) & 4,649 (7.3\%) & 3,873 (7.4\%) & 3,943 (7.5\%) & 3,587 (7.8\%) & 3,196 (7.7\%) \\ 
     16--20 & 15,720 (4.6\%) & 3,783 (4.3\%) & 2,820 (4.4\%) & 2,498 (4.8\%) & 2,368 (4.5\%) & 2,212 (4.8\%) & 2,039 (4.9\%) \\ 
     21--24 & 6,038 (1.8\%) & 1,516 (1.7\%) & 1,024 (1.6\%) & 1,014 (1.9\%) & 908 (1.7\%) & 841 (1.8\%) & 735 (1.8\%) \\ 
\bottomrule
\end{longtable}
\begin{minipage}{\linewidth}
\textsuperscript{\textit{a}} Suicide attempt in the 90 days following the mental health care visit.\\
\textsuperscript{\textit{b}} At least one injury, poisoning, or attempt in the past 12 months.\\
\textsuperscript{\textit{c}} PHQ-8 and PHQi9 measured on the day of the mental health care visit.\\


\end{minipage}

}
\clearpage

We provide the variable set numbers, along with their description, in Table~\ref{tab:varsets}. The category of ``diagnosis \& utilization variables'' includes: the PHQ-8 total score measured at the visit; inpatient, outpatient, or emergency/urgent care encounters with a mental health diagnosis (total days with a diagnosis code, number of months with a diagnosis code, and number of days divided by months enrolled, each in the last 3 months and 12 months); diagnoses of attention deficit disorder (ADD), alcohol use disorder, autism spectrum disorder, asthma, conduct disorder, dementia or cognitive disorder, depression, diabetes, substance use disorder, eating disorder, psychosis, pain, personality disorder, post-traumatic stress disorder, or traumatic brain injury (all measured both at the visit and total days with a diagnosis code, number of months with a diagnosis code, and number of days divided by months enrolled, each in the last 3 or 12 months); and prescription drug fill information on antidepressants, benzodiazepines, hypnotics, second-generation antipsychotics, first-generation antipsychotics, ADD drugs, lithium, and anticonvulsants (total days supply and total days supply divided by number of months enrolled for the 3 and 12 months prior to the visit).

\begin{table}[h]
    \centering
     \caption{Variable sets of interest for studying predictors of suicide risk, along with the comparator group for defining variable importance (``--'' denotes cases where variable importance is not estimated) and a description of the variable set.}
    \label{tab:varsets}
    \begin{tabular}{p{0.6in}p{0.75in}p{4.5in}}
        Variable set number & Comparator group for VIM & Description \\
        \hline
        1 & -- & No variables \\
        2 & 1 & PHQi9 \\
        3 & 1 & Age and sex (base set) \\
        4 & 3 & Age, sex, and PHQi9 \\
        5 & 3 & Age, sex, and prior self-harm variables \\
        6 & 5 & Age, sex, prior self-harm variables, and PHQi9 \\
        7 & -- & Demographic variables \\
        8 & -- & Demographic and prior self-harm variables (second possible base set) \\
        9 & -- & Demographic, prior self-harm, and diagnosis \& utilization variables (third possible base set) \\
        10 & 8 & Demographic and prior self-harm variables and Charlson score \\
        11 & 8 & Demographic and prior self-harm variables and PHQi9 \\
        12 & 9 & Demographic, prior self-harm, and diagnosis \& utilization variables, and Charlson score \\
        13 & 9 & Demographic, prior self-harm, and diagnosis \& utilization variables, and PHQi9 \\
        14 & 10 & Demographic and prior self-harm variables, Charlson score, and PHQi9 \\
        15 & 12 & All variables \\
        16 & -- & Age \\
        17 & -- & Age, sex, and race or ethnicity \\
        18 & 16 & Age and PHQi9 \\
        19 & 17 & Age, sex, race or ethnicity, and PHQi9 \\
    \end{tabular}
\end{table}

\subsection{Super learner specification}

We describe the super learner ensemble used in the analysis of the suicide risk data, including candidate tuning parameter values for each individual algorithm, in Table~\ref{tab:sl-algs_data}. 

\begin{table}
    \centering
    \caption{Candidate learners in the super learner ensemble used in the data analysis, along with their R implementation, tuning parameter values, and description of the tuning parameters. All tuning parameters besides those listed here are set to their default values. In particular, the random forests are grown with a subsampling fraction of 1; and the boosted trees are grown with a minimum of 10 observations per node. \\
    ${}^{\dagger}$: $p$ denotes the total number of predictors. }
    \label{tab:sl-algs_data}
    \begin{tabular}{c|ccc}
       Candidate Learner & R & Tuning Parameter & Tuning parameter  \\
       & Implementation & and possible values & description\\ \hline
        Random forests & \texttt{ranger} & \texttt{mtry} $= 1/2\sqrt{p}$ ${}^{\dagger}$ & Number of variables \\
        & \citep{rangerpkg} & & to possibly split \\
        & & & at in each node \\ 
        & & \texttt{num.trees} $= 500$ & Number of trees \\
        & & \texttt{min.node.size} $\in \{1, 10, 50\}$ & Minimum node size \\
        \hline
        Gradient boosted & \texttt{xgboost} & \texttt{max.depth} $ \in \{1,4\}$ &  Maximum tree depth\\
        trees & \citep{xgboostpkg} & \texttt{ntree} $=500$ & Number of iterations \\ 
        & & \texttt{shrinkage} $= 0.1$ & Shrinkage \\ \hline
        Lasso & \texttt{glmnet} & $\lambda$ & $\ell_1$ regularization  \\
        & \citep{glmnetpkg} & chosen via 10-fold CV & parameter \\ \hline
    \end{tabular}
\end{table} 

\clearpage

\subsection{Empirical results}

First, we present the estimated AUC and PPV and sensitivity at the 95th percentile of predicted risk for each variable set from Table~\ref{tab:varsets} in Tables~\ref{tab:auc_data_analysis_pred_perf_all_supp}--\ref{tab:sensitivity_data_analysis_pred_perf_all_supp}; we show the mean, minimum, and maximum AUC, with these summary statistics taken across timepoints (1--6) and prediction algorithms (logistic regression, lasso, random forests, gradient boosted trees, discrete super learner, and super learner).  As expected, the AUC for no variables is 0.5. The AUC for all variables (variable set 15) ranges from 0.729 to 0.852, with a mean AUC over all timepoints and algorithms of 0.81. This is in line with prior suicide risk prediction modelling studies. 

\begin{table}
\centering
\caption{Mean, minimum (Min.), and maximum (Max.) estimated area under the receiver operating characteristic curve (AUC) for each group of variables in the predictors of suicide risk analysis. Summary statistics are taken over all timepoints (1--6) and prediction algorithms (logistic regression, lasso, random forests, gradient boosted trees, discrete super learner, and super learner).\label{tab:auc_data_analysis_pred_perf_all_supp}}
\centering
\fontsize{10}{12}\selectfont
\begin{tabular}[t]{>{\raggedright\arraybackslash}p{1in}>{\raggedleft\arraybackslash}p{4in}rrr}
\toprule
\multicolumn{2}{c}{ } & \multicolumn{3}{c}{Estimated AUC} \\
\cmidrule(l{3pt}r{3pt}){3-5}
Variable set number & Description & Mean & Min. & Max.\\
\midrule
1 & No variables & 0.500 & 0.500 & 0.500\\
2 & PHQi9 & 0.612 & 0.512 & 0.662\\
3 & Age and sex & 0.671 & 0.620 & 0.707\\
4 & Age, sex, and PHQi9 & 0.693 & 0.620 & 0.747\\
5 & Age, sex, and prior self-harm variables & 0.728 & 0.677 & 0.811\\
6 & Age, sex, prior self-harm variables, and PHQi9 & 0.658 & 0.596 & 0.702\\
7 & Demographic variables & 0.692 & 0.626 & 0.761\\
8 & Demographic and prior self-harm variables & 0.736 & 0.681 & 0.821\\
9 & Demographic, prior self-harm, and diagnosis \& utilization variables & 0.808 & 0.703 & 0.881\\
10 & Demographic and prior self-harm variables and Charlson score & 0.738 & 0.690 & 0.822\\
11 & Demographic and prior self-harm variables and PHQi9 & 0.751 & 0.693 & 0.837\\
12 & Demographic, prior self-harm, and diagnosis \& utilization variables, and Charlson score & 0.811 & 0.726 & 0.879\\
13 & Demographic, prior self-harm, and diagnosis \& utilization variables, and PHQi9 & 0.814 & 0.726 & 0.892\\
14 & Demographic and prior self-harm variables, Charlson score, and PHQi9 & 0.752 & 0.697 & 0.839\\
15 & All variables & 0.814 & 0.725 & 0.880\\
16 & Age & 0.662 & 0.607 & 0.715\\
17 & Age and PHQi9 & 0.689 & 0.616 & 0.753\\
18 & Age, sex, and race or ethnicity & 0.666 & 0.617 & 0.709\\
19 & Age, sex, race or ethnicity, and PHQi9 & 0.688 & 0.620 & 0.747\\
\bottomrule
\end{tabular}
\end{table}

\begin{table}
\centering
\caption{Mean, minimum (Min.), and maximum (Max.) estimated positive predictive value (PPV) at the 95th percentile of predicted risk for each group of variables in the predictors of suicide risk analysis. Summary statistics are taken over all timepoints (1--6) and prediction algorithms (logistic regression, lasso, random forests, gradient boosted trees, discrete super learner, and super learner).\label{tab:ppv_data_analysis_pred_perf_all_supp}}
\centering
\fontsize{10}{12}\selectfont
\begin{tabular}[t]{>{\raggedright\arraybackslash}p{1in}>{\raggedleft\arraybackslash}p{4in}rrr}
\toprule
\multicolumn{2}{c}{ } & \multicolumn{3}{c}{Estimated PPV} \\
\cmidrule(l{3pt}r{3pt}){3-5}
Variable set number & Description & Mean & Min. & Max.\\
\midrule
1 & No variables & 0.001 & 0.000 & 0.002\\
2 & PHQi9 & 0.015 & 0.008 & 0.022\\
3 & Age and sex & 0.015 & 0.008 & 0.022\\
4 & Age, sex, and PHQi9 & 0.019 & 0.009 & 0.028\\
5 & Age, sex, and prior self-harm variables & 0.028 & 0.014 & 0.059\\
6 & Age, sex, prior self-harm variables, and PHQi9 & 0.014 & 0.004 & 0.024\\
7 & Demographic variables & 0.015 & 0.009 & 0.023\\
8 & Demographic and prior self-harm variables & 0.027 & 0.018 & 0.050\\
9 & Demographic, prior self-harm, and diagnosis \& utilization variables & 0.035 & 0.024 & 0.058\\
10 & Demographic and prior self-harm variables and Charlson score & 0.028 & 0.016 & 0.055\\
11 & Demographic and prior self-harm variables and PHQi9 & 0.030 & 0.018 & 0.055\\
12 & Demographic, prior self-harm, and diagnosis \& utilization variables, and Charlson score & 0.035 & 0.023 & 0.060\\
13 & Demographic, prior self-harm, and diagnosis \& utilization variables, and PHQi9 & 0.035 & 0.024 & 0.059\\
14 & Demographic and prior self-harm variables, Charlson score, and PHQi9 & 0.029 & 0.016 & 0.051\\
15 & All variables & 0.035 & 0.024 & 0.059\\
16 & Age & 0.010 & 0.002 & 0.015\\
17 & Age and PHQi9 & 0.017 & 0.005 & 0.024\\
18 & Age, sex, and race or ethnicity & 0.014 & 0.004 & 0.020\\
19 & Age, sex, race or ethnicity, and PHQi9 & 0.018 & 0.004 & 0.026\\
\bottomrule
\end{tabular}
\end{table}

\begin{table}
\centering
\caption{Mean, minimum (Min.), and maximum (Max.) estimated sensitivity at the 95th percentile of predicted risk for each group of variables in the predictors of suicide risk analysis. Summary statistics are taken over all timepoints (1--6) and prediction algorithms (logistic regression, lasso, random forests, gradient boosted trees, discrete super learner, and super learner).\label{tab:sensitivity_data_analysis_pred_perf_all_supp}}
\centering
\fontsize{10}{12}\selectfont
\begin{tabular}[t]{>{\raggedright\arraybackslash}p{1in}>{\raggedleft\arraybackslash}p{4in}rrr}
\toprule
\multicolumn{2}{c}{ } & \multicolumn{3}{c}{Estimated Sensitivity} \\
\cmidrule(l{3pt}r{3pt}){3-5}
Variable set number & Description & Mean & Min. & Max.\\
\midrule
1 & No variables & 0.133 & 0.000 & 0.400\\
2 & PHQi9 & 0.133 & 0.000 & 0.168\\
3 & Age and sex & 0.173 & 0.091 & 0.242\\
4 & Age, sex, and PHQi9 & 0.206 & 0.105 & 0.308\\
5 & Age, sex, and prior self-harm variables & 0.281 & 0.140 & 0.389\\
6 & Age, sex, prior self-harm variables, and PHQi9 & 0.159 & 0.053 & 0.258\\
7 & Demographic variables & 0.177 & 0.105 & 0.259\\
8 & Demographic and prior self-harm variables & 0.300 & 0.210 & 0.375\\
9 & Demographic, prior self-harm, and diagnosis \& utilization variables & 0.394 & 0.297 & 0.484\\
10 & Demographic and prior self-harm variables and Charlson score & 0.304 & 0.190 & 0.375\\
11 & Demographic and prior self-harm variables and PHQi9 & 0.326 & 0.180 & 0.421\\
12 & Demographic, prior self-harm, and diagnosis \& utilization variables, and Charlson score & 0.399 & 0.300 & 0.484\\
13 & Demographic, prior self-harm, and diagnosis \& utilization variables, and PHQi9 & 0.404 & 0.308 & 0.505\\
14 & Demographic and prior self-harm variables, Charlson score, and PHQi9 & 0.324 & 0.190 & 0.421\\
15 & All variables & 0.401 & 0.292 & 0.496\\
16 & Age & 0.117 & 0.025 & 0.176\\
17 & Age and PHQi9 & 0.186 & 0.053 & 0.259\\
18 & Age, sex, and race or ethnicity & 0.152 & 0.050 & 0.225\\
19 & Age, sex, race or ethnicity, and PHQi9 & 0.199 & 0.074 & 0.308\\
\bottomrule
\end{tabular}
\end{table}

\begin{figure}
    \centering
    \includegraphics[width=1\textwidth]{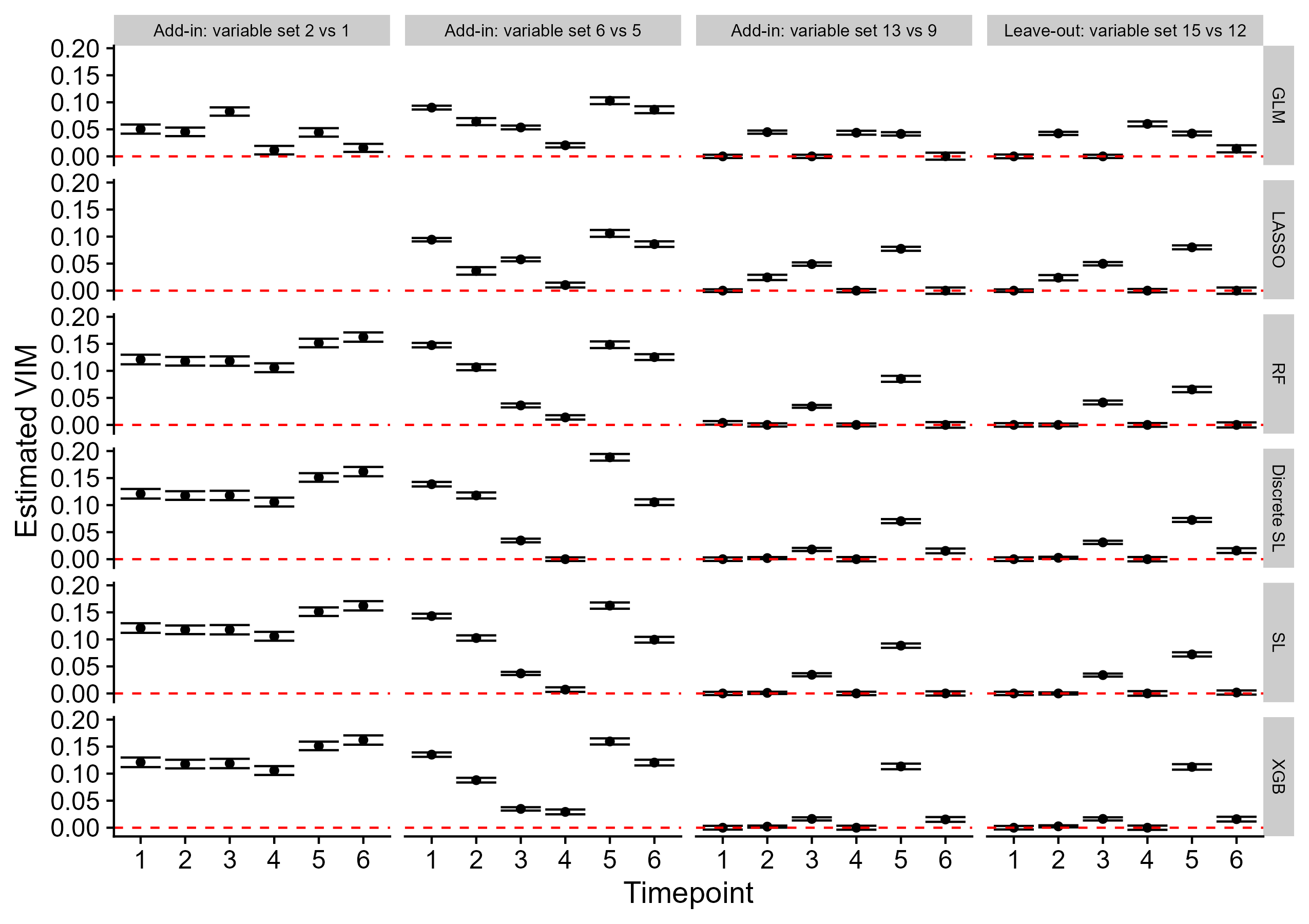}
    \caption{Variable importance estimates for PHQi9 measured using AUC, considering the importance of PHQi9 when compared to other variables. Comparisons are in the columns: PHQi9 vs no variables (comparing variable set 2 to 1), age and sex vs no variables (comparing variable set 3 to 1), PHQi9 vs age and sex (comparing variable set 4 to 3), prior self-harm variables vs age and sex (comparing variable set 5 to 3), PHQi9 vs age, sex, and prior self-harm variables (comparing variable set 6 to 5). Rows denote the estimator: logistic regression (GLM), lasso, random forests (RF), discrete super learner (Discrete SL), super learner (SL), and boosted trees (XGB). VIM values were truncated at zero.}
    \label{fig:auc_data_analysis_vim_over_time_all}
\end{figure}

\begin{figure}
    \centering
    \includegraphics[width=1\textwidth]{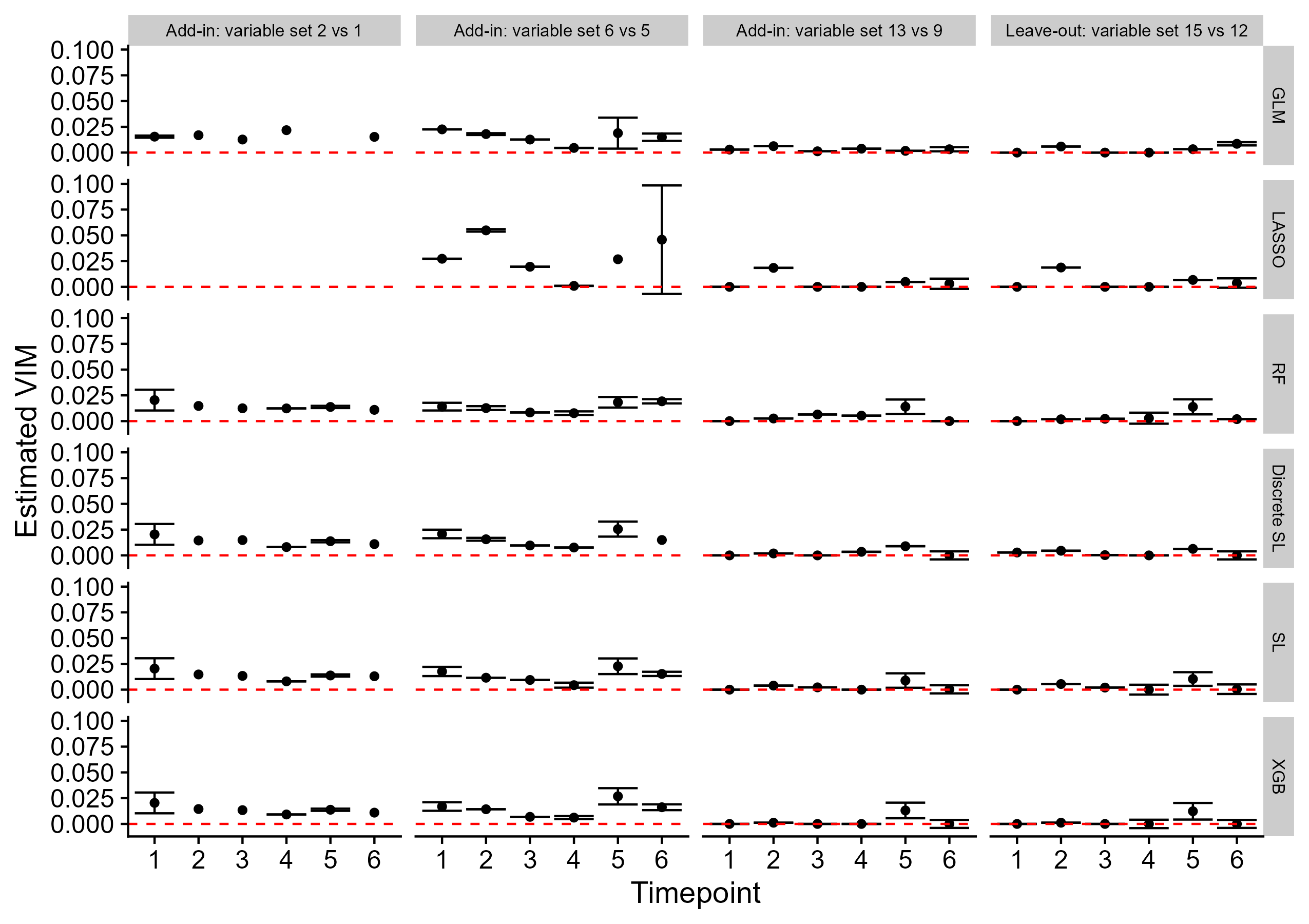}
    \caption{Variable importance estimates for PHQi9 measured using PPV at the 95th percentile, considering the importance of PHQi9 when compared to other variables. Comparisons are in the columns: PHQi9 vs no variables (comparing variable set 2 to 1), age and sex vs no variables (comparing variable set 3 to 1), PHQi9 vs age and sex (comparing variable set 4 to 3), prior self-harm variables vs age and sex (comparing variable set 5 to 3), PHQi9 vs age, sex, and prior self-harm variables (comparing variable set 6 to 5). Rows denote the estimator: logistic regression (GLM), lasso, random forests (RF), discrete super learner (Discrete SL), super learner (SL), and boosted trees (XGB). VIM values were truncated at zero.}
    \label{fig:ppv_data_analysis_vim_over_time_all}
\end{figure}

\begin{figure}
    \centering
    \includegraphics[width=1\textwidth]{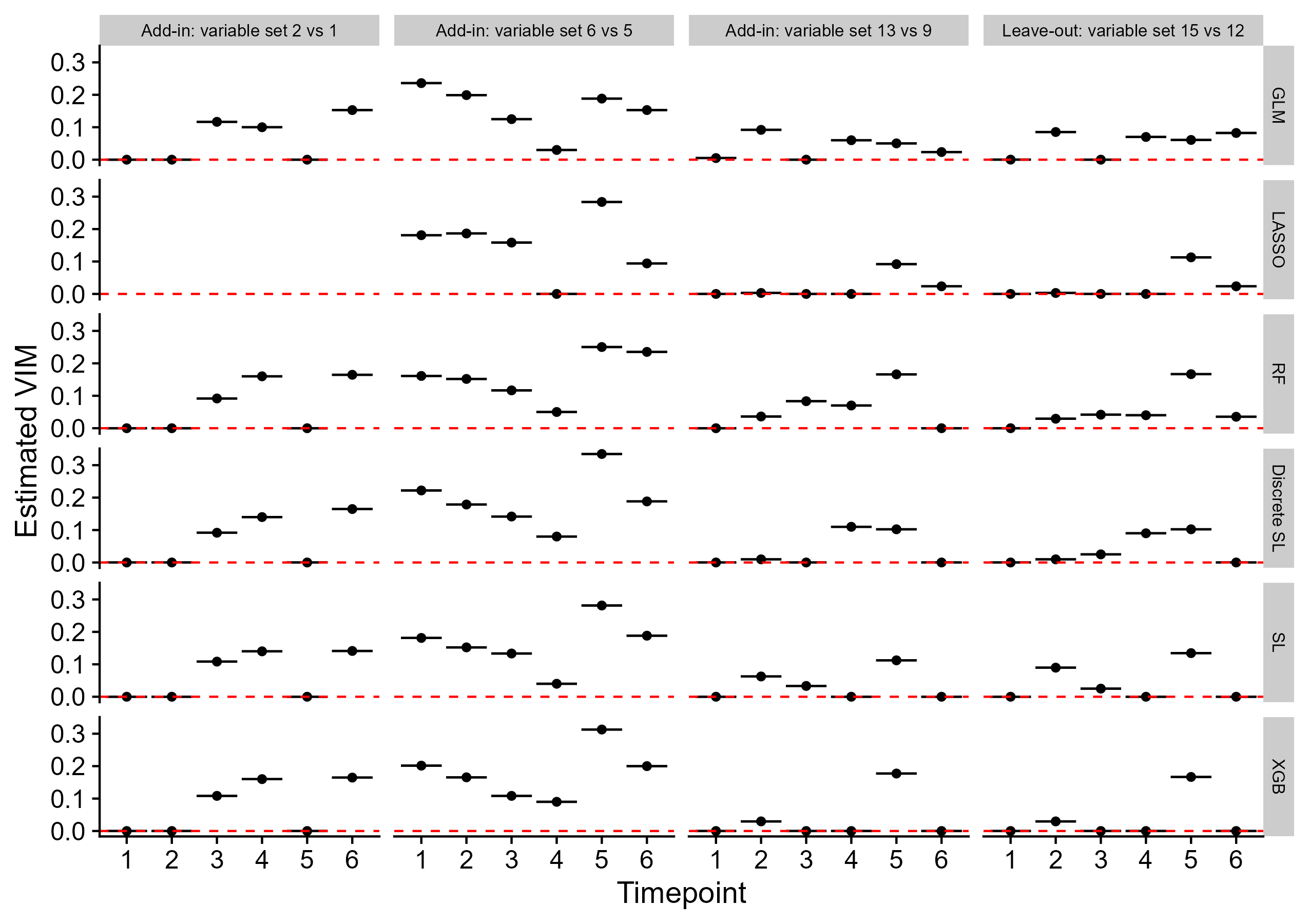}
    \caption{Variable importance estimates for PHQi9 measured using sensitivity at the 95th percentile, considering the importance of PHQi9 when compared to other variables. Comparisons are in the columns: PHQi9 vs no variables (comparing variable set 2 to 1), age and sex vs no variables (comparing variable set 3 to 1), PHQi9 vs age and sex (comparing variable set 4 to 3), prior self-harm variables vs age and sex (comparing variable set 5 to 3), PHQi9 vs age, sex, and prior self-harm variables (comparing variable set 6 to 5). Rows denote the estimator: logistic regression (GLM), lasso, random forests (RF), discrete super learner (Discrete SL), super learner (SL), and boosted trees (XGB). VIM values were truncated at zero.}
    \label{fig:sensitivity_data_analysis_vim_over_time_all}
\end{figure}

In Figures~\ref{fig:auc_data_analysis_vim_over_time_all}--\ref{fig:sensitivity_data_analysis_vim_over_time_all}, we present the estimated VIM for PHQi9 relative to various sets of variables for all variable sets and all estimators. Missing points reflect the failure of an algorithm to converge, which occurred for the lasso for variable set 11 (PHQi9 alone) since it contained only a single element. 

In Tables~\ref{tab:auc_data_analysis_vim_summaries_all}--\ref{tab:sensitivity_data_analysis_vim_summaries_all}, we present the estimated average VIM and slope of the linear trend in VIM for PHQi9 over the six timepoints studied here, measured using AUC, PPV, and sensitivity, respectively. The variable sets are provided in the main manuscript. As suggested by the results for the individual time points provided in the main manuscript, there are some differences across estimation procedures. However, the overall trends -- that the average importance is near zero for all comparisons besides the add-in importance of PHQi9 compared to no variables, and that the importance of PHQi9 relative to demographic and prior self-harm variables is slightly decreasing -- hold across procedures.

\begingroup\fontsize{9}{11}\selectfont

\begin{longtable}[t]{lrrrr}
\caption{Estimates of the average VIM (defined using AUC) and slope of the linear trend in VIM over the time series,  considering the importance of PHQi9 when compared to other variables. Comparisons are: PHQi9 vs no variables (comparing variable set 2 to 1), PHQi9 vs age, sex, and prior self-harm (comparing variable set 6 to 5), PHQi9 vs demographic, prior self-harm, diagnosis \& utilization variables (comparing variable set 13 to 9), PHQi9 vs all other variables (comparing variable set 15 to 12). Estimates are shown for logistic regression (GLM), lasso (LASSO), random forests (RF), the discrete super learner (Discrete SL), super learner (SL), and boosted trees (XGB). \label{tab:auc_data_analysis_vim_summaries_all}}\\
\toprule
VIM type: comparison & Estimate & SE & 95\% CI & p-value\\
\midrule
\endfirsthead
\caption[]{Estimates of the average VIM (defined using AUC) and slope of the linear trend in VIM over the time series, considering the importance of PHQi9 when compared to other variables. \textit{(continued)}}\\
\toprule
VIM type: comparison & Estimate & SE & 95\% CI & p-value\\
\midrule
\endhead

\endfoot
\bottomrule
\endlastfoot
\addlinespace[0.3em]
\multicolumn{5}{l}{\textbf{Average (GLM)}}\\
\hspace{1em}PHQi9 vs no variables & 0.042 & 0.020 & {}[0.002, 0.081] & 0.019\\
\hspace{1em}PHQi9 vs age, sex, and prior self-harm & 0.070 & 0.012 & {}[0.047, 0.093] & < 0.001\\
\hspace{1em}PHQi9 vs demographic, prior self-harm, and diagnosis \& utilization variables & 0.018 & 0.012 & {}[-0.006, 0.041] & 0.073\\
\hspace{1em}PHQi9 vs all other variables & 0.022 & 0.012 & {}[-0.002, 0.045] & 0.037\\
\addlinespace[0.3em]
\multicolumn{5}{l}{\textbf{Trend - slope (GLM)}}\\
\hspace{1em}PHQi9 vs no variables & -0.007 & 0.012 & {}[-0.030, 0.016] & 0.543\\
\hspace{1em}PHQi9 vs age, sex, and prior self-harm & 0.002 & 0.007 & {}[-0.012, 0.015] & 0.791\\
\hspace{1em}PHQi9 vs demographic, prior self-harm, and diagnosis \& utilization variables & 0.003 & 0.007 & {}[-0.011, 0.017] & 0.664\\
\hspace{1em}PHQi9 vs all other variables & 0.006 & 0.007 & {}[-0.007, 0.019] & 0.377\\
\addlinespace[0.3em]
\multicolumn{5}{l}{\textbf{Average (LASSO)}}\\
\hspace{1em}PHQi9 vs no variables & --- & --- & --- & \vphantom{1} ---\\
\hspace{1em}PHQi9 vs age, sex, and prior self-harm & 0.065 & 0.012 & {}[0.042, 0.088] & < 0.001\\
\hspace{1em}PHQi9 vs demographic, prior self-harm, and diagnosis \& utilization variables & 0.018 & 0.010 & {}[-0.002, 0.039] & 0.042\\
\hspace{1em}PHQi9 vs all other variables & 0.017 & 0.010 & {}[-0.004, 0.038] & 0.052\\
\addlinespace[0.3em]
\multicolumn{5}{l}{\textbf{Trend - slope (LASSO)}}\\
\hspace{1em}PHQi9 vs no variables & --- & --- & --- & ---\\
\hspace{1em}PHQi9 vs age, sex, and prior self-harm & 0.003 & 0.007 & {}[-0.010, 0.017] & 0.615\\
\hspace{1em}PHQi9 vs demographic, prior self-harm, and diagnosis \& utilization variables & 0.002 & 0.006 & {}[-0.009, 0.014] & 0.697\\
\hspace{1em}PHQi9 vs all other variables & 0.002 & 0.006 & {}[-0.009, 0.014] & 0.707\\
\addlinespace[0.3em]
\multicolumn{5}{l}{\textbf{Average (RF)}}\\
\hspace{1em}PHQi9 vs no variables & 0.129 & 0.019 & {}[0.093, 0.166] & < \vphantom{3} 0.001\\
\hspace{1em}PHQi9 vs age, sex, and prior self-harm & 0.096 & 0.012 & {}[0.074, 0.119] & < 0.001\\
\hspace{1em}PHQi9 vs demographic, prior self-harm, and diagnosis \& utilization variables & 0.016 & 0.010 & {}[-0.004, 0.035] & 0.056\\
\hspace{1em}PHQi9 vs all other variables & 0.008 & 0.010 & {}[-0.011, 0.027] & 0.203\\
\addlinespace[0.3em]
\multicolumn{5}{l}{\textbf{Trend - slope (RF)}}\\
\hspace{1em}PHQi9 vs no variables & 0.008 & 0.011 & {}[-0.012, 0.029] & 0.428\\
\hspace{1em}PHQi9 vs age, sex, and prior self-harm & 0.000 & 0.007 & {}[-0.013, 0.013] & 0.973\\
\hspace{1em}PHQi9 vs demographic, prior self-harm, and diagnosis \& utilization variables & 0.003 & 0.006 & {}[-0.008, 0.014] & 0.617\\
\hspace{1em}PHQi9 vs all other variables & 0.003 & 0.005 & {}[-0.008, 0.014] & 0.562\\
\addlinespace[0.3em]
\multicolumn{5}{l}{\textbf{Average (Discrete SL)}}\\
\hspace{1em}PHQi9 vs no variables & 0.129 & 0.019 & {}[0.093, 0.166] & < \vphantom{2} 0.001\\
\hspace{1em}PHQi9 vs age, sex, and prior self-harm & 0.096 & 0.012 & {}[0.073, 0.120] & < 0.001\\
\hspace{1em}PHQi9 vs demographic, prior self-harm, and diagnosis \& utilization variables & 0.016 & 0.010 & {}[-0.003, 0.034] & 0.053\\
\hspace{1em}PHQi9 vs all other variables & 0.015 & 0.010 & {}[-0.003, 0.034] & 0.054\\
\addlinespace[0.3em]
\multicolumn{5}{l}{\textbf{Trend - slope (Discrete SL)}}\\
\hspace{1em}PHQi9 vs no variables & 0.008 & 0.011 & {}[-0.013, 0.029] & 0.433\\
\hspace{1em}PHQi9 vs age, sex, and prior self-harm & 0.000 & 0.007 & {}[-0.013, 0.014] & 0.985\\
\hspace{1em}PHQi9 vs demographic, prior self-harm, and diagnosis \& utilization variables & 0.008 & 0.006 & {}[-0.002, 0.019] & 0.129\\
\hspace{1em}PHQi9 vs all other variables & 0.008 & 0.005 & {}[-0.003, 0.019] & 0.138\\
\addlinespace[0.3em]
\multicolumn{5}{l}{\textbf{Average (SL)}}\\
\hspace{1em}PHQi9 vs no variables & 0.129 & 0.019 & {}[0.093, 0.166] & < \vphantom{1} 0.001\\
\hspace{1em}PHQi9 vs age, sex, and prior self-harm & 0.092 & 0.012 & {}[0.069, 0.115] & < 0.001\\
\hspace{1em}PHQi9 vs demographic, prior self-harm, and diagnosis \& utilization variables & 0.018 & 0.009 & {}[-0.000, 0.036] & 0.026\\
\hspace{1em}PHQi9 vs all other variables & 0.016 & 0.009 & {}[-0.003, 0.034] & 0.047\\
\addlinespace[0.3em]
\multicolumn{5}{l}{\textbf{Trend - slope (SL)}}\\
\hspace{1em}PHQi9 vs no variables & 0.008 & 0.011 & {}[-0.013, 0.029] & \vphantom{1} 0.432\\
\hspace{1em}PHQi9 vs age, sex, and prior self-harm & -0.002 & 0.007 & {}[-0.015, 0.011] & 0.771\\
\hspace{1em}PHQi9 vs demographic, prior self-harm, and diagnosis \& utilization variables & 0.006 & 0.005 & {}[-0.004, 0.016] & 0.254\\
\hspace{1em}PHQi9 vs all other variables & 0.006 & 0.005 & {}[-0.004, 0.016] & 0.254\\
\addlinespace[0.3em]
\multicolumn{5}{l}{\textbf{Average (XGB)}}\\
\hspace{1em}PHQi9 vs no variables & 0.129 & 0.019 & {}[0.093, 0.166] & < 0.001\\
\hspace{1em}PHQi9 vs age, sex, and prior self-harm & 0.095 & 0.012 & {}[0.071, 0.118] & < 0.001\\
\hspace{1em}PHQi9 vs demographic, prior self-harm, and diagnosis \& utilization variables & 0.020 & 0.010 & {}[0.001, 0.039] & 0.021\\
\hspace{1em}PHQi9 vs all other variables & 0.019 & 0.010 & {}[0.000, 0.038] & 0.022\\
\addlinespace[0.3em]
\multicolumn{5}{l}{\textbf{Trend - slope (XGB)}}\\
\hspace{1em}PHQi9 vs no variables & 0.008 & 0.011 & {}[-0.013, 0.029] & 0.432\\
\hspace{1em}PHQi9 vs age, sex, and prior self-harm & 0.004 & 0.007 & {}[-0.010, 0.017] & 0.575\\
\hspace{1em}PHQi9 vs demographic, prior self-harm, and diagnosis \& utilization variables & 0.014 & 0.006 & {}[0.003, 0.025] & 0.010\\
\hspace{1em}PHQi9 vs all other variables & 0.014 & 0.006 & {}[0.003, 0.025] & 0.010\\*
\end{longtable}
\endgroup{}

\begingroup\fontsize{9}{11}\selectfont

\begin{longtable}[t]{lrrrr}
\caption{Estimates of the average VIM (defined using PPV) and slope of the linear trend in VIM over the time series,  considering the importance of PHQi9 when compared to other variables. Comparisons are: PHQi9 vs no variables (comparing variable set 2 to 1), PHQi9 vs age, sex, and prior self-harm (comparing variable set 6 to 5), PHQi9 vs demographic, prior self-harm, diagnosis \& utilization variables (comparing variable set 13 to 9), PHQi9 vs all other variables (comparing variable set 15 to 12). Estimates are shown for logistic regression (GLM), lasso (LASSO), random forests (RF), the discrete super learner (Discrete SL), super learner (SL), and boosted trees (XGB). \label{tab:ppv_data_analysis_vim_summaries_all}}\\
\toprule
VIM type: comparison & Estimate & SE & 95\% CI & p-value\\
\midrule
\endfirsthead
\caption[]{Estimates of the average VIM (defined using PPV) and slope of the linear trend in VIM over the time series, considering the importance of PHQi9 when compared to other variables. \textit{(continued)}}\\
\toprule
VIM type: comparison & Estimate & SE & 95\% CI & p-value\\
\midrule
\endhead

\endfoot
\bottomrule
\endlastfoot
\addlinespace[0.3em]
\multicolumn{5}{l}{\textbf{Average (GLM)}}\\
\hspace{1em}PHQi9 vs no variables & --- & --- & --- & \vphantom{3} ---\\
\hspace{1em}PHQi9 vs age, sex, and prior self-harm & 0.015 & 0.002 & {}[0.012, 0.019] & < 0.001\\
\hspace{1em}PHQi9 vs demographic, prior self-harm, and diagnosis \& utilization variables & 0.003 & 0.002 & {}[-0.001, 0.007] & 0.055\\
\hspace{1em}PHQi9 vs all other variables & 0.000 & 0.003 & {}[-0.007, 0.006] & 0.585\\
\addlinespace[0.3em]
\multicolumn{5}{l}{\textbf{Trend - slope (GLM)}}\\
\hspace{1em}PHQi9 vs no variables & --- & --- & --- & \vphantom{2} ---\\
\hspace{1em}PHQi9 vs age, sex, and prior self-harm & -0.001 & 0.001 & {}[-0.003, 0.001] & 0.225\\
\hspace{1em}PHQi9 vs demographic, prior self-harm, and diagnosis \& utilization variables & 0.000 & 0.001 & {}[-0.003, 0.002] & 0.817\\
\hspace{1em}PHQi9 vs all other variables & 0.000 & 0.001 & {}[-0.002, 0.003] & 0.746\\
\addlinespace[0.3em]
\multicolumn{5}{l}{\textbf{Average (LASSO)}}\\
\hspace{1em}PHQi9 vs no variables & --- & --- & --- & \vphantom{1} ---\\
\hspace{1em}PHQi9 vs age, sex, and prior self-harm & 0.029 & 0.003 & {}[0.022, 0.036] & < 0.001\\
\hspace{1em}PHQi9 vs demographic, prior self-harm, and diagnosis \& utilization variables & 0.003 & 0.005 & {}[-0.006, 0.013] & 0.241\\
\hspace{1em}PHQi9 vs all other variables & 0.004 & 0.005 & {}[-0.006, 0.013] & 0.225\\
\addlinespace[0.3em]
\multicolumn{5}{l}{\textbf{Trend - slope (LASSO)}}\\
\hspace{1em}PHQi9 vs no variables & --- & --- & --- & ---\\
\hspace{1em}PHQi9 vs age, sex, and prior self-harm & 0.000 & 0.002 & {}[-0.005, 0.004] & 0.896\\
\hspace{1em}PHQi9 vs demographic, prior self-harm, and diagnosis \& utilization variables & 0.000 & 0.003 & {}[-0.006, 0.004] & 0.784\\
\hspace{1em}PHQi9 vs all other variables & 0.000 & 0.003 & {}[-0.005, 0.005] & 0.883\\
\addlinespace[0.3em]
\multicolumn{5}{l}{\textbf{Average (RF)}}\\
\hspace{1em}PHQi9 vs no variables & 0.014 & 0.002 & {}[0.010, 0.018] & < \vphantom{1} 0.001\\
\hspace{1em}PHQi9 vs age, sex, and prior self-harm & 0.013 & 0.002 & {}[0.010, 0.017] & < \vphantom{1} 0.001\\
\hspace{1em}PHQi9 vs demographic, prior self-harm, and diagnosis \& utilization variables & 0.004 & 0.002 & {}[-0.000, 0.008] & 0.031\\
\hspace{1em}PHQi9 vs all other variables & 0.003 & 0.002 & {}[-0.001, 0.007] & 0.078\\
\addlinespace[0.3em]
\multicolumn{5}{l}{\textbf{Trend - slope (RF)}}\\
\hspace{1em}PHQi9 vs no variables & -0.001 & 0.001 & {}[-0.004, 0.001] & 0.201\\
\hspace{1em}PHQi9 vs age, sex, and prior self-harm & 0.001 & 0.001 & {}[-0.001, 0.003] & 0.225\\
\hspace{1em}PHQi9 vs demographic, prior self-harm, and diagnosis \& utilization variables & 0.001 & 0.001 & {}[-0.001, 0.004] & 0.356\\
\hspace{1em}PHQi9 vs all other variables & 0.002 & 0.001 & {}[-0.001, 0.004] & 0.140\\
\addlinespace[0.3em]
\multicolumn{5}{l}{\textbf{Average (Discrete SL)}}\\
\hspace{1em}PHQi9 vs no variables & 0.014 & 0.002 & {}[0.010, 0.017] & < \vphantom{1} 0.001\\
\hspace{1em}PHQi9 vs age, sex, and prior self-harm & 0.016 & 0.002 & {}[0.012, 0.020] & < 0.001\\
\hspace{1em}PHQi9 vs demographic, prior self-harm, and diagnosis \& utilization variables & 0.000 & 0.002 & {}[-0.004, 0.005] & 0.348\\
\hspace{1em}PHQi9 vs all other variables & 0.002 & 0.002 & {}[-0.003, 0.006] & 0.211\\
\addlinespace[0.3em]
\multicolumn{5}{l}{\textbf{Trend - slope (Discrete SL)}}\\
\hspace{1em}PHQi9 vs no variables & -0.002 & 0.001 & {}[-0.004, 0.001] & 0.152\\
\hspace{1em}PHQi9 vs age, sex, and prior self-harm & 0.000 & 0.001 & {}[-0.002, 0.002] & 0.967\\
\hspace{1em}PHQi9 vs demographic, prior self-harm, and diagnosis \& utilization variables & 0.001 & 0.001 & {}[-0.001, 0.004] & 0.285\\
\hspace{1em}PHQi9 vs all other variables & 0.000 & 0.001 & {}[-0.003, 0.002] & 0.741\\
\addlinespace[0.3em]
\multicolumn{5}{l}{\textbf{Average (SL)}}\\
\hspace{1em}PHQi9 vs no variables & 0.014 & 0.002 & {}[0.010, 0.018] & < 0.001\\
\hspace{1em}PHQi9 vs age, sex, and prior self-harm & 0.013 & 0.002 & {}[0.010, 0.017] & < 0.001\\
\hspace{1em}PHQi9 vs demographic, prior self-harm, and diagnosis \& utilization variables & 0.002 & 0.002 & {}[-0.003, 0.006] & 0.213\\
\hspace{1em}PHQi9 vs all other variables & 0.002 & 0.002 & {}[-0.002, 0.007] & 0.181\\
\addlinespace[0.3em]
\multicolumn{5}{l}{\textbf{Trend - slope (SL)}}\\
\hspace{1em}PHQi9 vs no variables & -0.001 & 0.001 & {}[-0.004, 0.001] & 0.288\\
\hspace{1em}PHQi9 vs age, sex, and prior self-harm & 0.000 & 0.001 & {}[-0.002, 0.002] & 0.647\\
\hspace{1em}PHQi9 vs demographic, prior self-harm, and diagnosis \& utilization variables & 0.000 & 0.001 & {}[-0.002, 0.003] & 0.631\\
\hspace{1em}PHQi9 vs all other variables & 0.000 & 0.001 & {}[-0.002, 0.003] & 0.615\\
\addlinespace[0.3em]
\multicolumn{5}{l}{\textbf{Average (XGB)}}\\
\hspace{1em}PHQi9 vs no variables & 0.014 & 0.002 & {}[0.010, 0.017] & < 0.001\\
\hspace{1em}PHQi9 vs age, sex, and prior self-harm & 0.015 & 0.002 & {}[0.011, 0.018] & < 0.001\\
\hspace{1em}PHQi9 vs demographic, prior self-harm, and diagnosis \& utilization variables & 0.001 & 0.002 & {}[-0.003, 0.006] & 0.285\\
\hspace{1em}PHQi9 vs all other variables & 0.000 & 0.002 & {}[-0.004, 0.005] & 0.355\\
\addlinespace[0.3em]
\multicolumn{5}{l}{\textbf{Trend - slope (XGB)}}\\
\hspace{1em}PHQi9 vs no variables & -0.002 & 0.001 & {}[-0.004, 0.001] & 0.170\\
\hspace{1em}PHQi9 vs age, sex, and prior self-harm & 0.000 & 0.001 & {}[-0.001, 0.003] & 0.367\\
\hspace{1em}PHQi9 vs demographic, prior self-harm, and diagnosis \& utilization variables & 0.000 & 0.001 & {}[-0.002, 0.004] & 0.489\\
\hspace{1em}PHQi9 vs all other variables & 0.000 & 0.001 & {}[-0.002, 0.003] & 0.560\\*
\end{longtable}
\endgroup{}

\begingroup\fontsize{9}{11}\selectfont

\begin{longtable}[t]{lrrrr}
\caption{Estimates of the average VIM (defined using sensitivity) and slope of the linear trend in VIM over the time series,  considering the importance of PHQi9 when compared to other variables. Comparisons are: PHQi9 vs no variables (comparing variable set 2 to 1), PHQi9 vs age, sex, and prior self-harm (comparing variable set 6 to 5), PHQi9 vs demographic, prior self-harm, diagnosis \& utilization variables (comparing variable set 13 to 9), PHQi9 vs all other variables (comparing variable set 15 to 12). Estimates are shown for logistic regression (GLM), lasso (LASSO), random forests (RF), the discrete super learner (Discrete SL), super learner (SL), and boosted trees (XGB). \label{tab:sensitivity_data_analysis_vim_summaries_all}}\\
\toprule
VIM type: comparison & Estimate & SE & 95\% CI & p-value\\
\midrule
\endfirsthead
\caption[]{Estimates of the average VIM (defined using sensitivity) and slope of the linear trend in VIM over the time series, considering the importance of PHQi9 when compared to other variables. \textit{(continued)}}\\
\toprule
VIM type: comparison & Estimate & SE & 95\% CI & p-value\\
\midrule
\endhead

\endfoot
\bottomrule
\endlastfoot
\addlinespace[0.3em]
\multicolumn{5}{l}{\textbf{Average (GLM)}}\\
\hspace{1em}PHQi9 vs no variables & -0.023 & 0.009 & {}[-0.041, -0.005] & 0.995\\
\hspace{1em}PHQi9 vs age, sex, and prior self-harm & 0.155 & 0.016 & {}[0.124, 0.187] & < 0.001\\
\hspace{1em}PHQi9 vs demographic, prior self-harm, and diagnosis \& utilization variables & 0.037 & 0.019 & {}[-0.001, 0.075] & 0.027\\
\hspace{1em}PHQi9 vs all other variables & 0.046 & 0.019 & {}[0.008, 0.083] & 0.008\\
\addlinespace[0.3em]
\multicolumn{5}{l}{\textbf{Trend - slope (GLM)}}\\
\hspace{1em}PHQi9 vs no variables & 0.034 & 0.006 & {}[0.024, 0.045] & < 0.001\\
\hspace{1em}PHQi9 vs age, sex, and prior self-harm & -0.016 & 0.009 & {}[-0.034, 0.003] & 0.096\\
\hspace{1em}PHQi9 vs demographic, prior self-harm, and diagnosis \& utilization variables & 0.001 & 0.011 & {}[-0.021, 0.023] & 0.927\\
\hspace{1em}PHQi9 vs all other variables & 0.012 & 0.011 & {}[-0.009, 0.034] & 0.264\\
\addlinespace[0.3em]
\multicolumn{5}{l}{\textbf{Average (LASSO)}}\\
\hspace{1em}PHQi9 vs no variables & --- & --- & --- & \vphantom{1} ---\\
\hspace{1em}PHQi9 vs age, sex, and prior self-harm & 0.139 & 0.015 & {}[0.109, 0.169] & < 0.001\\
\hspace{1em}PHQi9 vs demographic, prior self-harm, and diagnosis \& utilization variables & 0.007 & 0.020 & {}[-0.031, 0.045] & 0.362\\
\hspace{1em}PHQi9 vs all other variables & 0.009 & 0.020 & {}[-0.029, 0.047] & 0.323\\
\addlinespace[0.3em]
\multicolumn{5}{l}{\textbf{Trend - slope (LASSO)}}\\
\hspace{1em}PHQi9 vs no variables & --- & --- & --- & ---\\
\hspace{1em}PHQi9 vs age, sex, and prior self-harm & -0.011 & 0.009 & {}[-0.028, 0.006] & 0.222\\
\hspace{1em}PHQi9 vs demographic, prior self-harm, and diagnosis \& utilization variables & 0.014 & 0.011 & {}[-0.009, 0.036] & 0.229\\
\hspace{1em}PHQi9 vs all other variables & 0.015 & 0.011 & {}[-0.007, 0.038] & 0.183\\
\addlinespace[0.3em]
\multicolumn{5}{l}{\textbf{Average (RF)}}\\
\hspace{1em}PHQi9 vs no variables & 0.007 & 0.010 & {}[-0.014, 0.027] & 0.255\\
\hspace{1em}PHQi9 vs age, sex, and prior self-harm & 0.161 & 0.018 & {}[0.126, 0.195] & < 0.001\\
\hspace{1em}PHQi9 vs demographic, prior self-harm, and diagnosis \& utilization variables & 0.054 & 0.020 & {}[0.015, 0.093] & 0.003\\
\hspace{1em}PHQi9 vs all other variables & 0.045 & 0.020 & {}[0.006, 0.084] & 0.011\\
\addlinespace[0.3em]
\multicolumn{5}{l}{\textbf{Trend - slope (RF)}}\\
\hspace{1em}PHQi9 vs no variables & 0.056 & 0.006 & {}[0.044, 0.068] & < 0.001\\
\hspace{1em}PHQi9 vs age, sex, and prior self-harm & 0.017 & 0.010 & {}[-0.002, 0.036] & 0.079\\
\hspace{1em}PHQi9 vs demographic, prior self-harm, and diagnosis \& utilization variables & 0.015 & 0.012 & {}[-0.008, 0.038] & 0.194\\
\hspace{1em}PHQi9 vs all other variables & 0.023 & 0.012 & {}[-0.000, 0.045] & 0.052\\
\addlinespace[0.3em]
\multicolumn{5}{l}{\textbf{Average (Discrete SL)}}\\
\hspace{1em}PHQi9 vs no variables & 0.003 & 0.010 & {}[-0.017, 0.023] & 0.368\\
\hspace{1em}PHQi9 vs age, sex, and prior self-harm & 0.191 & 0.017 & {}[0.157, 0.225] & < 0.001\\
\hspace{1em}PHQi9 vs demographic, prior self-harm, and diagnosis \& utilization variables & 0.021 & 0.020 & {}[-0.018, 0.060] & 0.149\\
\hspace{1em}PHQi9 vs all other variables & 0.025 & 0.020 & {}[-0.014, 0.063] & 0.107\\
\addlinespace[0.3em]
\multicolumn{5}{l}{\textbf{Trend - slope (Discrete SL)}}\\
\hspace{1em}PHQi9 vs no variables & 0.055 & 0.006 & {}[0.044, 0.067] & < \vphantom{1} 0.001\\
\hspace{1em}PHQi9 vs age, sex, and prior self-harm & 0.007 & 0.010 & {}[-0.012, 0.026] & 0.485\\
\hspace{1em}PHQi9 vs demographic, prior self-harm, and diagnosis \& utilization variables & 0.014 & 0.012 & {}[-0.008, 0.037] & 0.215\\
\hspace{1em}PHQi9 vs all other variables & 0.014 & 0.012 & {}[-0.008, 0.037] & 0.213\\
\addlinespace[0.3em]
\multicolumn{5}{l}{\textbf{Average (SL)}}\\
\hspace{1em}PHQi9 vs no variables & 0.002 & 0.010 & {}[-0.018, 0.022] & 0.408\\
\hspace{1em}PHQi9 vs age, sex, and prior self-harm & 0.163 & 0.018 & {}[0.127, 0.198] & < 0.001\\
\hspace{1em}PHQi9 vs demographic, prior self-harm, and diagnosis \& utilization variables & 0.029 & 0.020 & {}[-0.011, 0.069] & 0.079\\
\hspace{1em}PHQi9 vs all other variables & 0.029 & 0.020 & {}[-0.011, 0.068] & 0.078\\
\addlinespace[0.3em]
\multicolumn{5}{l}{\textbf{Trend - slope (SL)}}\\
\hspace{1em}PHQi9 vs no variables & 0.051 & 0.006 & {}[0.040, 0.063] & < 0.001\\
\hspace{1em}PHQi9 vs age, sex, and prior self-harm & 0.009 & 0.010 & {}[-0.011, 0.029] & 0.355\\
\hspace{1em}PHQi9 vs demographic, prior self-harm, and diagnosis \& utilization variables & 0.008 & 0.012 & {}[-0.015, 0.031] & 0.477\\
\hspace{1em}PHQi9 vs all other variables & 0.006 & 0.012 & {}[-0.017, 0.029] & 0.631\\
\addlinespace[0.3em]
\multicolumn{5}{l}{\textbf{Average (XGB)}}\\
\hspace{1em}PHQi9 vs no variables & 0.010 & 0.010 & {}[-0.011, 0.030] & 0.181\\
\hspace{1em}PHQi9 vs age, sex, and prior self-harm & 0.180 & 0.018 & {}[0.145, 0.214] & < 0.001\\
\hspace{1em}PHQi9 vs demographic, prior self-harm, and diagnosis \& utilization variables & 0.025 & 0.020 & {}[-0.014, 0.064] & 0.104\\
\hspace{1em}PHQi9 vs all other variables & 0.018 & 0.020 & {}[-0.021, 0.057] & 0.180\\
\addlinespace[0.3em]
\multicolumn{5}{l}{\textbf{Trend - slope (XGB)}}\\
\hspace{1em}PHQi9 vs no variables & 0.055 & 0.006 & {}[0.044, 0.067] & < 0.001\\
\hspace{1em}PHQi9 vs age, sex, and prior self-harm & 0.012 & 0.010 & {}[-0.007, 0.031] & 0.228\\
\hspace{1em}PHQi9 vs demographic, prior self-harm, and diagnosis \& utilization variables & 0.011 & 0.012 & {}[-0.011, 0.034] & 0.322\\
\hspace{1em}PHQi9 vs all other variables & 0.010 & 0.012 & {}[-0.013, 0.032] & 0.400\\*
\end{longtable}
\endgroup{}

In Tables~\ref{tab:auc_data_analysis_vim_summaries_all_supp}--\ref{tab:sensitivity_data_analysis_vim_summaries_all_supp}, we show the estimated average VIM and slope of the linear trend in VIM for Charlson score and PHQi9 over the six timepoints for the remaining variable sets described in the main manuscript, measured using AUC, PPV, and sensitivity, respectively. The Charlson score has similar average VIM when compared to the demographic and prior self-harm variables as the PHQi9 when compared to the same variable set (Table~\ref{tab:auc_data_analysis_vim_summaries_all}), but the slope of the trend in VIM appears to be the opposite sign (slightly positive for the Charlson vs slightly negative for the PHQi9). The Charlson score and PHQi9 have similar importance when compared to variable set 4 (demographic and prior self-harm variables, diagnosis and utilization variables). The PHQi9 seems to have constant importance just above zero compared to demographic and prior self-harm variables and the Charlson score (variable set 9 vs 5).

\begingroup\fontsize{9}{11}\selectfont

\begin{longtable}[t]{lrrrr}
\caption{Estimates of the average VIM (defined using AUC) and slope of the linear trend in VIM over the time series,  considering the importance of PHQi9 or Charlson score when compared to other variables. Comparisons are: Age, sex, and prior self-harm vs no variables (comparing variable set 5 to 1), Charlson vs demographic and prior self-harm variables (comparing variable set 10 to 8), PHQi9 vs demographic and prior self-harm variables (comparing variable set 11 to 8), Charlson vs demographic, prior self-harm, and diagnosis \& utilization variables (comparing variable set 12 to 9), PHQi9 vs demographic, prior self-harm variables and Charlson score (comparing variable set 14 to 10), PHQi9 vs age (comparing variable set 18 to 16), PHQi9 vs age, sex, and race or ethnicity (comparing variable set 19 to 17). Estimates are shown for logistic regression (GLM), lasso (LASSO), random forests (RF), the discrete super learner (Discrete SL), super learner (SL), and boosted trees (XGB). \label{tab:auc_data_analysis_vim_summaries_all_supp}}\\
\toprule
VIM type: comparison & Estimate & SE & 95\% CI & p-value\\
\midrule
\endfirsthead
\caption[]{Estimates of the average VIM (defined using AUC) and slope of the linear trend in VIM over the time series,  considering the importance of PHQi9 or Charlson score when compared to other variables. \textit{(continued)}}\\
\toprule
VIM type: comparison & Estimate & SE & 95\% CI & p-value\\
\midrule
\endhead

\endfoot
\bottomrule
\endlastfoot
\addlinespace[0.3em]
\multicolumn{5}{l}{\textbf{Average (GLM)}}\\
\hspace{1em}Age, sex, and prior self-harm vs no variables & 0.069 & 0.012 & {}[0.046, 0.092] & < 0.001\\
\hspace{1em}Charlson vs demographic, prior self-harm variables & 0.022 & 0.012 & {}[-0.001, 0.045] & 0.032\\
\hspace{1em}PHQi9 vs demographic, prior self-harm variables & 0.017 & 0.012 & {}[-0.006, 0.041] & 0.072\\
\hspace{1em}Charlson vs demographic, prior self-harm, and diagnosis \& utilization variables & 0.022 & 0.012 & {}[-0.001, 0.046] & 0.033\\
\hspace{1em}PHQi9 vs demographic, prior self-harm variables and Charlson score & 0.016 & 0.012 & {}[-0.007, 0.039] & 0.084\\
\hspace{1em}PHQi9 vs age & 0.006 & 0.011 & {}[-0.016, 0.028] & 0.294\\
\hspace{1em}PHQi9 vs age, sex, and race or ethnicity & 0.004 & 0.012 & {}[-0.019, 0.028] & 0.359\\
\addlinespace[0.3em]
\multicolumn{5}{l}{\textbf{Trend - slope (GLM)}}\\
\hspace{1em}Age, sex, and prior self-harm vs no variables & 0.004 & 0.007 & {}[-0.010, 0.017] & 0.572\\
\hspace{1em}Charlson vs demographic, prior self-harm variables & 0.014 & 0.007 & {}[0.001, 0.028] & 0.034\\
\hspace{1em}PHQi9 vs demographic, prior self-harm variables & 0.012 & 0.007 & {}[-0.002, 0.025] & 0.095\\
\hspace{1em}Charlson vs demographic, prior self-harm, and diagnosis \& utilization variables & 0.005 & 0.007 & {}[-0.008, 0.019] & 0.454\\
\hspace{1em}PHQi9 vs demographic, prior self-harm variables and Charlson score & 0.013 & 0.007 & {}[-0.000, 0.026] & 0.056\\
\hspace{1em}PHQi9 vs age & 0.002 & 0.007 & {}[-0.011, 0.015] & 0.787\\
\hspace{1em}PHQi9 vs age, sex, and race or ethnicity & 0.000 & 0.007 & {}[-0.013, 0.014] & 0.982\\
\addlinespace[0.3em]
\multicolumn{5}{l}{\textbf{Average (LASSO)}}\\
\hspace{1em}Age, sex, and prior self-harm vs no variables & 0.063 & 0.012 & {}[0.040, 0.086] & < 0.001\\
\hspace{1em}Charlson vs demographic, prior self-harm variables & 0.016 & 0.012 & {}[-0.007, 0.039] & 0.082\\
\hspace{1em}PHQi9 vs demographic, prior self-harm variables & 0.015 & 0.012 & {}[-0.007, 0.038] & 0.093\\
\hspace{1em}Charlson vs demographic, prior self-harm, and diagnosis \& utilization variables & 0.017 & 0.011 & {}[-0.004, 0.037] & 0.055\\
\hspace{1em}PHQi9 vs demographic, prior self-harm variables and Charlson score & 0.012 & 0.012 & {}[-0.010, 0.035] & 0.140\\
\hspace{1em}PHQi9 vs age & 0.164 & 0.017 & {}[0.131, 0.198] & < 0.001\\
\hspace{1em}PHQi9 vs age, sex, and race or ethnicity & 0.004 & 0.011 & {}[-0.018, 0.027] & 0.352\\
\addlinespace[0.3em]
\multicolumn{5}{l}{\textbf{Trend - slope (LASSO)}}\\
\hspace{1em}Age, sex, and prior self-harm vs no variables & 0.005 & 0.007 & {}[-0.008, 0.019] & 0.429\\
\hspace{1em}Charlson vs demographic, prior self-harm variables & 0.007 & 0.007 & {}[-0.006, 0.021] & 0.269\\
\hspace{1em}PHQi9 vs demographic, prior self-harm variables & 0.008 & 0.007 & {}[-0.005, 0.021] & 0.238\\
\hspace{1em}Charlson vs demographic, prior self-harm, and diagnosis \& utilization variables & 0.002 & 0.006 & {}[-0.010, 0.013] & 0.795\\
\hspace{1em}PHQi9 vs demographic, prior self-harm variables and Charlson score & 0.009 & 0.007 & {}[-0.004, 0.022] & 0.168\\
\hspace{1em}PHQi9 vs age & 0.012 & 0.010 & {}[-0.008, 0.031] & 0.238\\
\hspace{1em}PHQi9 vs age, sex, and race or ethnicity & 0.000 & 0.007 & {}[-0.014, 0.012] & 0.928\\
\addlinespace[0.3em]
\multicolumn{5}{l}{\textbf{Average (RF)}}\\
\hspace{1em}Age, sex, and prior self-harm vs no variables & 0.076 & 0.012 & {}[0.052, 0.099] & < 0.001\\
\hspace{1em}Charlson vs demographic, prior self-harm variables & -0.009 & 0.011 & {}[-0.032, 0.013] & 0.786\\
\hspace{1em}PHQi9 vs demographic, prior self-harm variables & 0.016 & 0.011 & {}[-0.006, 0.038] & 0.076\\
\hspace{1em}Charlson vs demographic, prior self-harm, and diagnosis \& utilization variables & 0.010 & 0.010 & {}[-0.010, 0.029] & 0.166\\
\hspace{1em}PHQi9 vs demographic, prior self-harm variables and Charlson score & 0.010 & 0.011 & {}[-0.012, 0.031] & 0.193\\
\hspace{1em}PHQi9 vs age & 0.052 & 0.013 & {}[0.027, 0.076] & < 0.001\\
\hspace{1em}PHQi9 vs age, sex, and race or ethnicity & 0.036 & 0.012 & {}[0.012, 0.059] & 0.001\\
\addlinespace[0.3em]
\multicolumn{5}{l}{\textbf{Trend - slope (RF)}}\\
\hspace{1em}Age, sex, and prior self-harm vs no variables & -0.003 & 0.007 & {}[-0.017, 0.010] & 0.636\\
\hspace{1em}Charlson vs demographic, prior self-harm variables & 0.010 & 0.007 & {}[-0.003, 0.023] & 0.147\\
\hspace{1em}PHQi9 vs demographic, prior self-harm variables & 0.013 & 0.007 & {}[-0.000, 0.026] & 0.058\\
\hspace{1em}Charlson vs demographic, prior self-harm, and diagnosis \& utilization variables & 0.003 & 0.005 & {}[-0.008, 0.014] & 0.593\\
\hspace{1em}PHQi9 vs demographic, prior self-harm variables and Charlson score & 0.010 & 0.007 & {}[-0.003, 0.023] & 0.123\\
\hspace{1em}PHQi9 vs age & 0.000 & 0.007 & {}[-0.014, 0.013] & 0.948\\
\hspace{1em}PHQi9 vs age, sex, and race or ethnicity & -0.002 & 0.007 & {}[-0.016, 0.012] & 0.772\\
\addlinespace[0.3em]
\multicolumn{5}{l}{\textbf{Average (Discrete SL)}}\\
\hspace{1em}Age, sex, and prior self-harm vs no variables & 0.073 & 0.012 & {}[0.049, 0.097] & < 0.001\\
\hspace{1em}Charlson vs demographic, prior self-harm variables & 0.000 & 0.011 & {}[-0.023, 0.021] & 0.531\\
\hspace{1em}PHQi9 vs demographic, prior self-harm variables & 0.019 & 0.011 & {}[-0.003, 0.041] & 0.046\\
\hspace{1em}Charlson vs demographic, prior self-harm, and diagnosis \& utilization variables & 0.012 & 0.010 & {}[-0.007, 0.031] & 0.112\\
\hspace{1em}PHQi9 vs demographic, prior self-harm variables and Charlson score & 0.011 & 0.011 & {}[-0.010, 0.033] & 0.156\\
\hspace{1em}PHQi9 vs age & 0.046 & 0.012 & {}[0.022, 0.070] & < 0.001\\
\hspace{1em}PHQi9 vs age, sex, and race or ethnicity & 0.046 & 0.012 & {}[0.022, 0.070] & < 0.001\\
\addlinespace[0.3em]
\multicolumn{5}{l}{\textbf{Trend - slope (Discrete SL)}}\\
\hspace{1em}Age, sex, and prior self-harm vs no variables & -0.003 & 0.007 & {}[-0.017, 0.011] & 0.684\\
\hspace{1em}Charlson vs demographic, prior self-harm variables & 0.011 & 0.006 & {}[-0.002, 0.024] & 0.088\\
\hspace{1em}PHQi9 vs demographic, prior self-harm variables & 0.014 & 0.006 & {}[0.002, 0.026] & 0.025\\
\hspace{1em}Charlson vs demographic, prior self-harm, and diagnosis \& utilization variables & 0.006 & 0.006 & {}[-0.005, 0.017] & 0.270\\
\hspace{1em}PHQi9 vs demographic, prior self-harm variables and Charlson score & 0.013 & 0.006 & {}[0.001, 0.025] & 0.036\\
\hspace{1em}PHQi9 vs age & 0.003 & 0.007 & {}[-0.011, 0.017] & 0.636\\
\hspace{1em}PHQi9 vs age, sex, and race or ethnicity & 0.005 & 0.007 & {}[-0.009, 0.019] & 0.503\\
\addlinespace[0.3em]
\multicolumn{5}{l}{\textbf{Average (SL)}}\\
\hspace{1em}Age, sex, and prior self-harm vs no variables & 0.068 & 0.012 & {}[0.044, 0.091] & < 0.001\\
\hspace{1em}Charlson vs demographic, prior self-harm variables & -0.007 & 0.011 & {}[-0.029, 0.015] & 0.729\\
\hspace{1em}PHQi9 vs demographic, prior self-harm variables & 0.011 & 0.011 & {}[-0.011, 0.033] & 0.159\\
\hspace{1em}Charlson vs demographic, prior self-harm, and diagnosis \& utilization variables & 0.012 & 0.009 & {}[-0.007, 0.030] & 0.109\\
\hspace{1em}PHQi9 vs demographic, prior self-harm variables and Charlson score & 0.009 & 0.011 & {}[-0.012, 0.030] & 0.208\\
\hspace{1em}PHQi9 vs age & 0.037 & 0.012 & {}[0.014, 0.060] & < 0.001\\
\hspace{1em}PHQi9 vs age, sex, and race or ethnicity & 0.040 & 0.012 & {}[0.016, 0.063] & < 0.001\\
\addlinespace[0.3em]
\multicolumn{5}{l}{\textbf{Trend - slope (SL)}}\\
\hspace{1em}Age, sex, and prior self-harm vs no variables & -0.005 & 0.007 & {}[-0.018, 0.009] & 0.486\\
\hspace{1em}Charlson vs demographic, prior self-harm variables & 0.010 & 0.006 & {}[-0.003, 0.022] & 0.133\\
\hspace{1em}PHQi9 vs demographic, prior self-harm variables & 0.012 & 0.006 & {}[-0.000, 0.025] & 0.054\\
\hspace{1em}Charlson vs demographic, prior self-harm, and diagnosis \& utilization variables & 0.004 & 0.005 & {}[-0.007, 0.014] & 0.503\\
\hspace{1em}PHQi9 vs demographic, prior self-harm variables and Charlson score & 0.012 & 0.006 & {}[0.000, 0.024] & 0.050\\
\hspace{1em}PHQi9 vs age & 0.000 & 0.007 & {}[-0.013, 0.014] & 0.946\\
\hspace{1em}PHQi9 vs age, sex, and race or ethnicity & 0.000 & 0.007 & {}[-0.014, 0.014] & 0.997\\
\addlinespace[0.3em]
\multicolumn{5}{l}{\textbf{Average (XGB)}}\\
\hspace{1em}Age, sex, and prior self-harm vs no variables & 0.071 & 0.012 & {}[0.047, 0.095] & < 0.001\\
\hspace{1em}Charlson vs demographic, prior self-harm variables & -0.005 & 0.011 & {}[-0.027, 0.017] & 0.660\\
\hspace{1em}PHQi9 vs demographic, prior self-harm variables & 0.014 & 0.011 & {}[-0.007, 0.036] & 0.097\\
\hspace{1em}Charlson vs demographic, prior self-harm, and diagnosis \& utilization variables & 0.015 & 0.010 & {}[-0.005, 0.034] & 0.068\\
\hspace{1em}PHQi9 vs demographic, prior self-harm variables and Charlson score & 0.011 & 0.011 & {}[-0.010, 0.033] & 0.154\\
\hspace{1em}PHQi9 vs age & 0.046 & 0.012 & {}[0.021, 0.070] & < 0.001\\
\hspace{1em}PHQi9 vs age, sex, and race or ethnicity & 0.049 & 0.012 & {}[0.025, 0.073] & < 0.001\\
\addlinespace[0.3em]
\multicolumn{5}{l}{\textbf{Trend - slope (XGB)}}\\
\hspace{1em}Age, sex, and prior self-harm vs no variables & 0.000 & 0.007 & {}[-0.014, 0.013] & 0.932\\
\hspace{1em}Charlson vs demographic, prior self-harm variables & 0.011 & 0.006 & {}[-0.002, 0.023] & 0.086\\
\hspace{1em}PHQi9 vs demographic, prior self-harm variables & 0.014 & 0.006 & {}[0.001, 0.026] & 0.030\\
\hspace{1em}Charlson vs demographic, prior self-harm, and diagnosis \& utilization variables & 0.012 & 0.006 & {}[0.001, 0.023] & 0.032\\
\hspace{1em}PHQi9 vs demographic, prior self-harm variables and Charlson score & 0.014 & 0.006 & {}[0.002, 0.027] & 0.022\\
\hspace{1em}PHQi9 vs age & 0.002 & 0.007 & {}[-0.011, 0.016] & 0.736\\
\hspace{1em}PHQi9 vs age, sex, and race or ethnicity & 0.005 & 0.007 & {}[-0.009, 0.018] & 0.509\\*
\end{longtable}
\endgroup{}

\begingroup\fontsize{9}{11}\selectfont

\begin{longtable}[t]{lrrrr}
\caption{Estimates of the average VIM (defined using PPV) and slope of the linear trend in VIM over the time series,  considering the importance of PHQi9 or Charlson score when compared to other variables. Comparisons are: Age, sex, and prior self-harm vs no variables (comparing variable set 5 to 1), Charlson vs demographic and prior self-harm variables (comparing variable set 10 to 8), PHQi9 vs demographic and prior self-harm variables (comparing variable set 11 to 8), Charlson vs demographic, prior self-harm, and diagnosis \& utilization variables (comparing variable set 12 to 9), PHQi9 vs demographic, prior self-harm variables and Charlson score (comparing variable set 14 to 10), PHQi9 vs age (comparing variable set 18 to 16), PHQi9 vs age, sex, and race or ethnicity (comparing variable set 19 to 17). Estimates are shown for logistic regression (GLM), lasso (LASSO), random forests (RF), the discrete super learner (Discrete SL), super learner (SL), and boosted trees (XGB). \label{tab:ppv_data_analysis_vim_summaries_all_supp}}\\
\toprule
VIM type: comparison & Estimate & SE & 95\% CI & p-value\\
\midrule
\endfirsthead
\caption[]{Estimates of the average VIM (defined using PPV) and slope of the linear trend in VIM over the time series,  considering the importance of PHQi9 or Charlson score when compared to other variables. \textit{(continued)}}\\
\toprule
VIM type: comparison & Estimate & SE & 95\% CI & p-value\\
\midrule
\endhead

\endfoot
\bottomrule
\endlastfoot
\addlinespace[0.3em]
\multicolumn{5}{l}{\textbf{Average (GLM)}}\\
\hspace{1em}Age, sex, and prior self-harm vs no variables & 0.012 & 0.002 & {}[0.009, 0.016] & < 0.001\\
\hspace{1em}Charlson vs demographic, prior self-harm variables & 0.002 & 0.002 & {}[-0.002, 0.005] & 0.152\\
\hspace{1em}PHQi9 vs demographic, prior self-harm variables & 0.002 & 0.002 & {}[-0.001, 0.006] & 0.117\\
\hspace{1em}Charlson vs demographic, prior self-harm, and diagnosis \& utilization variables & 0.003 & 0.002 & {}[-0.001, 0.007] & 0.086\\
\hspace{1em}PHQi9 vs demographic, prior self-harm variables and Charlson score & 0.003 & 0.002 & {}[-0.000, 0.007] & 0.045\\
\hspace{1em}PHQi9 vs age & 0.012 & 0.001 & {}[0.009, 0.014] & < 0.001\\
\hspace{1em}PHQi9 vs age, sex, and race or ethnicity & 0.002 & 0.001 & {}[-0.000, 0.005] & 0.051\\
\addlinespace[0.3em]
\multicolumn{5}{l}{\textbf{Trend - slope (GLM)}}\\
\hspace{1em}Age, sex, and prior self-harm vs no variables & 0.000 & 0.001 & {}[-0.002, 0.002] & 0.992\\
\hspace{1em}Charlson vs demographic, prior self-harm variables & 0.002 & 0.001 & {}[-0.000, 0.004] & 0.062\\
\hspace{1em}PHQi9 vs demographic, prior self-harm variables & 0.002 & 0.001 & {}[-0.000, 0.004] & 0.054\\
\hspace{1em}Charlson vs demographic, prior self-harm, and diagnosis \& utilization variables & 0.000 & 0.001 & {}[-0.002, 0.002] & 0.973\\
\hspace{1em}PHQi9 vs demographic, prior self-harm variables and Charlson score & 0.003 & 0.001 & {}[0.001, 0.005] & 0.005\\
\hspace{1em}PHQi9 vs age & -0.002 & 0.001 & {}[-0.003, -0.000] & 0.043\\
\hspace{1em}PHQi9 vs age, sex, and race or ethnicity & -0.002 & 0.001 & {}[-0.003, -0.000] & 0.034\\
\addlinespace[0.3em]
\multicolumn{5}{l}{\textbf{Average (LASSO)}}\\
\hspace{1em}Age, sex, and prior self-harm vs no variables & 0.033 & 0.004 & {}[0.025, 0.040] & < 0.001\\
\hspace{1em}Charlson vs demographic, prior self-harm variables & 0.008 & 0.005 & {}[-0.001, 0.018] & 0.042\\
\hspace{1em}PHQi9 vs demographic, prior self-harm variables & 0.009 & 0.005 & {}[-0.001, 0.018] & 0.035\\
\hspace{1em}Charlson vs demographic, prior self-harm, and diagnosis \& utilization variables & 0.005 & 0.005 & {}[-0.005, 0.014] & 0.166\\
\hspace{1em}PHQi9 vs demographic, prior self-harm variables and Charlson score & 0.006 & 0.005 & {}[-0.004, 0.016] & 0.121\\
\hspace{1em}PHQi9 vs age & 0.013 & 0.002 & {}[0.009, 0.016] & < 0.001\\
\hspace{1em}PHQi9 vs age, sex, and race or ethnicity & 0.003 & 0.002 & {}[-0.001, 0.007] & 0.090\\
\addlinespace[0.3em]
\multicolumn{5}{l}{\textbf{Trend - slope (LASSO)}}\\
\hspace{1em}Age, sex, and prior self-harm vs no variables & 0.000 & 0.002 & {}[-0.005, 0.004] & 0.856\\
\hspace{1em}Charlson vs demographic, prior self-harm variables & 0.005 & 0.003 & {}[-0.001, 0.010] & 0.077\\
\hspace{1em}PHQi9 vs demographic, prior self-harm variables & 0.005 & 0.003 & {}[-0.000, 0.010] & 0.075\\
\hspace{1em}Charlson vs demographic, prior self-harm, and diagnosis \& utilization variables & 0.000 & 0.003 & {}[-0.005, 0.005] & 0.862\\
\hspace{1em}PHQi9 vs demographic, prior self-harm variables and Charlson score & 0.003 & 0.003 & {}[-0.003, 0.008] & 0.372\\
\hspace{1em}PHQi9 vs age & -0.002 & 0.001 & {}[-0.004, 0.000] & 0.085\\
\hspace{1em}PHQi9 vs age, sex, and race or ethnicity & -0.002 & 0.001 & {}[-0.005, 0.000] & 0.093\\
\addlinespace[0.3em]
\multicolumn{5}{l}{\textbf{Average (RF)}}\\
\hspace{1em}Age, sex, and prior self-harm vs no variables & 0.010 & 0.002 & {}[0.007, 0.014] & < 0.001\\
\hspace{1em}Charlson vs demographic, prior self-harm variables & -0.002 & 0.002 & {}[-0.006, 0.002] & 0.837\\
\hspace{1em}PHQi9 vs demographic, prior self-harm variables & 0.002 & 0.002 & {}[-0.002, 0.006] & 0.158\\
\hspace{1em}Charlson vs demographic, prior self-harm, and diagnosis \& utilization variables & 0.002 & 0.002 & {}[-0.002, 0.007] & 0.137\\
\hspace{1em}PHQi9 vs demographic, prior self-harm variables and Charlson score & 0.000 & 0.002 & {}[-0.003, 0.005] & 0.359\\
\hspace{1em}PHQi9 vs age & 0.005 & 0.002 & {}[0.002, 0.008] & 0.002\\
\hspace{1em}PHQi9 vs age, sex, and race or ethnicity & 0.004 & 0.002 & {}[0.000, 0.007] & 0.012\\
\addlinespace[0.3em]
\multicolumn{5}{l}{\textbf{Trend - slope (RF)}}\\
\hspace{1em}Age, sex, and prior self-harm vs no variables & 0.001 & 0.001 & {}[-0.001, 0.003] & 0.224\\
\hspace{1em}Charlson vs demographic, prior self-harm variables & 0.002 & 0.001 & {}[-0.000, 0.004] & 0.071\\
\hspace{1em}PHQi9 vs demographic, prior self-harm variables & 0.003 & 0.001 & {}[0.000, 0.005] & 0.027\\
\hspace{1em}Charlson vs demographic, prior self-harm, and diagnosis \& utilization variables & 0.001 & 0.001 & {}[-0.001, 0.004] & 0.275\\
\hspace{1em}PHQi9 vs demographic, prior self-harm variables and Charlson score & 0.002 & 0.001 & {}[-0.001, 0.004] & 0.142\\
\hspace{1em}PHQi9 vs age & 0.000 & 0.001 & {}[-0.002, 0.002] & 0.851\\
\hspace{1em}PHQi9 vs age, sex, and race or ethnicity & 0.000 & 0.001 & {}[-0.002, 0.002] & 0.818\\
\addlinespace[0.3em]
\multicolumn{5}{l}{\textbf{Average (Discrete SL)}}\\
\hspace{1em}Age, sex, and prior self-harm vs no variables & 0.011 & 0.002 & {}[0.008, 0.015] & < 0.001\\
\hspace{1em}Charlson vs demographic, prior self-harm variables & 0.002 & 0.002 & {}[-0.002, 0.006] & 0.224\\
\hspace{1em}PHQi9 vs demographic, prior self-harm variables & 0.004 & 0.002 & {}[-0.000, 0.008] & 0.030\\
\hspace{1em}Charlson vs demographic, prior self-harm, and diagnosis \& utilization variables & 0.000 & 0.002 & {}[-0.005, 0.004] & 0.535\\
\hspace{1em}PHQi9 vs demographic, prior self-harm variables and Charlson score & 0.003 & 0.002 & {}[-0.001, 0.007] & 0.049\\
\hspace{1em}PHQi9 vs age & 0.007 & 0.002 & {}[0.003, 0.011] & < 0.001\\
\hspace{1em}PHQi9 vs age, sex, and race or ethnicity & 0.006 & 0.002 & {}[0.003, 0.010] & < 0.001\\
\addlinespace[0.3em]
\multicolumn{5}{l}{\textbf{Trend - slope (Discrete SL)}}\\
\hspace{1em}Age, sex, and prior self-harm vs no variables & 0.000 & 0.001 & {}[-0.002, 0.002] & 0.810\\
\hspace{1em}Charlson vs demographic, prior self-harm variables & 0.001 & 0.001 & {}[-0.001, 0.003] & 0.399\\
\hspace{1em}PHQi9 vs demographic, prior self-harm variables & 0.000 & 0.001 & {}[-0.002, 0.003] & 0.743\\
\hspace{1em}Charlson vs demographic, prior self-harm, and diagnosis \& utilization variables & 0.000 & 0.001 & {}[-0.002, 0.004] & 0.561\\
\hspace{1em}PHQi9 vs demographic, prior self-harm variables and Charlson score & 0.000 & 0.001 & {}[-0.002, 0.003] & 0.761\\
\hspace{1em}PHQi9 vs age & 0.000 & 0.001 & {}[-0.002, 0.002] & 0.967\\
\hspace{1em}PHQi9 vs age, sex, and race or ethnicity & 0.000 & 0.001 & {}[-0.002, 0.002] & 0.699\\
\addlinespace[0.3em]
\multicolumn{5}{l}{\textbf{Average (SL)}}\\
\hspace{1em}Age, sex, and prior self-harm vs no variables & 0.009 & 0.002 & {}[0.005, 0.012] & < 0.001\\
\hspace{1em}Charlson vs demographic, prior self-harm variables & 0.000 & 0.002 & {}[-0.003, 0.005] & 0.323\\
\hspace{1em}PHQi9 vs demographic, prior self-harm variables & 0.003 & 0.002 & {}[-0.001, 0.007] & 0.091\\
\hspace{1em}Charlson vs demographic, prior self-harm, and diagnosis \& utilization variables & 0.000 & 0.002 & {}[-0.004, 0.005] & 0.401\\
\hspace{1em}PHQi9 vs demographic, prior self-harm variables and Charlson score & 0.003 & 0.002 & {}[-0.001, 0.007] & 0.078\\
\hspace{1em}PHQi9 vs age & 0.007 & 0.002 & {}[0.003, 0.010] & < 0.001\\
\hspace{1em}PHQi9 vs age, sex, and race or ethnicity & 0.005 & 0.002 & {}[0.002, 0.008] & 0.001\\
\addlinespace[0.3em]
\multicolumn{5}{l}{\textbf{Trend - slope (SL)}}\\
\hspace{1em}Age, sex, and prior self-harm vs no variables & 0.000 & 0.001 & {}[-0.001, 0.003] & 0.457\\
\hspace{1em}Charlson vs demographic, prior self-harm variables & 0.002 & 0.001 & {}[0.000, 0.005] & 0.049\\
\hspace{1em}PHQi9 vs demographic, prior self-harm variables & 0.002 & 0.001 & {}[-0.000, 0.005] & 0.065\\
\hspace{1em}Charlson vs demographic, prior self-harm, and diagnosis \& utilization variables & 0.001 & 0.001 & {}[-0.001, 0.004] & 0.372\\
\hspace{1em}PHQi9 vs demographic, prior self-harm variables and Charlson score & 0.001 & 0.001 & {}[-0.001, 0.004] & 0.230\\
\hspace{1em}PHQi9 vs age & 0.000 & 0.001 & {}[-0.002, 0.002] & 0.910\\
\hspace{1em}PHQi9 vs age, sex, and race or ethnicity & 0.000 & 0.001 & {}[-0.001, 0.002] & 0.683\\
\addlinespace[0.3em]
\multicolumn{5}{l}{\textbf{Average (XGB)}}\\
\hspace{1em}Age, sex, and prior self-harm vs no variables & 0.010 & 0.002 & {}[0.006, 0.013] & < 0.001\\
\hspace{1em}Charlson vs demographic, prior self-harm variables & 0.000 & 0.002 & {}[-0.004, 0.004] & 0.436\\
\hspace{1em}PHQi9 vs demographic, prior self-harm variables & 0.003 & 0.002 & {}[-0.001, 0.007] & 0.067\\
\hspace{1em}Charlson vs demographic, prior self-harm, and diagnosis \& utilization variables & 0.000 & 0.002 & {}[-0.004, 0.005] & 0.459\\
\hspace{1em}PHQi9 vs demographic, prior self-harm variables and Charlson score & 0.004 & 0.002 & {}[-0.000, 0.008] & 0.035\\
\hspace{1em}PHQi9 vs age & 0.006 & 0.002 & {}[0.002, 0.009] & < 0.001\\
\hspace{1em}PHQi9 vs age, sex, and race or ethnicity & 0.005 & 0.002 & {}[0.002, 0.009] & < 0.001\\
\addlinespace[0.3em]
\multicolumn{5}{l}{\textbf{Trend - slope (XGB)}}\\
\hspace{1em}Age, sex, and prior self-harm vs no variables & 0.000 & 0.001 & {}[-0.002, 0.002] & 0.667\\
\hspace{1em}Charlson vs demographic, prior self-harm variables & 0.001 & 0.001 & {}[-0.001, 0.004] & 0.268\\
\hspace{1em}PHQi9 vs demographic, prior self-harm variables & 0.001 & 0.001 & {}[-0.001, 0.004] & 0.318\\
\hspace{1em}Charlson vs demographic, prior self-harm, and diagnosis \& utilization variables & 0.000 & 0.001 & {}[-0.002, 0.003] & 0.535\\
\hspace{1em}PHQi9 vs demographic, prior self-harm variables and Charlson score & 0.000 & 0.001 & {}[-0.001, 0.003] & 0.442\\
\hspace{1em}PHQi9 vs age & 0.000 & 0.001 & {}[-0.003, 0.001] & 0.312\\
\hspace{1em}PHQi9 vs age, sex, and race or ethnicity & 0.000 & 0.001 & {}[-0.002, 0.002] & 0.830\\*
\end{longtable}
\endgroup{}

\begingroup\fontsize{9}{11}\selectfont

\begin{longtable}[t]{lrrrr}
\caption{Estimates of the average VIM (defined using sensitivity) and slope of the linear trend in VIM over the time series,  considering the importance of PHQi9 or Charlson score when compared to other variables. Comparisons are: Age, sex, and prior self-harm vs no variables (comparing variable set 5 to 1), Charlson vs demographic and prior self-harm variables (comparing variable set 10 to 8), PHQi9 vs demographic and prior self-harm variables (comparing variable set 11 to 8), Charlson vs demographic, prior self-harm, and diagnosis \& utilization variables (comparing variable set 12 to 9), PHQi9 vs demographic, prior self-harm variables and Charlson score (comparing variable set 14 to 10), PHQi9 vs age (comparing variable set 18 to 16), PHQi9 vs age, sex, and race or ethnicity (comparing variable set 19 to 17). Estimates are shown for logistic regression (GLM), lasso (LASSO), random forests (RF), the discrete super learner (Discrete SL), super learner (SL), and boosted trees (XGB). \label{tab:sensitivity_data_analysis_vim_summaries_all_supp}}\\
\toprule
VIM type: comparison & Estimate & SE & 95\% CI & p-value\\
\midrule
\endfirsthead
\caption[]{Estimates of the average VIM (defined using sensitivity) and slope of the linear trend in VIM over the time series,  considering the importance of PHQi9 or Charlson score when compared to other variables. \textit{(continued)}}\\
\toprule
VIM type: comparison & Estimate & SE & 95\% CI & p-value\\
\midrule
\endhead

\endfoot
\bottomrule
\endlastfoot
\addlinespace[0.3em]
\multicolumn{5}{l}{\textbf{Average (GLM)}}\\
\hspace{1em}Age, sex, and prior self-harm vs no variables & 0.130 & 0.016 & {}[0.099, 0.161] & < 0.001\\
\hspace{1em}Charlson vs demographic, prior self-harm variables & 0.027 & 0.018 & {}[-0.008, 0.062] & 0.065\\
\hspace{1em}PHQi9 vs demographic, prior self-harm variables & 0.031 & 0.018 & {}[-0.004, 0.066] & 0.043\\
\hspace{1em}Charlson vs demographic, prior self-harm, and diagnosis \& utilization variables & 0.035 & 0.019 & {}[-0.003, 0.072] & 0.035\\
\hspace{1em}PHQi9 vs demographic, prior self-harm variables and Charlson score & 0.041 & 0.018 & {}[0.006, 0.076] & 0.010\\
\hspace{1em}PHQi9 vs age & 0.106 & 0.012 & {}[0.083, 0.129] & < 0.001\\
\hspace{1em}PHQi9 vs age, sex, and race or ethnicity & 0.019 & 0.014 & {}[-0.008, 0.047] & 0.086\\
\addlinespace[0.3em]
\multicolumn{5}{l}{\textbf{Trend - slope (GLM)}}\\
\hspace{1em}Age, sex, and prior self-harm vs no variables & -0.002 & 0.009 & {}[-0.019, 0.016] & 0.867\\
\hspace{1em}Charlson vs demographic, prior self-harm variables & 0.019 & 0.010 & {}[-0.001, 0.039] & 0.062\\
\hspace{1em}PHQi9 vs demographic, prior self-harm variables & 0.020 & 0.010 & {}[-0.000, 0.040] & 0.054\\
\hspace{1em}Charlson vs demographic, prior self-harm, and diagnosis \& utilization variables & 0.004 & 0.011 & {}[-0.018, 0.026] & 0.711\\
\hspace{1em}PHQi9 vs demographic, prior self-harm variables and Charlson score & 0.032 & 0.010 & {}[0.012, 0.052] & 0.002\\
\hspace{1em}PHQi9 vs age & -0.021 & 0.007 & {}[-0.035, -0.008] & 0.002\\
\hspace{1em}PHQi9 vs age, sex, and race or ethnicity & -0.016 & 0.008 & {}[-0.032, -0.000] & 0.047\\
\addlinespace[0.3em]
\multicolumn{5}{l}{\textbf{Average (LASSO)}}\\
\hspace{1em}Age, sex, and prior self-harm vs no variables & 0.128 & 0.015 & {}[0.098, 0.158] & < 0.001\\
\hspace{1em}Charlson vs demographic, prior self-harm variables & 0.006 & 0.018 & {}[-0.029, 0.041] & 0.363\\
\hspace{1em}PHQi9 vs demographic, prior self-harm variables & 0.004 & 0.018 & {}[-0.031, 0.039] & 0.409\\
\hspace{1em}Charlson vs demographic, prior self-harm, and diagnosis \& utilization variables & 0.024 & 0.020 & {}[-0.015, 0.063] & 0.112\\
\hspace{1em}PHQi9 vs demographic, prior self-harm variables and Charlson score & -0.009 & 0.018 & {}[-0.044, 0.025] & 0.703\\
\hspace{1em}PHQi9 vs age & -0.001 & 0.010 & {}[-0.020, 0.018] & 0.545\\
\hspace{1em}PHQi9 vs age, sex, and race or ethnicity & 0.009 & 0.013 & {}[-0.017, 0.035] & 0.254\\
\addlinespace[0.3em]
\multicolumn{5}{l}{\textbf{Trend - slope (LASSO)}}\\
\hspace{1em}Age, sex, and prior self-harm vs no variables & 0.003 & 0.009 & {}[-0.014, 0.020] & 0.701\\
\hspace{1em}Charlson vs demographic, prior self-harm variables & 0.017 & 0.010 & {}[-0.003, 0.037] & 0.103\\
\hspace{1em}PHQi9 vs demographic, prior self-harm variables & 0.023 & 0.010 & {}[0.003, 0.043] & 0.026\\
\hspace{1em}Charlson vs demographic, prior self-harm, and diagnosis \& utilization variables & 0.017 & 0.012 & {}[-0.005, 0.040] & 0.132\\
\hspace{1em}PHQi9 vs demographic, prior self-harm variables and Charlson score & 0.010 & 0.010 & {}[-0.010, 0.030] & 0.317\\
\hspace{1em}PHQi9 vs age & 0.025 & 0.006 & {}[0.014, 0.036] & < 0.001\\
\hspace{1em}PHQi9 vs age, sex, and race or ethnicity & -0.020 & 0.007 & {}[-0.035, -0.005] & 0.008\\
\addlinespace[0.3em]
\multicolumn{5}{l}{\textbf{Average (RF)}}\\
\hspace{1em}Age, sex, and prior self-harm vs no variables & 0.124 & 0.017 & {}[0.090, 0.157] & < 0.001\\
\hspace{1em}Charlson vs demographic, prior self-harm variables & -0.008 & 0.019 & {}[-0.045, 0.028] & 0.672\\
\hspace{1em}PHQi9 vs demographic, prior self-harm variables & 0.033 & 0.019 & {}[-0.004, 0.070] & 0.041\\
\hspace{1em}Charlson vs demographic, prior self-harm, and diagnosis \& utilization variables & 0.039 & 0.020 & {}[0.001, 0.078] & 0.024\\
\hspace{1em}PHQi9 vs demographic, prior self-harm variables and Charlson score & 0.019 & 0.019 & {}[-0.018, 0.056] & 0.157\\
\hspace{1em}PHQi9 vs age & 0.051 & 0.016 & {}[0.021, 0.082] & < 0.001\\
\hspace{1em}PHQi9 vs age, sex, and race or ethnicity & 0.038 & 0.016 & {}[0.006, 0.070] & 0.009\\
\addlinespace[0.3em]
\multicolumn{5}{l}{\textbf{Trend - slope (RF)}}\\
\hspace{1em}Age, sex, and prior self-harm vs no variables & 0.018 & 0.009 & {}[-0.001, 0.036] & 0.061\\
\hspace{1em}Charlson vs demographic, prior self-harm variables & 0.020 & 0.011 & {}[-0.002, 0.041] & 0.070\\
\hspace{1em}PHQi9 vs demographic, prior self-harm variables & 0.028 & 0.011 & {}[0.006, 0.050] & 0.012\\
\hspace{1em}Charlson vs demographic, prior self-harm, and diagnosis \& utilization variables & 0.016 & 0.012 & {}[-0.007, 0.038] & 0.170\\
\hspace{1em}PHQi9 vs demographic, prior self-harm variables and Charlson score & 0.018 & 0.011 & {}[-0.004, 0.040] & 0.111\\
\hspace{1em}PHQi9 vs age & -0.003 & 0.009 & {}[-0.020, 0.014] & 0.736\\
\hspace{1em}PHQi9 vs age, sex, and race or ethnicity & -0.003 & 0.009 & {}[-0.020, 0.015] & 0.772\\
\addlinespace[0.3em]
\multicolumn{5}{l}{\textbf{Average (Discrete SL)}}\\
\hspace{1em}Age, sex, and prior self-harm vs no variables & 0.132 & 0.017 & {}[0.099, 0.165] & < 0.001\\
\hspace{1em}Charlson vs demographic, prior self-harm variables & 0.013 & 0.019 & {}[-0.023, 0.050] & 0.238\\
\hspace{1em}PHQi9 vs demographic, prior self-harm variables & 0.049 & 0.019 & {}[0.012, 0.086] & 0.005\\
\hspace{1em}Charlson vs demographic, prior self-harm, and diagnosis \& utilization variables & 0.017 & 0.020 & {}[-0.022, 0.056] & 0.198\\
\hspace{1em}PHQi9 vs demographic, prior self-harm variables and Charlson score & 0.045 & 0.019 & {}[0.008, 0.082] & 0.009\\
\hspace{1em}PHQi9 vs age & 0.066 & 0.016 & {}[0.035, 0.097] & < 0.001\\
\hspace{1em}PHQi9 vs age, sex, and race or ethnicity & 0.064 & 0.016 & {}[0.032, 0.096] & < 0.001\\
\addlinespace[0.3em]
\multicolumn{5}{l}{\textbf{Trend - slope (Discrete SL)}}\\
\hspace{1em}Age, sex, and prior self-harm vs no variables & 0.006 & 0.009 & {}[-0.013, 0.024] & 0.547\\
\hspace{1em}Charlson vs demographic, prior self-harm variables & 0.008 & 0.011 & {}[-0.013, 0.030] & 0.433\\
\hspace{1em}PHQi9 vs demographic, prior self-harm variables & 0.011 & 0.011 & {}[-0.010, 0.033] & 0.304\\
\hspace{1em}Charlson vs demographic, prior self-harm, and diagnosis \& utilization variables & 0.013 & 0.012 & {}[-0.009, 0.036] & 0.256\\
\hspace{1em}PHQi9 vs demographic, prior self-harm variables and Charlson score & 0.010 & 0.011 & {}[-0.012, 0.031] & 0.377\\
\hspace{1em}PHQi9 vs age & 0.000 & 0.009 & {}[-0.016, 0.018] & 0.928\\
\hspace{1em}PHQi9 vs age, sex, and race or ethnicity & -0.006 & 0.009 & {}[-0.024, 0.011] & 0.478\\
\addlinespace[0.3em]
\multicolumn{5}{l}{\textbf{Average (SL)}}\\
\hspace{1em}Age, sex, and prior self-harm vs no variables & 0.098 & 0.017 & {}[0.064, 0.132] & < 0.001\\
\hspace{1em}Charlson vs demographic, prior self-harm variables & 0.016 & 0.019 & {}[-0.021, 0.053] & 0.203\\
\hspace{1em}PHQi9 vs demographic, prior self-harm variables & 0.039 & 0.019 & {}[0.001, 0.076] & 0.021\\
\hspace{1em}Charlson vs demographic, prior self-harm, and diagnosis \& utilization variables & 0.014 & 0.020 & {}[-0.026, 0.054] & 0.251\\
\hspace{1em}PHQi9 vs demographic, prior self-harm variables and Charlson score & 0.041 & 0.019 & {}[0.004, 0.079] & 0.015\\
\hspace{1em}PHQi9 vs age & 0.073 & 0.016 & {}[0.041, 0.104] & < 0.001\\
\hspace{1em}PHQi9 vs age, sex, and race or ethnicity & 0.050 & 0.017 & {}[0.017, 0.083] & 0.002\\
\addlinespace[0.3em]
\multicolumn{5}{l}{\textbf{Trend - slope (SL)}}\\
\hspace{1em}Age, sex, and prior self-harm vs no variables & 0.009 & 0.010 & {}[-0.010, 0.028] & 0.376\\
\hspace{1em}Charlson vs demographic, prior self-harm variables & 0.020 & 0.011 & {}[-0.002, 0.041] & 0.072\\
\hspace{1em}PHQi9 vs demographic, prior self-harm variables & 0.022 & 0.011 & {}[0.001, 0.044] & 0.044\\
\hspace{1em}Charlson vs demographic, prior self-harm, and diagnosis \& utilization variables & 0.012 & 0.012 & {}[-0.012, 0.035] & 0.330\\
\hspace{1em}PHQi9 vs demographic, prior self-harm variables and Charlson score & 0.013 & 0.011 & {}[-0.008, 0.035] & 0.229\\
\hspace{1em}PHQi9 vs age & -0.005 & 0.009 & {}[-0.023, 0.013] & 0.606\\
\hspace{1em}PHQi9 vs age, sex, and race or ethnicity & 0.002 & 0.009 & {}[-0.017, 0.020] & 0.841\\
\addlinespace[0.3em]
\multicolumn{5}{l}{\textbf{Average (XGB)}}\\
\hspace{1em}Age, sex, and prior self-harm vs no variables & 0.120 & 0.017 & {}[0.086, 0.154] & < 0.001\\
\hspace{1em}Charlson vs demographic, prior self-harm variables & 0.017 & 0.019 & {}[-0.019, 0.054] & 0.177\\
\hspace{1em}PHQi9 vs demographic, prior self-harm variables & 0.049 & 0.019 & {}[0.011, 0.086] & 0.005\\
\hspace{1em}Charlson vs demographic, prior self-harm, and diagnosis \& utilization variables & 0.014 & 0.020 & {}[-0.025, 0.052] & 0.246\\
\hspace{1em}PHQi9 vs demographic, prior self-harm variables and Charlson score & 0.056 & 0.019 & {}[0.019, 0.093] & 0.002\\
\hspace{1em}PHQi9 vs age & 0.051 & 0.016 & {}[0.020, 0.082] & < 0.001\\
\hspace{1em}PHQi9 vs age, sex, and race or ethnicity & 0.055 & 0.017 & {}[0.023, 0.088] & < 0.001\\
\addlinespace[0.3em]
\multicolumn{5}{l}{\textbf{Trend - slope (XGB)}}\\
\hspace{1em}Age, sex, and prior self-harm vs no variables & 0.011 & 0.010 & {}[-0.008, 0.030] & 0.266\\
\hspace{1em}Charlson vs demographic, prior self-harm variables & 0.013 & 0.011 & {}[-0.009, 0.034] & 0.251\\
\hspace{1em}PHQi9 vs demographic, prior self-harm variables & 0.014 & 0.011 & {}[-0.008, 0.036] & 0.200\\
\hspace{1em}Charlson vs demographic, prior self-harm, and diagnosis \& utilization variables & 0.009 & 0.011 & {}[-0.014, 0.031] & 0.439\\
\hspace{1em}PHQi9 vs demographic, prior self-harm variables and Charlson score & 0.011 & 0.011 & {}[-0.011, 0.033] & 0.319\\
\hspace{1em}PHQi9 vs age & -0.006 & 0.009 & {}[-0.024, 0.011] & 0.474\\
\hspace{1em}PHQi9 vs age, sex, and race or ethnicity & -0.002 & 0.009 & {}[-0.020, 0.016] & 0.802\\*
\end{longtable}
\endgroup{}

\end{document}